\documentclass[fleqn, usenatbib]{mnras} 
\usepackage{newtxtext, newtxmath} 	% Required for MNRAS 
\usepackage[T1]{fontenc} 
\usepackage{graphicx}
\usepackage{amsmath}
\usepackage{mathtools}
\usepackage{multicol}
\usepackage{bm}
\usepackage{pdflscape}
\usepackage{natbib}
\usepackage[section]{placeins}
\usepackage{lipsum}
\usepackage{etoolbox} 
\usepackage{tabularx} 
\usepackage{xcolor} 
\usepackage[toc, page]{appendix} 
\usepackage{subfiles} 
\usepackage{geometry} 
\hypersetup{ % all blue links 
	colorlinks		= true, 
	urlcolor 		= blue, 
	linkcolor 		= blue, 
	citecolor 		= blue 
} 

\newcommand{\ddfrac}[2]{\frac{\displaystyle #1}{\displaystyle #2}} 
\newcommand{\refp}[1]{(\ref{#1})} 
\newcommand{\msun}{\ensuremath{\text{M}_\odot}} 
\newcommand{\persqkpc}{\ensuremath{\text{kpc}^{-2}}} 
\newcommand{\rgal}{\ensuremath{R_\text{gal}}} 
\newcommand{\absz}{\ensuremath{\left|z\right|}} 
\newcommand{\afe}{[$\alpha$/Fe]} 
\newcommand{\ofe}{[O/Fe]} 
\newcommand{\feh}{[Fe/H]} 
\newcommand{\oh}{[O/H]} 
\newcommand{\hsim}{\texttt{h277}} 
\newcommand{\vice}{\texttt{VICE}} 

\providecommand{\noopsort}[1]{} 

\defcitealias{Kroupa2001}{Kroupa} 
\defcitealias{Salpeter1955}{Salpeter} 

\title[Stellar Migration and Chemical Enrichment]{Stellar Migration and 
Chemical Enrichment in the Milky Way Disc: A Hybrid Model} 

\author[J.W. Johnson et al.]{James W. Johnson,$^{1}$\thanks{
	Contanct e-mail: \href{mailto:
	johnson.7419@osu.edu}{johnson.7419@osu.edu}} 
	David H. Weinberg,$^{1, 2, 3}$ 
	Fiorenzo Vincenzo,$^{1, 2}$ 
	Jonathan C. Bird,$^{4}$ 
	\newauthor 
	Sarah R. Loebman,$^{5}$ 
	Alyson M. Brooks,$^{6}$ 
	Thomas R. Quinn,$^{7}$ 
	Charlotte R. Christensen,$^{8}$ 
	\newauthor 
	and Emily J. Griffith$^{1}$ 
	\\ \null \\ 
	$^{1}$ Department of Astronomy, The Ohio State University, 
	140 W. 18th Ave., Columbus, OH, 43210, USA 
	\\ 
	$^{2}$ Center for Cosmology and Astroparticle Physics (CCAPP), 
	The Ohio State University, 191 W. Woodruff Ave., Columbus, OH, 43210, USA 
	\\ 
	$^{3}$ Institute for Advanced Study, 1 Einstein Dr., Princeton, NJ, 08540, 
	USA 
	\\ 
	$^{4}$ Department of Physics \& Astronomy, Vanderbilt University, 
	2301 Vanderbilt Place, Nashville, TN, 37235, USA 
	\\ 
	$^{5}$ Department of Physics, University of California Merced, 
	5200 North Lake Rd., Merced, CA, 95343, USA 
	\\ 
	$^{6}$ Department of Physics \& Astronomy, Rutgers University, 136 
	Frelinghuysen Rd., Piscataway, NJ, 08854, USA 
	\\ 
	$^{7}$ Department of Astronomy, University of Washington, Box 351580, 
	Seattle, WA, 98195, USA 
	\\ 
	$^{8}$ Department of Physics, Grinnell College, 1116 8th Ave., Grinnell, 
	IA, 50112, USA 
} 

\date{Accepted XXX; Received YYY; in original form ZZZ} 
\pubyear{2021} 

\begin{document} 
\label{firstpage} 
\pagerange{\pageref{firstpage}--\pageref{lastpage}} 
\maketitle

\begin{abstract} 
We develop a hybrid model of galactic chemical evolution that combines a 
multi-ring computation of chemical enrichment with a prescription for stellar 
migration and the vertical distribution of stellar populations informed by a 
cosmological hydrodynamic disc galaxy simulation. 
Our fiducial model adopts empirically motivated forms of the star formation law 
and star formation history, with a gradient in outflow mass loading tuned to 
reproduce the observed metallicity gradient. 
With this approach, the model reproduces many of the striking qualitative 
features of the Milky Way disc's abundance structure: 
(i) the dependence of the [O/Fe]-[Fe/H] distribution on radius~\rgal~and 
midplane distance~\absz; 
(ii) the changing shapes of the [O/H] and [Fe/H] distributions 
with~\rgal~and~\absz; 
(iii) a broad distribution of [O/Fe] at sub-solar metallicity and changes in 
the [O/Fe] distribution with~\rgal,~\absz, and [Fe/H]; 
(iv) a tight correlation between [O/Fe] and stellar age for [O/Fe]~$>$~0.1; 
(v) a population of young and intermediate-age~$\alpha$-enhanced stars caused 
by migration-induced variability in the Type Ia supernova rate; 
(vi) non-monotonic age-[O/H] and age-[Fe/H] relations, with large scatter and a 
median age of~$\sim$4 Gyr near solar metallicity. 
Observationally motivated models with an enhanced star formation rate~$\sim$2 
Gyr ago improve agreement with the observed age-[Fe/H] and age-[O/H] relations, 
but worsen agreement with the observed age-[O/Fe] relation. 
None of our models predict an [O/Fe] distribution with the distinct bimodality 
seen in the observations, suggesting that more dramatic evolutionary pathways 
are required. 
All code and tables used for our models are publicly available through the 
\texttt{Versatile Integrator for Chemical Evolution} (\texttt{VICE}; 
\url{https://pypi.org/project/vice}). 
\end{abstract}


\section{Introduction} 
\label{sec:intro} 

The orbits of stars are not fixed. The considerable intrinsic scatter in 
age-abundance relations of local disc stars~\citep{Edvardsson1993} and the high 
metallicity of the sun relative to nearby stars of similar age 
\citep*{Wielen1996} provided early evidence that stars in the Milky Way disc 
can migrate several kiloparsecs from the Galactocentric radius at which they 
formed. 
Interest in radial migration as an important element of galactic chemical 
evolution (GCE) grew further with the demonstration by~\citet{Sellwood2002} 
that resonant interactions with transient spiral perturbations could 
change stars' orbital guiding centre radii without increasing orbital 
eccentricity, and with subsequent studies showing ubiquitous radial migration 
in numerical simulations of disc galaxies (e.g.~\citealp{Roskar2008a, 
Roskar2008b, Loebman2011, Minchev2011};~\citealp*{Bird2012}; 
\citealp{Bird2013};~\citealp*{Grand2012a, Grand2012b, Kubryk2013}). 
\citet{Schoenrich2009a,Schoenrich2009b} developed the first detailed GCE 
models incorporating radial migration, describing it with a flexible, 
dynamically motivated parameterization constrained simultaneously with other 
GCE parameters when fitting to observations. A number of subsequent studies 
have incorporated radial migration using similar analytic or parameterized 
models (e.g.~\citealp{Bilitewski2012, Hayden2015};~\citealp*{Kubryk2015a, 
Kubryk2015b};~\citealp{Feuillet2018};~\citealp*{Sharma2020}), and 
\citet{Frankel2018, Frankel2019, Frankel2020} have used stellar abundances, 
ages, and kinematics to constrain radial migration empirically. 
\par 
In this paper we construct evolutionary models for the Milky Way disc that 
combine a classic multi-ring GCE approach (e.g.,~\citealp{Matteucci1989, 
Wyse1989, Prantzos1995}) with stellar migration predicted by a 
hydrodynamic simulation of disc galaxy formation from cosmological initial 
conditions. 
% Our methodology is similar to that of~\citet*{Minchev2013, Minchev2014}, but 
% the logic runs in the opposite direction of~\citet{Buck2021} who recently 
% applied chemical evolution models with~\texttt{Chempy}~\citep*{Rybizki2017} to 
% the N-body+SPH code~\texttt{Gasoline2}~\citep{Wadsley2017}. 
Our methodology is similar to that of~\citet*{Minchev2013, Minchev2014} and has 
similar motivations. 
% \citet{Buck2021} recently explored variations of chemical evolution model 
% assumptions in hydrodynamical simulations by applying~\texttt{Chempy} 
% \citep*{Rybizki2017} to the N-body+SPH code~\texttt{Gasoline2} 
% \citep{Wadsley2017}. 
The use of a cosmological simulation that agrees with many observed properties 
of the Milky Way assures that our stellar migration scenario is physically 
plausible, including any correlations of migration in time and space that might 
be difficult to capture in a parameterized description. 
Most hydrodynamic cosmological simulations include metal enrichment, and direct 
comparison between the predicted and observed abundance patterns can provide 
valuable insights into the accuracy of the simulations and the possible origin 
of the observed element structure (e.g., ~\citealt{Mackereth2017, Grand2018, 
Buck2020b, Vincenzo2020, Buck2021}). 
\par 
However, many ingredients of the simulations' enrichment recipes are 
uncertain, and metal transport and mixing within the interstellar medium (ISM) 
are sensitive to numerical resolution and to details of the hydrodynamics and 
star formation algorithms. 
Our hybrid approach allows us to consider many 
choices of uncertain GCE parameters, tuning them to reproduce some 
observations while leaving others as independent empirical tests. 
This flexible approach also allows us to isolate the impact of different 
GCE model ingredients and to zero-in on the ways that stellar migration 
influences the predicted chemical evolution. 
In exchange for this exploratory freedom, the hybrid model is not fully 
self-consistent, instead adopting its own accretion, star formation, and 
outflow histories rather than the simulation's. Although our methodology can 
be applied with any given choice of cosmological simulation, we make use of 
only one in the present paper. 
One could however apply the same method to multiple simulations to 
predict a statistical distribution of outcomes. 
\par 
We focus our predictions and observational comparisons on oxygen, a 
representative~$\alpha$-element producted almost exclusively by core 
collapse supernovae (CCSN), and iron, which at solar abundances has roughly 
equal contributions from CCSN and Type Ia supernovae (SN Ia). We will consider 
other elements with other nucleosynthetic sources in future work, but 
observed trends of~\feh~and~\afe\footnote{
	We follow standard notation where $[X/Y] = \log_{10}(X/Y) - 
	\log_{10}(X/Y)_\odot$. Different observational studies use different 
	$\alpha$-elements (or combinations thereof) in abundance ratios, and we 
	will generally use~\ofe~and~\afe~synonymously. 
} in the Milky Way disc already display a number of striking features, 
including: 
\begin{itemize} 
	\item At sub-solar~\feh, the distribution of~\afe~is bimodal, with 
	``high-$\alpha$'' and ``low-$\alpha$'' sequences typically separated by 
	0.1-0.4 dex (e.g., 
	\citealp{Fuhrmann1998};~\citealp*{Bensby2003};~\citealp{Adibekyan2012, 
	Vincenzo2021a}). 

	\item The location of the high-$\alpha$ and low-$\alpha$ sequences is 
	nearly independent of position in the disc, but the relative number of 
	stars in these sequences and the distributions of those stars in 
    \feh~changes systematically with Galactocentric radius~$\rgal$~and midplane 
	distance~$\absz$~\citep{Nidever2014, Hayden2015, Weinberg2019}. 

	\item In addition to an overall radial gradient, the shape of the 
	\feh~distribution for stars with~$\absz < 0.5$ kpc changes from negatively 
	skewed in the inner disc to roughly symmetric at the solar neighbourhood 
	to positively skewed in the outer disc~\citep{Hayden2015, Weinberg2019}. 

	\item With increasing~$\absz$,~\feh~distributions become more symmetric 
	and less dependent on~$\rgal$~\citep{Hayden2015}. 

	\item The age-metallicity relation (AMR) is broad, with a wide range of 
	\feh~at fixed stellar age and vice versa in the solar neighbourhood 
	\citep{Edvardsson1993} and beyond~\citep{Feuillet2019}. The trend of 
	median age with~\feh~or~\oh~is non-monotonic, with solar metallicity 
	stars being younger on average than both metal-poor~\textit{and} 
	metal-rich stars~\citep{Feuillet2018, Feuillet2019, Lu2021}. 

	\item The trend of stellar age with~\afe~is much tighter than the trend 
	with~\feh, becoming broad near~\afe~$\approx$~0~\citep{Feuillet2018, 
	Feuillet2019}. Although most stars with~\afe~$\geq$~0.1 are old, 
	observations have revealed a significant population of~$\alpha$-rich stars 
	that appear to be young or intermediate age~\citep{Chiappini2015, 
	Martig2015, Martig2016, Warfield2021}. Some of these stars may have been 
	``rejuvinated'' by stellar mergers or mass-transfer events 
	\citep{Izzard2018, SilvaAguirre2018}, and the question of what fraction 
	are truly much younger than the median age-\afe~relation remains open 
	\citep{Hekker2019}. 
\end{itemize} 
Many of these results have emerged most clearly from the Apache Point 
Observatory Galaxy Evolution Experiment (APOGEE;~\citealp{Majewski2017}) of 
the Sloan Digital Sky Survey (SDSS-III:~\citealp{Eisenstein2011}; 
SDSS-IV:~\citealp{Blanton2017}), sometimes confirming and extending trends 
suggested by earlier observational data. We will assess the degree to which 
models with fairly conventional GCE assumptions coupled to simulation-based 
radial and vertical migration of stars can explain, or fail to explain, these 
observations. 
\par 
Relative to~\citet{Minchev2013, Minchev2014}, our base GCE model has many 
differences of detail, the most important being our inclusion of outflows, 
which is in turn connected to our different choice of oxygen and iron yields. 
Our simulation, the galaxy~\hsim~from the~\citet{Christensen2012} 
suite evolved with the N-body+SPH code~\texttt{GASOLINE}~\citep{Wadsley2004}, 
is fully cosmological, while the simulation used by~\citet{Minchev2013, 
Minchev2014} has a more idealized geometry with merger and accretion history 
drawn from a larger cosmological volume~\citep{Martig2012}. 
The~\citet{Minchev2013, Minchev2014} simulation has a fairly strong, 
long-lived bar while~\hsim~has only a weak, transient bar, and this difference 
could have some impact on radial migration. 
Another methodological difference, which turns out to be important for some 
observables, is that we track enrichment from stellar populations as they 
migrate (see~\S~\ref{sec:obs_comp:age_alpha} below), while~\citet{Minchev2013, 
Minchev2014} assume that populations enrich only the radial zone in which they 
were born. 
\par 
Previous studies have shown that~\hsim~and other disc galaxies evolved with 
similar physics have realistic rotation curves~\citep{Governato2012, 
Christensen2014a, Christensen2014b}, stellar mass~\citep{Munshi2013}, 
metallicity~\citep{Christensen2016}, dwarf satellite populations 
\citep{Zolotov2012, Brooks2014}, and HI properties~\citep{Brooks2017}. 
Most directly relevant to this study,~\citet{Bird2020} demonstrate that 
\hsim~accurately reproduces the observed relation between stellar age and 
vertical velocity dispersion~$\sigma_z$. 
This relation arises as a consequence of ``upside-down'' disc formation in 
which the star-forming gas layer becomes thinner with time as well as the 
dynamical heating of stars as they age (\citealp*{Bournaud2009a, 
Bournaud2009b, Forbes2012};~\citealp{Bird2013};~\citealp*{Vincenzo2019}; 
\citealp{Yu2021}). 
\citet{Schoenrich2009a} distinguish between the radial mixing caused by 
``blurring'' of stars on moderately eccentric orbits and ``churning'' that 
changes the guiding centre radii of their orbits. 
Both phenomena occur in our simulation and we do not attempt to separate them, 
simply using the terms ``migration'' or ``mixing'' to refer to the combined 
effect. 
In addition to radial migration, we use the~\hsim~predictions for the vertical 
locations (i.e., midplane distances~$\absz$) of stars at the present day. 
Our GCE model assumes that the gas disc is vertically well mixed, so a stellar 
population's birth abundances depend only on~$\rgal$~and time. 
Vertical gradients arise because older populations have larger~$\sigma_z$~and 
thus larger average~$\absz$, and also because radial migration is 
coupled to changes in~$\sigma_z$~\citep*{Solway2012}. 
The good match to the observed age-velocity relation found by~\citet{Bird2020} 
allows us to use vertical trends of abundance ditributions as a further test 
of our chemical evolution model. 
\par 
We describe the~\hsim simulation further in~\S~\ref{sec:methods:h277} and our 
implementation of radial migration in~\S~\ref{sec:methods:migration}. 
We describe the base GCE model in~\S\S~\ref{sec:methods:sfhs} - 
\ref{sec:methods:summary}. Distinctive features of our GCE model are the use 
of a radially dependent outflow mass loading~$\eta(\rgal)$~to tune the 
metallicity gradient, our implementation of a star formation law motivated by 
spatially resolved studies of nearby galaxies and high-redshift studies of its 
time-dependence, and our use of mean radial age trends of disc galaxies to set 
the radial dependence of the star formation history (SFH). 
Our fiducial model adopts a smooth SFH with an ``inside-out'' radial trend in 
which star formation proceeds more rapidly in the inner Galaxy. 
Motivated by the observational analyses of~\citet{Isern2019} and 
\citet{Mor2019}, we also consider models with a burst of star formation 
centred~$\approx$~2 Gyr in the past, similar to the one-zone models 
investigated by~\citet{Johnson2020}. 
Other authors have suggested multiple bursts in the Milky Way's SFH 
(e.g.,~\citealp*{Lian2020a, Lian2020b};~\citealp{RuizLara2020, 
Sysoliatina2021}), perhaps triggered by satellite interactions, while others 
have advocated a two-phase SFH to explain the~\afe dichotomy (e.g., 
\citealp*{Chiappini1997};~\citealp{Haywood2016, Spitoni2019a, Buck2020a, 
Khoperskov2021}). We do not investigate these more complex SFHs here, but we 
plan to do so in future work.

\section{Methods} 
\label{sec:methods} 
To fulfill the goals of this paper, we develop and make use of newly released 
features within the~\texttt{Versatile Integrator for Chemical Evolution} 
(\vice;~\citealp{Johnson2020}), an open-source~\texttt{python} package 
available for~\texttt{Unix} system architectures (for further details, see 
Appendix~\ref{sec:vice}). 
These features are designed to handle models such as these with a wide range of 
flexibility. 
We reserve a description of~\vice's migration algorithm and our simulation 
parameters for~\S~\ref{sec:methods:migration}, first describing our sample of 
star particles from the hydrodynamic simulation. 
\par 
Although we make use of a hydrodynamic simulation to drive stellar migration, 
we do~\textit{not} make use of its SFH. 
While our study employs similar techniques as~\citet{Minchev2013}, ours differs 
slightly in that they take a single SFH which is similar, but not identical to 
that of their N-body+SPH simulation; in this paper we present a handful of 
assumptions about the SFH, which do not necessarily resemble that of~\hsim. 
An alternative to modeling radial migration based on a 
hydrodynamical simulation is to invoke dynamical arguments; 
\citet{Schoenrich2009a} and~\citet{Sharma2020} take such an approach. 
This method, however, introduces free parameters which then require fitting to 
data. 
An advantage of our technique is that there are no free parameters describing 
radial migration introduced to the model; it is unclear how much 
fitting radial migration parameters could bias the model into agreement with 
parts of the data not involved in the fitting process. 
Instead we adopt a physically motivated migration model evolved from 
cosmological initial conditions. 
We rely on a single realization of this numerical model and are thus subject to 
the differences between the dynamical history of this simulated galaxy and that 
of the Milky Way. 
However, one could compare the predictions made by our chemical evolution 
models when applied to different hydrodynamical simulations, a direction we 
plan to pursue in future work. 
We emphasize that the hydrodynamical simulation~\textit{only} informs the 
mixing processes in our models, and there is no N-body integration involved in 
our models aside from that which was used to run the~\hsim~simulation in the 
first place. 

\subsection{The Hydrodynamical Simulation} 
\label{sec:methods:h277} 
In this paper we make use of star particles from the~\texttt{h277} simulation 
\citep{Christensen2012, Zolotov2012, Loebman2012, Loebman2014, Brooks2014}. 
Recently employed to study the stellar age-velocity relationship, a synopsis of 
its detailed simulation parameters and cosmological model can be found in~\S~2 
of~\citet{Bird2020}. 
We do not repeat these details here, instead focusing on how we vet the sample 
of star particles for use in our chemical evolution models. 
\par 
The parameters of stars that we use in our analysis are the birth and final 
radii and the final midplane distance. 
\hsim~did not record the exact birth radius of each star particle; however, 
each star particle does have an accurate age at each snapshot. 
The orbital radii of stars that are sufficiently young in their first snapshot 
should be good approximations of their birth radii. 
We therefore restrict our sample to those star particles with an age at first 
snapshot of~$\leq$~150 Myr, and we adopt their Galactocentric radius at that 
time as their birth radius. 
We have repeated our analysis with a maximum age at first snapshot of 50 
Myr and found similar results, indicating that these time intervals are short 
enough to not impact our conclusions. 
We adopt the 150 Myr interval because it provides a larger number of star 
particles to sample from. 
Of the star particles that remain after imposing this cut, the oldest has an 
age of 13.23 Gyr at the present day (i.e. at the simulation's final output). 
Our GCE model can only apply on timescales as long as or shorter than the full 
range of ages of the sample of star particles; we therefore subtract 0.5 Gyr 
from the formation times of all star particles, allowing~$T$~= 0 in our models 
to correspond to~$T$~= 0.5 Gyr in~\hsim. 
As a consequence, our disc models trace the chemical evolution of the Galaxy 
out to a lookback time of~$\sim$13.2 Gyr, or a redshift of~$z \approx$~9, 
placing the onset of star formation at that time. 
\par 
We further restrict our sample of star particles to only those with both 
formation and final radii of~$\rgal \leq$~20 kpc, and to have formed 
within~$\absz\leq$~3 kpc of the disc midplane. 
These criteria are intended to restrict our sample to star particles that 
formed within the disc and can therefore be described by a disc GCE model. 
While it is possible that some star particles formed in a dwarf galaxy happen 
to satisfy our geometrical cuts at the star's formation time, these particles 
are few in number, and are only relevant at large~$\rgal$ and high 
ages: a region of parameter space where few stars form anyway. 
\par 
Based on a kinematic decomposition performed on the present-day phase space 
distribution of the~\texttt{h277} star particles conducted with the 
\texttt{analysis.decomp} routine within the~\texttt{pynbody} package 
\citep{pynbody}, we classify each star particle as having thin disc, thick 
disc, bulge, pseudobulge, or halo-like kinematics.
Details on the decomposition process can be found in~\citet{Brook2012} 
and~\citet{Bird2013}. 
We include the entire sample in~\vice's public code base, but only make use of 
those with disc-like kinematics in this paper. 
Our geometric selection yields 3,152,211 star particles in total, 1,751,765 of 
which have disc-like kinematics and are included in our sample. 
% Based on a kinematic decomposition performed on the present-day phase space 
% distribution of the~\texttt{h277} star particles, we include all those with 
% bulge, pseudobulge, and disc-like kinematics, excluding halo stars. 
% While we are not modeling the evolution of the bulge here, these star particles 
% are overwhelmingly located at~$\rgal\leq$~3 kpc at the present day 
% and therefore do not enter our analysis and observational comparisons below. 
% These cuts yield a sample of 3,102,519 star particles from~\hsim. 
\par 
\hsim~had a weak and transient bar during its evolution, but it does not have 
one at~$z$~= 0. 
This is a noteworthy difference between our model and that of 
\citet{Minchev2013}, and by extension the~\citet{Minchev2014, Minchev2017} 
studies as well, because they selected a hydrodynamic simulation of a galaxy 
specifically so that it would have a strong bar at~$z$~= 0. 
This could mean that the dynamical history of our model Galaxy differs from 
that of~\citet{Minchev2013} and perhaps the Milky Way itself. 
However, the difference is likely within the uncertainties of the current 
understanding of the Milky Way's dynamical history. 
Although an investigation of the impact of bar evolution on stellar migration 
and thus chemical evolution is outside the scope of this paper, it is an 
interesting question that can be probed by simply swapping the~\hsim~data 
within~\vice~for another simulation, then rerunning our numerical models and 
comparing the results. 

% fig 1 
\begin{figure*} 
\centering 
\includegraphics[scale = 0.32]{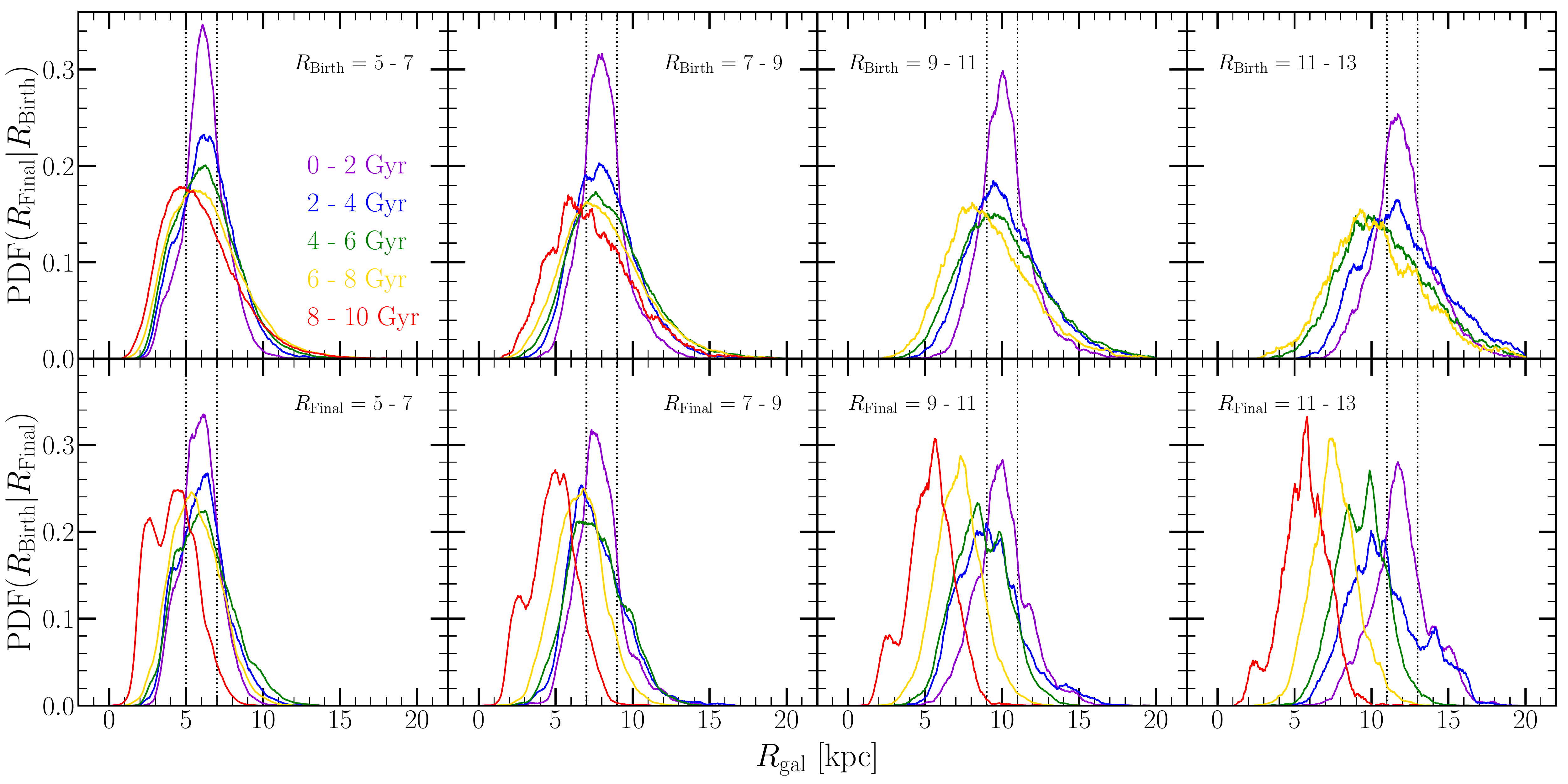} 
\caption{Radial distributions of our sample of star particles from 
\texttt{h277}. In the top row, we show distributions of~\textit{final} radius 
in bins of birth radius and age, and in the bottom row, we show distributions 
of~\textit{birth} radius in bins of final radius and age. Each bin in 
Galactocentric radius is shown in its own panel, denoted in text at the top of 
each panel and by vertical black dashed lines. We colour-code the distributions 
according to the age of the star particles, denoted by the legend in the upper 
left panel. We smooth all distributions with a box-car width of 0.5 kpc to 
improve visual clarity. We omit the distributions for 8 - 10 Gyr old stars born 
in the 9 - 11 and 11 - 13 kpc bins due to an insufficient number of star 
particles with which to calculate the distribution. } 
\label{fig:h277_decomposition} 
\end{figure*} 

In the top row of Fig.~\ref{fig:h277_decomposition}, we plot the distributions 
of final radius in bins of birth radius and age for our sample of star 
particles. 
Conversely, the bottom row shows distributions of birth radii in bins of final 
radius and age. 
Focusing on the top row of panels, we note that for star particles born at any 
radius and time, the distribution of final radius is still peaked near the 
birth radius, but the peak moves slightly inward with increasing age. 
The tails of the distributions toward larger~$\rgal$ are nearly 
age-independent, while the tails toward smaller~$\rgal$ are not. 
This suggests that radial migration inward and outward occur on different 
timescales in~\texttt{h277}, specifically that inward migration is slower than 
outward migration. 
By extension, this suggests that the two may be tied to different physical 
processes. 
Alternatively, it may simply be that stars that migrate to the outer Galaxy are 
no longer subject to strong dynamical perturbations while stars that move 
inward can still experience strong orbital disturbances. 
\par 
Focusing on the bottom row of panels in Fig.~\ref{fig:h277_decomposition}, we 
note that the modes of the birth radius distributions show a much stronger 
dependence on age than the modes of the final radius distributions. 
At any Galactocentric radius at the present day, the youngest stars are 
overwhelmingly born at comparable radii, while the oldest stars are 
overwhelmingly born at smaller radii. 
This trend is most noticeable at large~$\rgal$. 
The differences between the final radius and birth radius distributions can be 
understood by considering the radial gradient of stellar surface density: there 
are more stars at small radius to move outward than vice versa, so roughly 
symmetric evolution of~$R_\text{final}$ produces strongly asymmetric evolution 
of~$R_\text{birth}$. 
\par 
Taking~$\left|\Delta \rgal\right| \geq$~500 pc 
between birth and final radii as the criterion for migration inward or outward, 
we find as global percentages in our sample that 27\% of star particles 
migrated inward, 29\% migrated outward, and the remaining 44\% stayed near 
their birth radius. 
As one can see from the top panels of Fig.~\ref{fig:h277_decomposition}, a 
large fraction of migration away from birth radius has already occurred by the 
time stars are~$\sim$2 Gyr old. 
If the SN Ia delay-time distribution (DTD) is a~$t^{-1.1} \approx t^{-1}$ 
power-law as suggested by observational results~\citep[e.g.][]{Maoz2012, 
Maoz2017}, then we expect similar numbers of SN Ia events to occur with delay 
times between 0.1 - 1 Gyr and 1 - 10 Gyr. 
With an extended DTD and the timescales for migration implied by 
Fig.~\ref{fig:h277_decomposition}, SN Ia progenitors can migrate significant 
distances before exploding. 
This effect has largely been neglected by GCE studies to date on the grounds 
that radial migration is a slow process, and thus the majority of 
nucleosynthesis should occur near a star's birth radius (e.g. as assumed in 
\citealp{Minchev2013}, and the application of the~\citealp*{Weinberg2017} 
analytic models in~\citealp{Feuillet2018}). 
However, we show below that radial migration within the timescale of the SN Ia 
DTD can have an important impact on some aspects of chemical evolution. 

\subsection{Radial Migration} 
\label{sec:methods:migration} 
As in previous studies~\citep[e.g.][]{Matteucci1989, Schoenrich2009a, 
Minchev2013, Sharma2020}, in this paper we model the Milky Way as a series of 
concentric rings\footnote{
	For clarity, we use the term ``ring'' to refer to a computational zone of 
	our calculation (100 pc in radial range) and the term ``annulus'' to refer 
	to a larger radial range (typically 2 kpc). 
} with a uniform width~$\Delta \rgal$. 
To run numerical simulations of these models, we develop and make use 
of~\vice's~\texttt{milkyway} object, designed specifically for such an 
approach. 
The~\texttt{milkyway} object is a subclass of a more general object named 
\texttt{multizone}; at its core a~\texttt{multizone} object is an array of 
\texttt{singlezone} objects, which are designed to handle one-zone models of 
GCE and were the focus of~\citet{Johnson2020},~\vice's initial release paper. 
The~\texttt{multizone} object affords users full control over which zone any 
individual stellar population is in at all timesteps following its formation as 
well as the ability to move gas between any two zones with any time dependence. 
In principle this should allow for arbitrarily complex zone configurations and 
migration prescriptions. 
The~\texttt{milkyway} object is a user-friendly extension of the 
\texttt{multizone} base class, which enforces an annular zone configuration as 
we take here. 
As defaults, it adopts the stellar migration model detailed in this section, 
our star formation law discussed in~\S~\ref{sec:methods:sfe}, and the scaling 
of the outflow mass loading factor~$\eta$ with radius~\rgal~parameterized 
in~\S~\ref{sec:methods:outflows}. 
\par 
As in hydrodynamical simulations, star particles in~\vice~are stand-ins for 
entire stellar populations. 
They are said to be in a given zone if their radius is between the inner and 
outer edges of the ring. 
At all times, their nucleosynthetic products and returned envelopes are placed 
in the ISM of the ring that they are in~\textit{at that time}. 
\vice~forms a fixed number of stellar populations per zone per timestep, and it 
allows their masses to vary to account for variations in the SFR. 
The total mass of stars formed in a given zone and timestep is divided evenly 
among the corresponding stellar populations, which can then experience 
different stellar migration histories. 
\par 
The final radius of a stellar population is then determined based on the birth 
and final radii of star particles in the hydrodynamical simulation. 
Describing the Galaxy as a series of concentric rings,~\vice's~
\texttt{milkyway} object assumes stellar populations are born at the centres of 
each ring. 
For a stellar population born at a time~$T$ and Galactocentric radius 
$\rgal$, it first searches for star particles from~\texttt{h277} that 
formed at~$T \pm$~250 Myr and~$\rgal \pm$~250 pc. 
It then randomly selects a star particle from this subsample to act as an 
\textit{analogue}. 
This stellar population then adopts the change in orbital radius 
$\Delta \rgal$ of its analogue, and moves from its birth radius to the 
implied final radius at~$T$ = 13.2 Gyr with an assumed time dependence (see 
below). 
If no candidate analogues are found,~\vice~widens the search to~$T \pm$~500 Myr 
and~$\rgal \pm$~500 pc. 
If still no analogue is found, then it finds the star particle with the 
smallest difference in birth radius still within a birth time of $T \pm$~500 
Myr, and assigns it as the analogue. 
While this prescription allows stellar populations to be assigned analogues 
with significantly different birth radii, this is only an issue for small~$T$ 
and large~$\rgal$ where there are few star particles from~\hsim, 
and where few stars form in nature anyway due to the inside-out growth of 
galaxies~\citep[e.g.][]{Bird2013}. 
Furthermore, due to the similarity of the histograms in the top row of 
Fig.~\ref{fig:h277_decomposition}, we expect taking~$\Delta \rgal$ from 
a star particle that formed at a similar time but different birth radius in 
these instances to be accurate enough for our purposes. 
When an~\hsim~star particle is assigned as an analogue, it is~\textit{not} 
thrown out of the sample of candidate analogues, in theory allowing a star 
particle to act as an analogue for multiple stellar populations. Because these 
populations will have similar~$R_\text{form}$~and~$T_\text{form}$, they will 
have similar but not identical abundances. 

% fig 2 
\begin{figure} 
\centering 
\includegraphics[scale = 0.45]{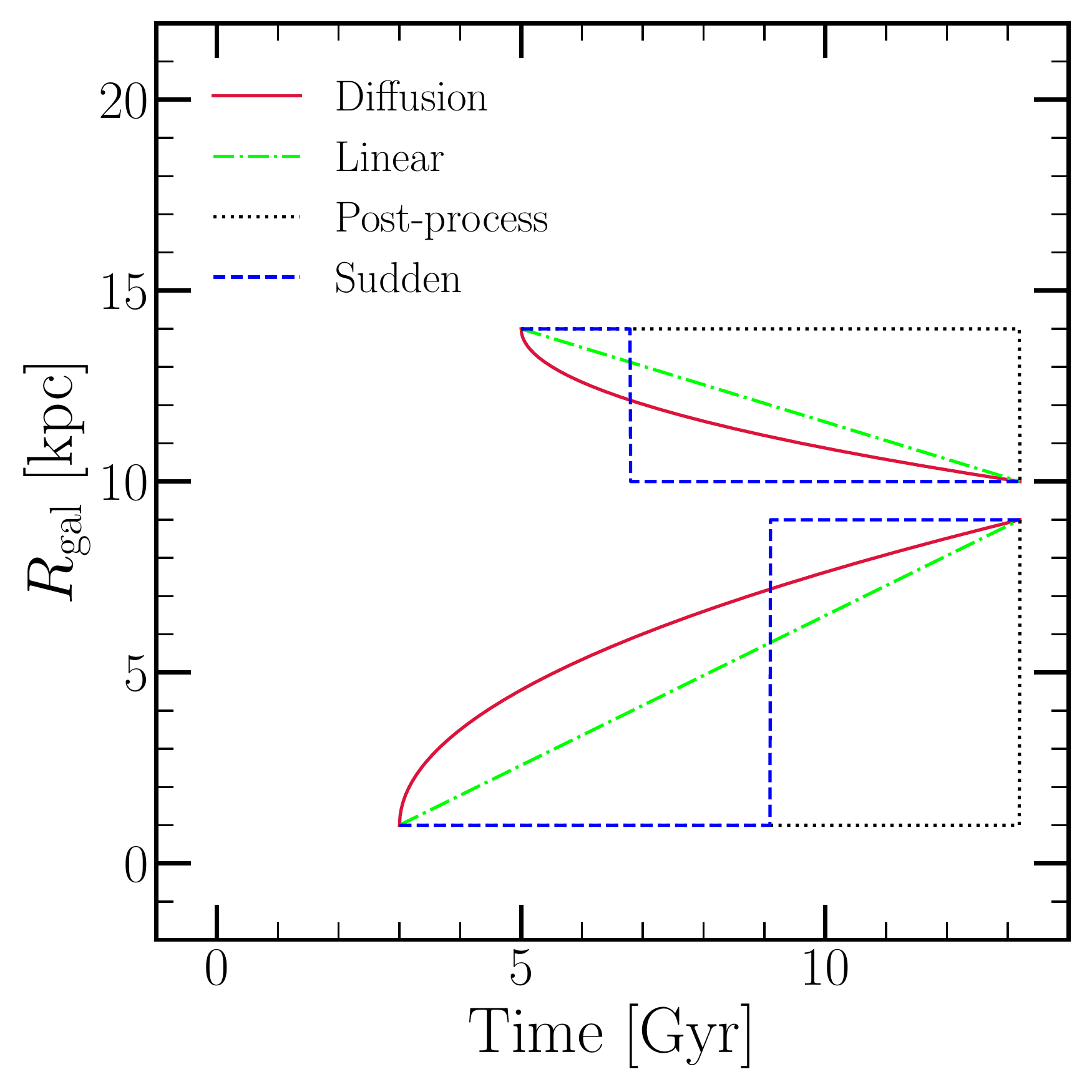} 
\caption{A diagram illustrating how Galactocentric radius changes with time for 
two stellar populations under our migration schema: diffusion (crimson, solid), 
linear (lime, dot-dashed), post-process (black, dotted), and sudden (blue, 
dashed). Here the initial and final radii and birth times are chosen at random 
for illustrative purposes. With the initial and final Galactocentric radii of 
a stellar population, its birth time, and one of these assumptions regarding 
the time-dependence of radial migration, the Galactocentric radius at all times 
is known. Diffusion is our default model. } 
\label{fig:migration_schema} 
\end{figure} 

We neglect radial gas flows in the present paper~\citep{Bilitewski2012, 
Lacey1985}, instead focusing our comparisons on the time-dependence with which 
stars migrate. 
We consider four models describing the evolution between a star particle's 
birth and final radius: 
\begin{itemize} 
	\item \textbf{Post-Processing}: Stars stay where they are born until the 
	final timestep, at which point they instantly migrate to their final 
	radius. 
	This retains the assumption that stars do not contribute to nucleosynthesis 
	beyond their birth radius as employed in previous studies 
	\citep[e.g.][]{Minchev2013}. 
	In this scenario, the ISM of each ring is treated as a one-zone model 
	independent of all other zones. We illustrate this case with a black dotted 
	line in Fig.~\ref{fig:migration_schema}. 

	\item \textbf{Sudden}: A random number is drawn from a uniform distribution 
	between a stellar population's time of birth and the present day. 
	That time is taken to be the time of instantaneous migration to the 
	present-day annulus. 
	This emulates a scenario in which a single dynamical interaction rapidly 
	changes a star's orbital radius, and it can be thought of mathematically as 
	a generalization of the post-processing scenario. 
	We illustrate this case with a blue dashed line in 
	Fig.~\ref{fig:migration_schema}. 

	\item \textbf{Diffusion}: Stars move to their final radii in a continuous, 
	time-dependent manner, with displacement~$\propto \sqrt{\text{age}}$. 
	This scenario corresponds to diffusion of angular momentum by a random 
	walk, similar to the assumption used by~\citet{Frankel2018, Frankel2020}. 
	We illustrate this case with a red solid line in 
	Fig.~\ref{fig:migration_schema}. 

	\item \textbf{Linear}: A simple variation of the diffusion model in which 
	the migration to the final radius scales linearly with age rather than with 
	$\sqrt{\text{age}}$. We illustrate this case with a green dot-dashed line 
	in Fig.~\ref{fig:migration_schema}. 
\end{itemize} 
The diffusion model has the clearest physical motivation, and it is our default 
assumption used in all cases unless otherwise noted. 
The other models provide illustrative contrasts and allow us to investigate our 
model's sensitivity to the details of radial migration. 
We do not distinguish between ``blurring'' and ``churning'' 
\citep{Schoenrich2009a}, terms frequenly used to refer to a star's epicyclic 
motions and changes in the guiding centre of its orbit, respectively. 
% We do not distinguish between ``blurring'' and ``churning'', terms frequently 
% used to refer to a star's epicyclic motions and changes in the guiding centre 
% of its orbit~\citep[e.g.][]{Sellwood2002, Schoenrich2009a}. 
Both effects, induced by a wide variety of underlying causes such as molecular 
cloud scattering~\citep{Mihalas1981, Jenkins1990, Jenkins1992}, orbital 
resonances with spiral arms or bars~\citep{Sellwood2002, Minchev2011}, and 
satellite perturbations~\citep{Bird2012}, are present in the~\hsim~simulation. 
\par 
Our GCE model assumes that the star-forming ISM is vertically and azimuthally 
mixed within each radial annulus. 
The abundances assigned to a stellar population depend on its birth radius and 
time, but they do not depend on its distance from the plane. 
Nonetheless, as shown below, the abundance patterns of our simulation exhibit 
clear vertical gradients because older stellar populations have thicker 
vertical distributions~\citep{Bird2020}. 
Radial migration is also coupled to vertical dynamics in complex ways 
\citep{Solway2012, Minchev2012b}. 
We will see that these effects already suffice to explain many of the observed 
vertical trends of Milky Way disc abundances.

% \subsection{Nucleosynthetic Yields, Outflows, and Recycling} 
% \label{sec:methods:yields} 
\subsection{Nucleosynthetic Yields} 
\label{sec:methods:yields} 
We focus our analysis on alpha and iron-peak elements, taking oxygen (O) and 
iron (Fe) as the representative cases. 
The dominant enrichment channels of interest in our models are thus CCSN and 
SN Ia~\citep{Johnson2019}. 
We would expect similar results for other alpha (e.g. Ne, Mg, Si) and 
iron-peak elements (e.g., Cr, Ni), with quantitative differences reflective of 
their relative yields. 
Odd-$Z$ iron-peak elements (e.g., V, Mn, Co) could behave somewhat distinctly 
because of metallicity-dependent yields. 
\par 
CCSN enrichment happens immediately following the formation of progenitor stars 
in~\vice. 
This is an adequate approximation, because the lifetimes of massive stars are 
short compared to the relevant timescales for galaxy evolution. 
For the most massive stars, the lifetimes are comparable to the timestep size 
we adopt in our numerical integrations. 
This assumption implies a linear relationship between the CCSN enrichment 
rate and the SFR: 
\begin{equation} 
\dot{M}_x^\text{CC} = y_x^\text{CC}\dot{M}_\star 
\end{equation} 
where~$y_x^\text{CC}$ is the CCSN yield of some element~$x$. Physically, this 
quantity represents the mass of an element~$x$ ejected to the ISM from all 
CCSN events associated with a stellar population in units of the stellar 
population's initial mass. 
For example, if~$y_x^\text{CC} = 0.01$, a hypothetical 100~\msun~stellar 
population would add 1~\msun~of~$x$ to the ISM immediately. 
In this paper, we adopt~$y_\text{O}^\text{CC}$ = 0.015 and 
$y_\text{Fe}^\text{CC}$~= 0.0012 from~\citet{Johnson2020}, who in turn adopt 
these values from~\citet{Weinberg2017}. 
\par 
SN Ia nucleosynthesis products are injected according to a~$t^{-1.1}$ DTD with 
a minimum delay time of~$t_\text{D}$ = 150 Myr. 
This is the default DTD in~\vice, which was also adopted by 
\citet{Johnson2020}, and is suggested by recent observational results comparing 
the cosmic SN Ia rate to the cosmic SFH~\citep{Maoz2012, Maoz2017}. 
In a one-zone model at times~$t > t_\text{D}$, the enrichment rate of some 
element~$x$~can be expressed as the product of some yield~$y_x^\text{Ia}$ and 
the SFH weighted by the DTD: 
\begin{subequations}\begin{align} 
\dot{M}_x^\text{Ia} &= y_x^\text{Ia}\langle\dot{M}_\star\rangle_\text{Ia} \\ 
&= y_x^\text{Ia}\ddfrac{
	\int_0^t \dot{M}_\star(t') R_\text{Ia}(t - t') dt' 
}{
	\int_{t_\text{D}}^{t_\text{max}} R_\text{Ia}(t')dt' 
} 
\label{eq:mdot_ia} 
\end{align}\end{subequations} 
where~$R_\text{Ia}$ is the DTD itself, which has units of~$M_\odot^{-1}$ 
Gyr$^{-1}$. 
Like the CCSN yield,~$y_x^\text{Ia}$ is the mass of some element~$x$ produced 
by SNe Ia over the time interval~$t_\text{D} - t_\text{max}$, in units of the 
stellar population's initial mass. 
It can also be expressed as an integral over the DTD: 
\begin{equation} 
y_x^\text{Ia} = m_x^\text{Ia} \int_{t_\text{D}}^{t_\text{max}} R_\text{Ia}(t') 
dt' = m_x^\text{Ia}\frac{N_\text{Ia}}{M_\star} 
\label{eq:y_x_ia} 
\end{equation} 
where~$m_x^\text{Ia}$ is the average mass of the element~$x$ produced in a 
single SN Ia event and the integral evaluates to the mean number of SN Ia 
events~$N_\text{Ia}$ per mass of stars formed~$M_\star$. 
\vice~forces~$t_\text{max}$~= 15 Gyr always, though provided one is consistent 
with equations~\refp{eq:mdot_ia} and~\refp{eq:y_x_ia}, the results are 
independent of~$t_\text{max}$ because the integrals cancel. 
Extending this formalism to multi-zone models is simple; rather than an 
integral over the star formation history of a given annulus, the rate becomes 
a summation over all stellar populations that are in a given zone at some time: 
\begin{equation} 
\dot{M}_x^\text{Ia} = y_x^\text{Ia} \ddfrac{
	\sum_i M_i R_\text{Ia}(\tau_i) 
}{
	\int_{t_\text{D}}^{t_\text{max}} R_\text{Ia}(t')dt' 
} 
\label{eq:mdot_ia_multizone} 
\end{equation} 
where~$M_i$~and~$\tau_i$~are the mass and age of the~$i$'th stellar population, 
respectively. 
\par 
Initially, we adopted~$y_\text{O}^\text{Ia}$~= 0 and~$y_\text{Fe}^\text{Ia}$~= 
0.0017 from~\citet{Johnson2020}, who in turn adopt these values from 
\citet{Weinberg2017}. 
However, we found that the e-folding timescales of star formation in our models 
are sufficiently long (see discussion in~\S~\ref{sec:methods:sfhs}) that our 
fiducial, inside-out SFH model predicted [O/Fe]~$\approx$~+0.05 for young 
stars. 
We therefore multiply~$y_\text{Fe}^\text{Ia}$~by~$10^{0.1}$, adopting 
$y_\text{Fe}^\text{Ia}$~= 0.00214 so that our fiducial model predicts a 
late-time [O/Fe] ratio in better agreement with observations. 
Changes at this level are within the uncertainties of SN Ia rates and yields, 
so we consider it reasonable to adjust the yields empirically to reproduce 
observed abundances. 
\par 
Our IMF-averaged O and Fe yields are based on a~\citet{Kroupa2001} IMF combined 
with supernova nucleosynthesis models in which most~$M > 8~\msun$ stars explode 
as a CCSN~\citep[e.g.][]{Chieffi2004, Chieffi2013}. 
Recent studies have strongly suggested that many high mass stars instead 
collapse directly to a black hole (see theoretical discussion by, e.g., 
\citealp{Pejcha2015, Sukhbold2016, Ertl2016}, and observational evidence from 
\citealp*{Gerke2015};~\citealp{Adams2017, Basinger2020}). 
Our yields would be lower in a scenario with extensive black hole formation 
and/or a steeper high mass IMF (Griffith et al. 2021). 
These effects could also introduce a metallicity dependence if the landscape of 
black hole formation changes with metallicity. 
The strong increase of the specific SN Ia rate seen at low galaxy masses 
\citep{Brown2019} provides circumstantial evidence of a higher~$R_\text{Ia}$ 
normalization at low metallicity; the stellar close binary fraction also 
depends on metallicity (\citealp{Badenes2018};~\citealp*{Moe2019}). 
However, there is presently no solid empirical basis for adopting metallicity 
dependent O and Fe yields over the range relevant to this paper (roughly 
-0.8~$\leq$~\feh~$\leq$~0.4), and some empirical evidence that any metallicity 
trends in this range are weak~\citep{Weinberg2019}. 
\vice~has capabilities to compute IMF-averaged yields for a flexible 
description of the massive star explodability landscape, as described by 
Griffith et al. (2021) and the~\vice~documentation, but we do not use this 
methodology here. 
We expect that most of our results would be largely unchanged if we were to 
lower all three yields ($y_\text{O}^\text{CC}$, $y_\text{Fe}^\text{CC}$, 
$y_\text{Fe}^\text{Ia}$) by the same factor and adjust our adopted outflow 
mass loading prescription to compensate (see below). 
\par 
Both AGB star enrichment and the recycling of previously produced metals in 
this paper proceed as they did in~\citet{Johnson2020}, with the caveat that the 
mass is added to the ring that a stellar population is in at a given time, 
which may or may not be the ring it was born in. 
Recycling proceeds according to the~\citet{Kalirai2008} initial-remnant mass 
relation assuming a~\citet{Kroupa2001} IMF and a mass-lifetime relationship of 
$\tau = 1.1\tau_\odot(M/M_\odot)^{-3.5}$, where~$\tau_\odot$~= 10 Gyr is the 
main sequence lifetime of the sun and the factor of 1.1 accounts for the 
post-main sequence lifetime. 
\vice~includes AGB enrichment in all models; here we adopt the net yields 
sampled on a table of stellar initial mass and metallicity from the FRUITY 
database \citep{Cristallo2011}. 
However, the AGB star yields of O and Fe are tiny compared to their 
supernova yields, so they have negligible impact on the results presented in 
this paper. 

\subsection{Outflows} 
\label{sec:methods:outflows} 
\begin{figure} 
\centering 
\includegraphics[scale = 0.45]{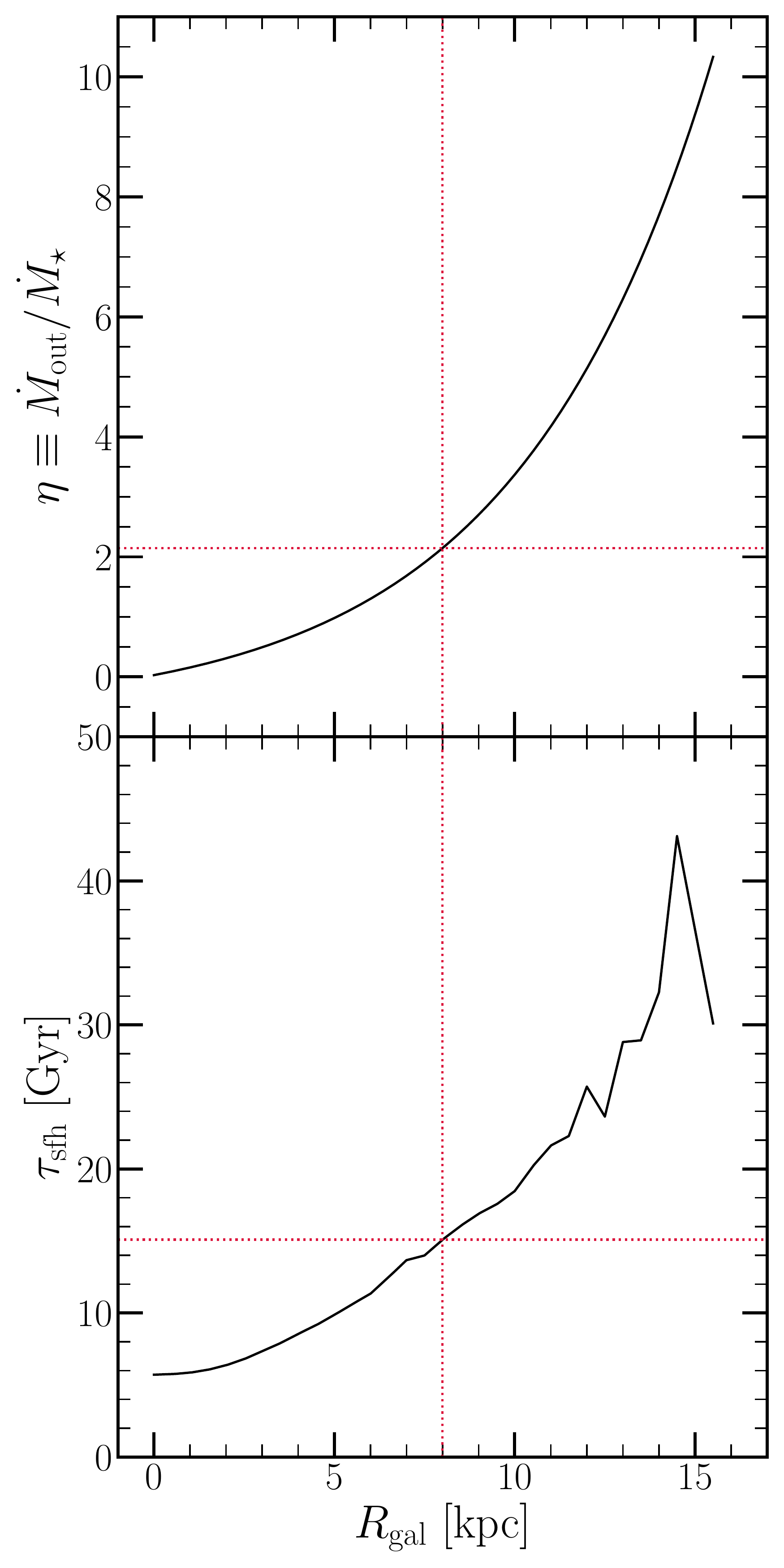} 
\caption{\textbf{Top}: Our implemented scaling of the mass loading factor 
$\eta$~with Galactocentric radius (black) as defined by equation 
\refp{eq:eta_rgal}.~\textbf{Bottom}: The e-folding timescales of the star 
formation histories of our model galaxies (black). These values come from a 
fit to the~$\Sigma_\star$-age relation in bins of~$R/R_\text{e}$~for 
$10^{10.5}$~-~$10^{11}$~\msun~Sa/Sb Hubble type spiral galaxies as reported by 
\citet[][see discussion in~\S~\ref{sec:methods:sfhs}]{Sanchez2020}. The 
horizontal and vertical red dashed lines in both panels highlight a mass 
loading factor of~$\eta \approx$~2.15 and a star formation timescale of 
$\tau_\text{sfh} \approx$~15 Gyr at an assumed radius of the sun of 
$R_\odot$ = 8 kpc. } 
\label{fig:eta_tau_sfh} 
\end{figure}

\citet{Weinberg2017} demonstrate that, to first order, the nucleosynthetic 
yields of a given element and the strength of outflowing winds determine the 
late-time equilibrium abundance in the ISM, with a secondary dependence on the 
SFH. 
We retain their characterization of outflows here, in terms of a mass-loading 
factor~$\eta$~describing the ratio of the mass outflow to the SFR: 
\begin{equation} 
\eta \equiv \frac{\dot{M}_\text{out}}{\dot{M}_\star}. 
\end{equation} 
We adopt a scaling of~$\eta$~with~$\rgal$~such that the late-time 
equilibrium abundance as a function of radius describes a metallicity 
gradient in agreement with observations. 
For a constant SFR, the equilibrium abundance of an~$\alpha$-element, produced 
by CCSNe with a metallicity-independent yield, is given by 
\begin{equation} 
Z_{\alpha,\text{eq}} = \frac{y_\alpha^\text{CC}}{1 + \eta(\rgal) - r}, 
\end{equation} 
where~$r$~is the recycling parameter ($\approx$0.4 for the sake of this scaling 
with a~\citealp{Kroupa2001} IMF; see discussion in~\S~2.2 
of \citealp{Weinberg2017}). Solving for~$\eta(\rgal)$ yields 
\begin{equation} 
\eta(\rgal) = \frac{y_\alpha^\text{CC}}{Z_{\alpha,\text{eq}}} + r - 1 = 
\frac{y_\alpha^\text{CC}}{Z_{\alpha,\odot}}10^{-\text{mode([$\alpha$/H])}
(\rgal)} + r - 1, 
\end{equation} 
where mode([$\alpha$/H])($\rgal$) denotes the mode of the stellar 
[$\alpha$/H] distribution at a radius~$\rgal$, which we assume to correspond 
to the equilibrium abundance at that radius. 
Recent studies of the disc metallicity gradient with APOGEE find values of 
-0.09 dex/kpc to -0.06 dex/kpc~\citep[e.g.][]{Frinchaboy2013, Hayden2014, 
Weinberg2019}, consistent with earlier studies. 
Here we adopt a slope of -0.08 dex/kpc and set mode([$\alpha$/H]) to be 
$\sim$+0.3 at~$\rgal$ = 4 kpc, producing mode([$\alpha$/H])~$\approx$ 0 at 
$\rgal$ = 7 - 9 kpc. This results in the following form for~$\eta$ as a 
function of Galactocentric radius: 
\begin{equation} 
\eta(\rgal) = \frac{y_\alpha^\text{CC}}{Z_{\alpha,\odot}} 
10^{(-0.08\text{ kpc}^{-1})(\rgal - 4\text{ kpc}) + 0.3} + r - 1, 
\label{eq:eta_rgal} 
\end{equation} 
where we adopt our CCSN yield of O for~$y_\alpha^\text{CC}$ and the solar 
photospheric abundance of O of~$Z_{\text{O},\odot}$ = 0.00572 based on 
\citet{Asplund2009}. 
We plot this adopted scaling in the top panel of Fig.~\ref{fig:eta_tau_sfh}, 
highlighting a value of~$\sim$2.15 for the solar circle with a red dotted line. 
In Fig.~\ref{fig:metallicity_gradient} below, we show that our full model 
including a time-dependent SFH and radial migration produces stellar and gas 
phase gradients similar but not identical to those of equation 
\refp{eq:eta_rgal}. 
In Figs.~\ref{fig:mdf_3panel_fe} and~\ref{fig:mdf_3panel_o} below, we show that 
the model achieves qualitative agreement with the metallicity gradient observed 
by APOGEE. 

% fig 3 

\subsection{Star Formation Histories} 
\label{sec:methods:sfhs} 
\vice~computes models in either ``infall'', ``star formation'', or ``gas'' 
mode, referring to which component of the evolutionary history the user has 
specified. 
The starburst models of~\citet{Johnson2020} ran in infall mode, 
meaning that the gas infall rate is specified and the SFR follows from the gas 
surface density and adopted star formation law. 
Here we run~\vice~in star formation mode so that we achieve a specified form of 
the SFHs in our models. 
In Appendix~\ref{sec:normalize_sfh}, we explain how we normalize the parameters 
of our SFHs to produce a realistic model Galaxy at the present day. 
In short, we take a unitless description of the time-dependence of the SFH at a 
given Galactocentric radius, denoted~$f(t|\rgal)$, and a unitless description 
of the present day stellar surface density gradient, denoted~$g(\rgal)$. 
We integrate~$f(t|\rgal)$ with time for each annulus, assuming~$\rgal$ to 
correspond to the centre of the zone, and attach a prefactor to 
$f(t|\rgal)$ in each annulus such that the desired gradient is achieved 
with a total stellar mass similar to the Milky Way. 
This procedure neglects the impact of radial migration, assuming that it does 
not significantly alter the form of~$g(\rgal)$. 
We demonstrate that this assumption holds 
in~\S~\ref{sec:methods:surface_density_gradient}, in which we also detail our 
adopted form of~$g(\rgal)$. 
The equation derived in Appendix~\ref{sec:normalize_sfh} can be used to 
calculate these prefactors for alternative models of Milky Way-like galaxies. 
\par 
In the present paper, we consider four forms of the SFH, which we dub 
``constant'', ``inside-out'', ``late-burst'', and ``outer-burst''. 
They are defined as follows: 
\begin{itemize} 
	\item \textbf{Constant}: The SFH at a given radius is time-independent, 
	\begin{equation} 
	f_\text{C}(t|\rgal) = 1. 
	\label{eq:constant_sfh} 
	\end{equation} 
	This case is of theoretical interest because it quantifies the effect of 
	stellar migration while removing the impact of a time-dependent SFH. 

	\item \textbf{Inside-Out}: This is our fiducial SFH, 
	\begin{equation} 
	f_\text{IO}(t|\rgal) = (1 - e^{-t / \tau_\text{rise}}) 
	e^{-t / \tau_\text{sfh}}. 
	\label{eq:insideout_sfh} 
	\end{equation} 
	We adopt this mathematical form over the somewhat more common 
	$te^{-t/\tau_\text{sfh}}$ because it allows separate control over the 
	rising and falling phase of the SFH. 
	Equation~\refp{eq:insideout_sfh} has a maximum near~$\tau_\text{rise}$, 
	which in this paper we set to 2 Gyr at all radii. 
	This form produces a peak in star formation at lookback times of~$\sim$11 
	Gyr, roughly corresponding to a redsfhit of~$z \approx$ 2.5. 
	We discuss the choice of~$\tau_\text{sfh}$ below. 

	\item \textbf{Late-Burst}: In this model, the inside-out SFH is modified to 
	exhibit a recent, slow ``burst'' in star formation described by a Gaussian 
	in time: 
	\begin{equation} 
	f_\text{LB}(t|\rgal) = f_\text{IO}(t|\rgal) 
	\left(1 + A_be^{-(t - t_b)^2/2\sigma_b^2}\right), 
	\label{eq:lateburst_sfh} 
	\end{equation} 
	where $A_b$ is a dimensionless parameter describing the strength of the 
	starburst,~$t_b$ is the time of the local maximum in the SFH during the 
	burst, and~$\sigma_b$ is the width of the Gaussian describing it. 
	To approximate the findings of~\citet{Mor2019} and~\citet{Isern2019}, we 
	adopt~$A_b$ = 1.5,~$t_b$ = 11.2 Gyr, and~$\sigma_b$ = 1 Gyr, finding 
	that~$A_b$ = 1.5 with a declining~$f_\text{IO}(t|\rgal)$ produces a local 
	maximum SFR that is a factor of~$\sim$2 larger than the preceding local 
	minimum. 

	\item \textbf{Outer-Burst}: A variation of the late-burst model in which 
	only zones at~$\rgal \geq$~6 kpc experience the starburst, with inner 
	regions following the inside-out SFH. 
	Because the empirical evidence for elevated recent star formation comes 
	from local observations, it is useful to investigate the case where it is 
	confined to the outer Galaxy. 
	In their hydrodynamical simulation of a Milky Way-like galaxy, 
	\citet{Vincenzo2020} find that satellite perturbations enhance gas 
	accretion preferentially in the outer regions. 
\end{itemize} 
The inside-out model is our fiducial choice, and it is the model shown in 
figures unless otherwise specified. 
The constant SFR model allows us to investigate the impact of migration when 
the time-dependence of the SFH is removed, and the two burst models allow us to 
explore the consequences of elevated recent star formation supported by some 
recent data. 
More complex scenarios with multiple bursts induced by repeated satellite 
passages~\citep[e.g.][]{Lian2020a, Lian2020b, RuizLara2020, Sysoliatina2021} 
can be modeled easily in~\vice, but we do not explore them here. 
\par 
We derive the~$\tau_\text{sfh}-\rgal$ relation from the data of 
\citet{Sanchez2020}, who presents the stellar surface density~$\Sigma_\star$ as 
a function of age in bins of~$R/R_\text{e}$ for MaNGA 
galaxies~\citep{Bundy2015}, where $R_\text{e}$ is the half-light radius. 
Here we take the~$M_\star = 10^{10.5} - 10^{11}~\msun$~bin for Sa/Sb spirals 
and simultaneously fit the normalization and e-folding timescale 
$\tau_\text{sfh}$ of our~$f_\text{IO}(t|\rgal)$ form to the data. 
Although the normalization is irrelevant to our models and determined via the 
method outlined in Appendix~\ref{sec:normalize_sfh}, we adopt the resulting 
$\tau_\text{sfh}-\rgal$ relation in our models. 
Our adopted stellar surface density gradient 
(see~\S~\ref{sec:methods:surface_density_gradient}) implies a present-day 
half-mass radius near 4 kpc. 
The findings of~\citet{Garcia-Benito2017} and~\citet{GonzalezDelgado2014} 
suggest that half-light radii are marginally larger than half-mass radii. 
Based on equation (4) of~\citet{GonzalezDelgado2014} relating the two for 
circular apertures, we expect our model Galaxy to have a 
half-light radius near 5 kpc. 
We therefore adopt~$R_\text{e}$ = 5 kpc to convert the 
$\tau_\text{sfh}-\rgal/R_\text{e}$ relation resulting from our fit to the 
\citet{Sanchez2020} data into a~$\tau_\text{sfh}-\rgal$ relation. 
\par 
We illustrate this relationship in the bottom panel of Fig. 
\ref{fig:eta_tau_sfh}. 
The resulting timescales are long, particularly for the outer Galaxy. 
With a red dotted line, we highlight a value of~$\tau_\text{sfh} \approx$~15 
Gyr at an assumed orbital radius of the sun of $R_\odot$ = 8 kpc. 
The long timescales reflect the fact that the 
$\Sigma_\star(\tau_\text{lookback})$ profiles in Fig. 11 of~\citet{Sanchez2020} 
are fairly flat. 
For comparison, we have also considered the assumption that the Galactic SFH 
may have resembled the cosmic SFH by running models with e-folding timescales 
of a~$\sim$few Gyr~\citep[e.g.][]{Madau2014}. 
We find similar results in these cases, suggesting that our qualitative 
conclusions are not sensitive to the exact values of~$\tau_\text{sfh}$. 

% fig 4 
\begin{figure*} 
\centering 
\includegraphics[scale = 0.32]{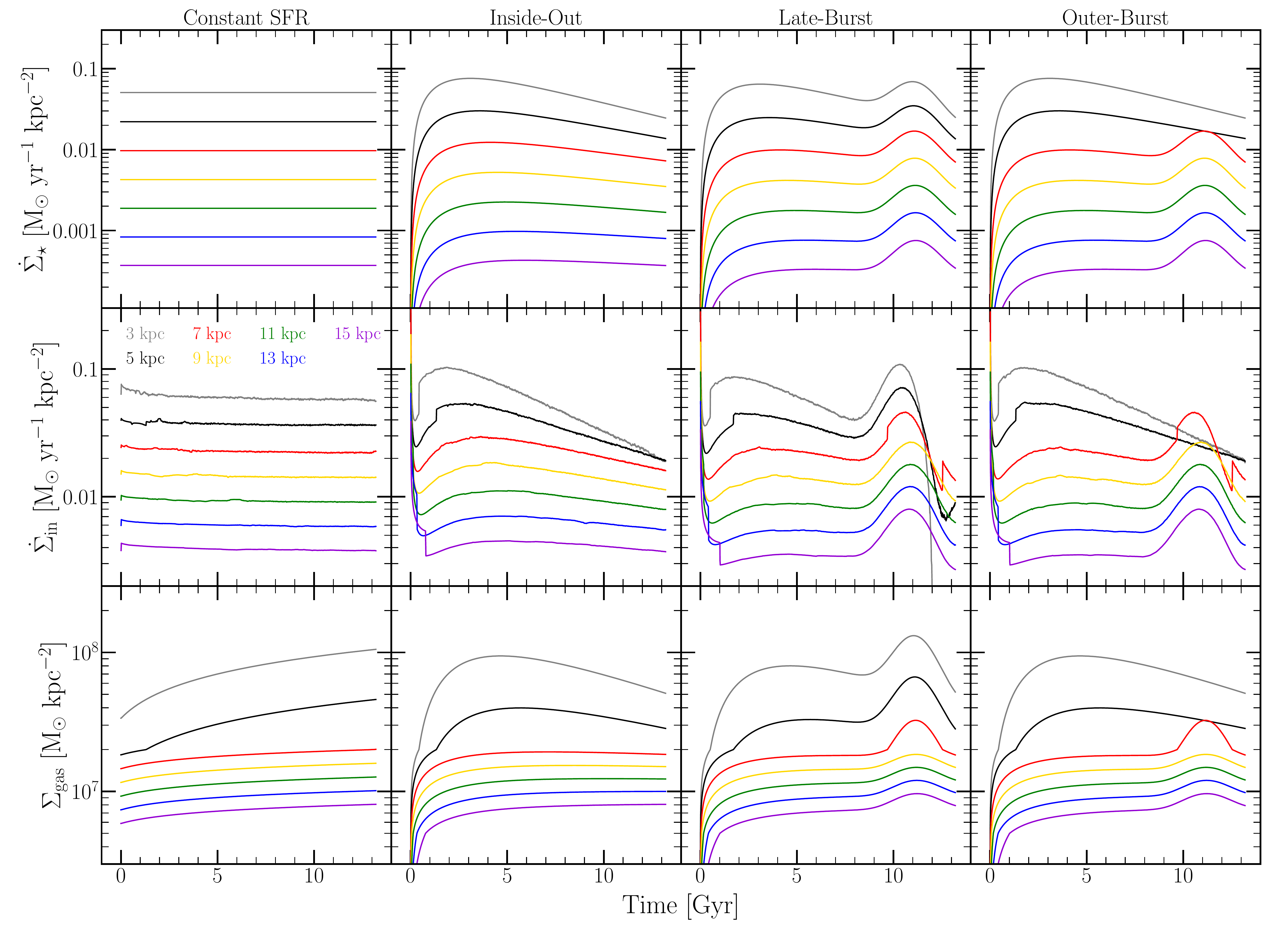} 
\caption{The surface densities of star formation~$\dot{\Sigma}_\star$ (top 
row), infall $\dot{\Sigma}_\text{in}$ (middle row), and gas~$\Sigma_\text{g}$ 
(bottom row) as functions of simulation time for our four fiducial SFHs: 
constant (far left), inside-out (left middle), late-burst (right middle), and 
outer-burst (far right). We plot curves for the rings whose inner radii are 
3 kpc (grey), 5 kpc (black), 7 kpc (red), 9 kpc (yellow), 11 kpc (green), 13 
kpc (blue), and 15 kpc (purple) (see equations~\refp{eq:constant_sfh}, 
\refp{eq:insideout_sfh}, and~\refp{eq:lateburst_sfh} for the mathematical 
definition of each SFH). 
}
\label{fig:evol} 
\end{figure*} 

We plot the resulting SFHs of our models in the top row of Fig.~\ref{fig:evol} 
for a handful of radii. 
Because of the long~$\tau_\text{sfh}$, the SFR at most radii has fallen only 
modestly from its 2 Gyr peak in the inside-out model. 
At all timesteps, the gas supply is known via the star formation efficiency 
timescale~$\tau_\star$ (see~\S~\ref{sec:methods:sfe}), which is illustrated in 
the bottom row of Fig.~\ref{fig:evol}. 
\vice~automatically calculates the implied infall rate by comparing the amount 
of gas lost to outflows and star formation in a given timestep to that which is 
required to sustain the user-specified level of star formation at the next 
timestep; we assume the infalling gas to be of zero metallicity at all times. 
This quantity is shown in the middle row of Fig.~\ref{fig:evol}. 

\subsection{Star Formation Efficiency} 
\label{sec:methods:sfe} 
The term ``star formation efficiency'' (SFE) is somewhat overloaded in the 
literature. 
In the star formation and feedback community, it usually refers to the fraction 
of a molecular cloud's mass that will eventually be converted into stars. 
In the chemical evolution literature, however, it typically refers to the 
inverse timescale relating the SFR within some star forming reservoir to the 
mass of gas in that region: 
$\tau_\star \equiv \Sigma_\text{g}/\dot{\Sigma}_\star$. 
High (Low) values of~$\tau_\star$ indicate slow (fast) conversion of gas and 
thus low (high) SFE; when we refer to SFE here, we mean the definition based on 
this timescale. 
In the star formation and feedback literature,~$\tau_\star$ is sometimes 
referred to as the ``depletion time'', though here we follow the terminology of 
\citet{Weinberg2017} who call it the ``star formation efficiency timescale.'' 
\par 
Based on the findings of~\citet{Kennicutt1998}, it is common practice in the 
chemical evolution literature to adopt a single power-law describing the 
relationship between the surface densities of gas and star formation 
$\Sigma_\text{g}$ and $\dot{\Sigma}_\star$, often referred to as the star 
formation law or the Kennicutt-Schmidt relation: 
\begin{equation} 
\dot{\Sigma}_\star \propto \Sigma_\text{g}^N. 
\end{equation} 
\citet{Kennicutt1998} finds~$N = 1.4 \pm 0.15$ relating the total 
$\dot{\Sigma}_\star$~and~$\Sigma_\text{g}$~within the disc across a sample of 
quiescent spiral galaxies and infrared and circumnuclear starbursts. 
However, recent studies have found evidence that much of the observed scatter 
in this relation is physical in origin~\citep{delosReyes2019} and that there 
are significant breaks in both the power-law index and zero-points 
\citep{Kennicutt2021}. 
Some of the uncertainty surrounding the details of the star formation law is a 
consequence of the ongoing debate about the CO-to-H$_2$ conversion factor 
(\citealp{Kennicutt2012};~\citealp*{Liu2015}). 
Although~\citet{Ellison2021} demonstrate that there are significant 
galaxy-to-galaxy variations in the star formation law,~\citet{delosReyes2019} 
argue that the mean trend is still a reasonable recipe for Galaxy evolution 
models. 
However, for our purposes we also need the dependence of~$\dot{\Sigma}_\star$ 
on~$\Sigma_\text{g}$ within a galaxy. 
\par 
\citet{Krumholz2018a} compare theoretically motivated star formation laws to 
the observations of~\citet{Bigiel2010} and~\citet{Leroy2013} (see their Fig. 2). 
We find that the following by-eye fit to the power-law index~$N$~is a 
reasonable description of the aggregate data: 
\begin{equation} 
N = \begin{cases} 
1.0 & (\Sigma_\text{g} \geq \Sigma_{\text{g},2})~, \\ 
3.6 & (\Sigma_{\text{g},1} \leq \Sigma_\text{g} \leq \Sigma_{\text{g},2})~, \\ 
1.7 & (\Sigma_\text{g} \leq \Sigma_{\text{g},1})~, 
\end{cases} 
\label{eq:sf_law_indeces} 
\end{equation} 
where~$\Sigma_{\text{g},1} = 5\times10^6$~\msun~\persqkpc~and 
$\Sigma_{\text{g},2} = 2\times10^7$~\msun~\persqkpc. 
The apparent linearity of the relationship above 
$\sim2\times10^7$~\msun~\persqkpc~suggests that in this regime, star formation 
proceeds at the highest efficiency, and that 
$\tau_\star \equiv \Sigma_\text{g}/\dot{\Sigma}_\star$~= constant. 
The observational results of~\citet{Leroy2013} and~\citet{Kennicutt2021} would 
suggest that these are the surface densities at which the molecular fraction 
$f_\text{mol} = M_{\text{H}_2} / (M_{\text{H}_2} + M_\text{HI}) \approx$~1. 
We therefore adopt the assumption that above 
$\Sigma_\text{g} = 2\times10^7$~\msun~\persqkpc,~$\tau_\star$ reaches its 
minimum value, and increases with decreasing~$f_\text{mol}$. 
We denote this value as~$\tau_\text{mol}$, the value of~$\tau_\star$ for a gas 
reservoir with~$f_\text{mol}$~= 1. 
This identification, combined with our three-component power-law index~$N$ 
results in the following final form for our adopted star formation law: 
\begin{equation} 
\dot{\Sigma}_\star = \begin{cases} 
\Sigma_\text{g} \tau_\text{mol}^{-1} & 
(\Sigma_\text{g} \geq \Sigma_{\text{g},2})~, 
\\ 
\Sigma_\text{g} \tau_\text{mol}^{-1} \left(\frac{
	\Sigma_\text{g}
}{
	\Sigma_{\text{g},2} 
}\right)^{2.6} & 
(\Sigma_{\text{g},1} \leq \Sigma_\text{g} \leq \Sigma_{\text{g},2})~, 
\\ 
\Sigma_\text{g} \tau_\text{mol}^{-1} \left(\frac{
	\Sigma_{\text{g},1} 
}{
	\Sigma_{\text{g},2} 
}\right)^{2.6}\left(\frac{
	\Sigma_\text{g}
}{
	\Sigma_{\text{g},1} 
}\right)^{0.7} & 
(\Sigma_\text{g} \leq \Sigma_{\text{g},1})~. 
\end{cases} 
\label{eq:sf_law} 
\end{equation} 
We choose the power-law indices such that this formalism is consistent 
with equation~\refp{eq:sf_law_indeces}, and prefactors are added to ensure 
piece-wise continuity. 
In implementation,~\vice~requires the~$\tau_\star-\dot{\Sigma}_\star$ relation 
when running in star formation mode and the~$\tau_\star-\Sigma_\text{g}$ 
relation when running in infall and gas modes. 
Both follow algebraically from this relationship given the substitution 
$\tau_\star \equiv \Sigma_\text{g} / \dot{\Sigma}_\star$. 
\par 
Based on the observed Kennicutt-Schmidt relation at different redshifts, 
\citet{Tacconi2018} suggest that~$\tau_\text{mol}$ should scale with redshift 
$z$ and with the deviation from the star forming main sequence~$\delta$MS via 
$\tau_\text{mol} \propto (1 + z)^{-0.6}\delta\text{MS}^{-0.44}$. 
We do not account for the effect of~$\delta$MS in our models, but we do 
incorporate the redshift dependence. 
For redshifts of~$z \lesssim$~3, encompassing most of the timesteps in 
our models, a reasonable approximation to the~$t - z$ relation assuming 
typical~$\Lambda$CDM cosmology is given by: 
\begin{equation} 
\frac{t}{t_0} \approx (1 + z)^{-5/4}, 
\end{equation} 
where~$t_0$ is the present-day age of the universe, and~$t$ is not simulation 
time but the age of the universe. Plugging this relation into the 
\citet{Tacconi2018} scaling yields 
\begin{equation} 
\tau_\text{mol} = \tau_{\text{mol},0}(t/t_0)^{12/25} \approx 
\tau_{\text{mol},0}(t/t_0)^{1/2}, 
\end{equation} 
where~$\tau_{\text{mol},0}$ is simply~$\tau_\text{mol}$ at the present day. We 
generalize this formula to the following form: 
\begin{equation} 
\tau_\text{mol} = \tau_{\text{mol},0}(t/t_0)^\gamma 
\label{eq:tau_mol}
\end{equation} 
In this paper we present models which adopt~$\tau_{\text{mol},0}$ = 2 Gyr 
\citep{Leroy2008, Leroy2013, Tacconi2018} and~$\gamma$ = 1/2 based on this 
argument. 
We have also run simulations witht~$\tau_{\text{mol},0}$ = 1 
Gyr and with~$\gamma$ = 0 (a time-independent~$\tau_\text{mol}$), as well as 
combinations of the two, and found similar results. 
In all timesteps and annnuli,~\vice~infers~$\Sigma_\text{g}$ from 
$\dot{\Sigma}_\star$ given our equations~\refp{eq:sf_law} and~\refp{eq:tau_mol}. 

% fig 5 
\begin{figure} 
\centering 
\includegraphics[scale = 0.45]{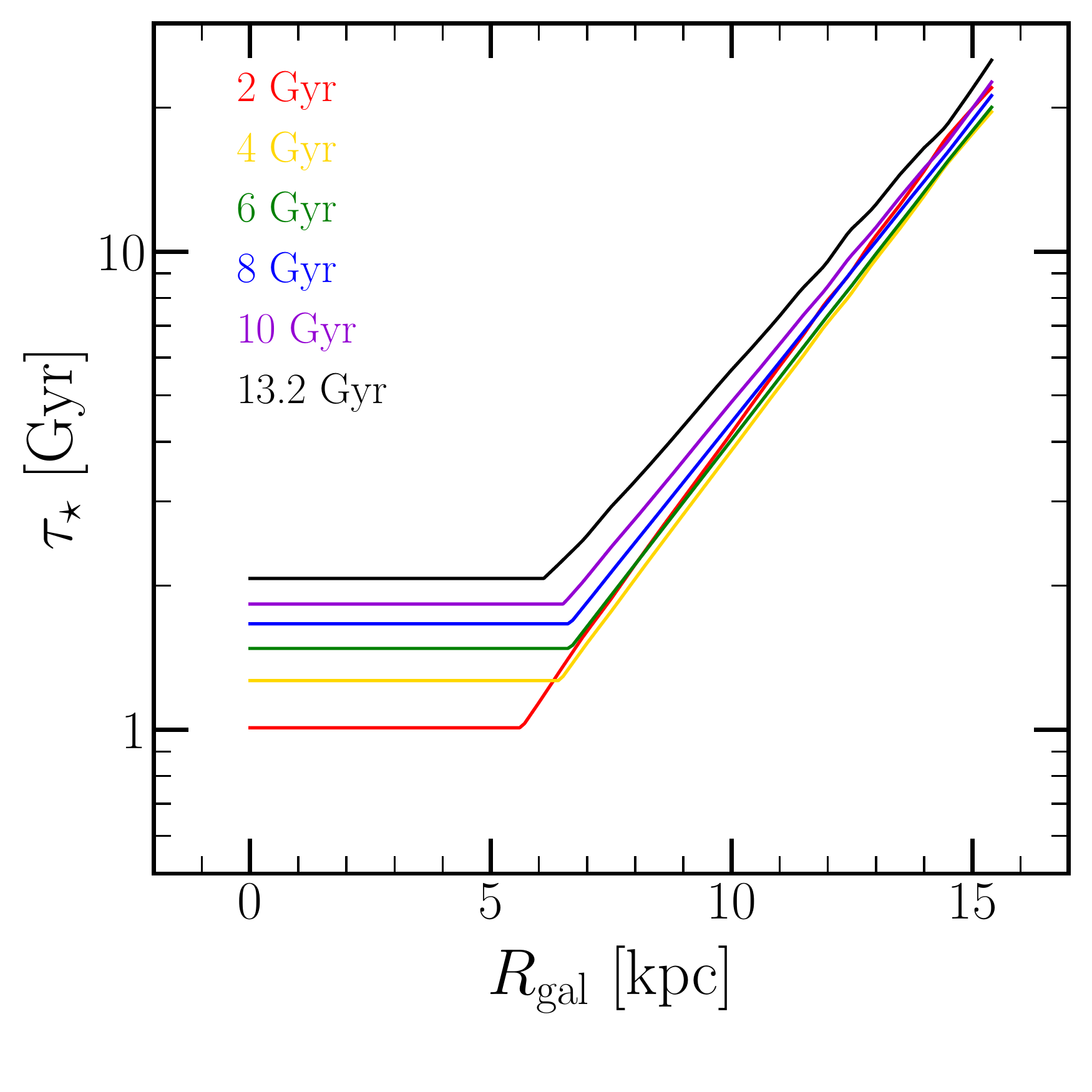} 
\caption{The star formation efficiency timescale~$\tau_\star$ as a function of 
Galactocentric radius at simulation times of 2 Gyr (red), 4 Gyr (yellow), 
6 Gyr (green), 8 Gyr (blue), 10 Gyr (purple), and 13.2 Gyr (the present day, 
black) predicted by our fiducial model. } 
\label{fig:sfe} 
\end{figure} 

In Fig.~\ref{fig:sfe}, we plot~$\tau_\star$ as a function of~$\rgal$ at 
six different time stamps predicted by our fiducial, inside-out SFH model. 
At~$\rgal \lesssim$ 6 kpc,~$\tau_\star$ is near~$\tau_\text{mol}$ at all 
times, implying a molecular fraction of unity at these radii. 
Although this prediction is likely unrealistic because 21-cm line observations 
suggest the presence of neutral hydrogen as close to the Galactic centre as 
several hundred pc~\citep{Kalberla2009}, we find in practice that changing our 
assumptions about the star formation law does not impact our conclusions. 
In exploratory work for this paper, we investigated purely linear, purely 
power-law, and broken power-law characterizations, finding similar predictions 
in all cases. 
In general we find that the detailed form of the SFH, and to some extent the 
time-dependence of radial migration, exert much greater power in establishing 
the model predictions than does the star formation law. 
Nonetheless it is an interesting puzzle that a 
$\dot{\Sigma}_\star - \Sigma_\text{g}$ relation informed by the observed 
population-averaged trends and the normalization of SFHs implied by the stellar 
mass of the Milky Way predicts results in tension with the observed HI 
distribution. 
Because the star formation law has minimal impact on our results, we do not 
pursue this question further here. 

\subsection{Surface Density Gradient} 
\label{sec:methods:surface_density_gradient} 

As discussed in~\S~\ref{sec:methods:sfhs}, Appendix~\ref{sec:normalize_sfh} 
presents the recipe by which we select a unitless function describing the 
stellar surface density gradient~$g(\rgal)$. 
In setting the normalization, our model ensures that the integral of~$g(\rgal)$ 
over the surface area of the disc predicts a total stellar mass in agreement 
with that observed for the Milky Way. 
For this value, we adopt~$M_\star^\text{MW} = 5.17~\times~10^{10}~\msun$ from 
\citet[][$\pm 1.11\times10^{10}~\msun$]{Licquia2015}. 
This is the total disc mass only; when the bulge is included, the total 
becomes~$(6.08~\pm~1.17)~\times~10^{10}~\msun$. 
Since we are modeling only the disc populations here, we omit the contribution 
from the bulge to the total mass budget. 
% For this value, we adopt 
% $M_\star^\text{MW} = (5.17 \pm 1.11)\times10^{10}$~\msun~\citep{Licquia2015}. 
% This is the total disc mass of the Galaxy;~\citet{Licquia2015} report 
% $(6.08 \pm 1.17)\times10^{10}~\msun$~when the bulge is considered. 
% Since we are modeling only the disc populations here, we omit the contribution 
% the bulge would have to the total mass budget. 
% This is the total stellar mass of the Galaxy, including the bulge; 
% \citet{Licquia2015} report~$(5.17 \pm 1.11)\times10^{10}$~\msun~as the mass of 
% the disc alone. 
% Although we are not modeling the bulge here, our model extends to~$\rgal$ = 0 
% and our sample of candidate analogue star particles from~\hsim includes those 
% with bulge-like kinematics. 
\par
\begin{figure} 
\centering 
\includegraphics[scale = 0.45]{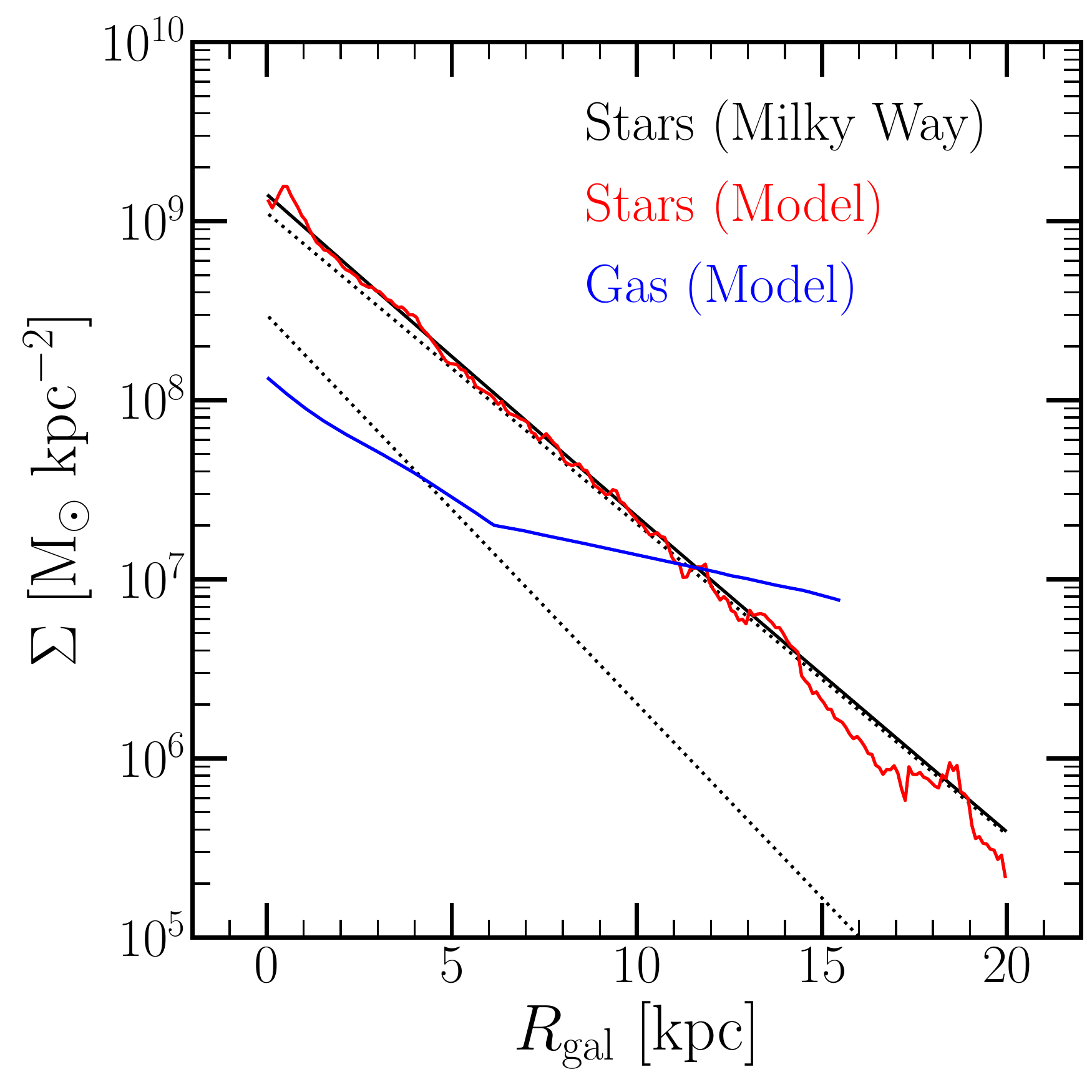} 
\caption{The surface density of gas (blue) and stars (red) as predicted by our 
fiducial model. The dotted black lines denote thin and thick disc 
profiles with scale lengths of~$R_t$ = 2.5 kpc and~$R_T$ = 2.0 kpc, 
respectively, with a ratio of~$\Sigma_T/\Sigma_t$ = 0.27 at~\rgal~= 0 (i.e. the 
thin disc profile has the higher normalization). 
The solid black line denotes the sum of the two; this is the stellar surface 
density gradient of the Milky Way as reported by~\citet{Bland-Hawthorn2016}, 
renormalized according to our adopted total stellar mass of 
$(5.17 \pm 1.11)\times10^{10}$~\msun~\citep{Licquia2015}. } 
\label{fig:surface_density} 
\end{figure}

 
We adopt a double exponential form for~$g(\rgal)$, describing the thin and 
thick disc components of the Galaxy: 
\begin{equation} 
g(\rgal) = e^{-\rgal/R_t} + \frac{\Sigma_T}{\Sigma_t} 
e^{-\rgal/R_T}~, 
\label{eq:gradient} 
\end{equation} 
where~$R_t$~and~$R_T$~are the scale radii of the thin and thick discs, 
respectively, and~$\Sigma_T/\Sigma_t$~is the ratio of their surface densities 
at~$\rgal$~= 0. 
We adopt~$R_t$~= 2.5 kpc,~$R_T$~= 2.0 kpc, and~$\Sigma_T/\Sigma_t$~= 0.27 based 
on the findings of~\citet{Bland-Hawthorn2016}. 
We plot the single exponential forms of each disc component as dotted black 
lines in Fig.~\ref{fig:surface_density}, with the solid black line denoting the 
sum of the two. 

% fig 6 

We plot the resultant surface density gradients from our fiducial, inside-out 
SFH model in Fig.~\ref{fig:surface_density} as well, with red denoting the 
stellar gradient and the gas in blue. 
The stellar gradient very nearly follows our target gradient (equation 
\ref{eq:gradient}) denoted by the solid black line. 
Stellar migration has not altered the overall form of the gradient, simply 
introducing scatter around the adopted trend. 
Although there are slight enhancements at small~$\rgal$ and deficits at large 
$\rgal$, the agreement is excellent in the regions of the Galaxy where the 
gradient is best constrained observationally. 
The~$\rgal$ > 15.5 kpc populations are composed entirely of stars that 
migrated there, since that is the radius at which we shut off star formation. 
The gas gradient shows a break in the scale radius near~$\rgal \approx$ 
6 kpc. This is a consequence of our adopted star formation law; at 
$\Sigma_\text{g} = 2\times10^7$~\msun~\persqkpc~the relation changes from 
linear at higher densities to~$\dot{\Sigma}_\star \sim \Sigma_\text{g}^{3.6}$ 
at lower densities (see discussion in~\S~\ref{sec:methods:sfe}). 
\par 
Although we adopt the~\citet{Licquia2015} disc stellar mass of the Milky Way 
here, we find similar results when taking a value which differs even by an 
order of magnitude. 
If we used a linear star formation law, then our chemical evolution would be 
independent of the mass normalization as it is in corresponding one-zone 
models (\citealp*{Spitoni2017};~\citealp{Weinberg2017, Belfiore2019}). 
The mass normalization enters our calculation because it affects the transition 
between the linear and non-linear regimes of our star formation law 
(equation~\ref{eq:sf_law}), but the impact on abundance evolution is minimal. 

\subsection{Summary} 
\label{sec:methods:summary} 
In summary, our fiducial model has an inside-out SFH with e-folding timescales 
derived from the observations of~\citet[][see discussion 
in~\S~\ref{sec:methods:sfhs}]{Sanchez2020}. 
Radial migration proceeds in a manner in which our model stellar populations 
have a change in radius~$\Delta \rgal$ informed from the~\texttt{h277} 
hydrodynamical simulation~\citep[][see discussion 
in~\S~\ref{sec:methods:h277}]{Christensen2012, Zolotov2012, Loebman2012, 
Loebman2014, Brooks2014}. 
In the baseline model, stars move to their final radius with a 
$\sqrt{\text{age}}$~dependence~\citep[][see discussion in 
\ref{sec:methods:migration}]{Frankel2018,Frankel2020}. 
Using~\vice~to calculate abundances for O and Fe in this paper, our supernova 
yields are adopted from~\citet{Johnson2020}, who in turn take these values from 
\citet{Weinberg2017} (see~\S~\ref{sec:methods:yields}). 
Outflows are characterized such that the equilibrium abundance of oxygen under 
a constant SFH follows an abundance gradient in agreement with observational 
results in the Milky Way (see~\S~\ref{sec:methods:outflows}). 
Our star formation law is based on the~\citet{Bigiel2010} and 
\citet{Leroy2013} data presented in comparison with theoretical models in 
\citet[][see~\S~\ref{sec:methods:sfe}]{Krumholz2018a}. 
To describe the stellar surface density gradient, we adopt the two-exponential 
form describing the thin and thick discs from 
\citet[][see~\S~\ref{sec:methods:surface_density_gradient}]{Bland-Hawthorn2016}. 
We adopt the~\citet{Kroupa2001} IMF throughout this paper. 
\par 
Our selection of star particles from~\hsim~yields a sample of 1,751,765 
candidate analogues with disc-like kinematics at the present day (see 
discussion in~\S~\ref{sec:methods:h277}). 
We take~$\Delta\rgal$ = 100 pc as the width of each annulus from~\rgal~= 0 to 
20 kpc and a timestep size of~$\Delta T$ = 10 Myr from~$T$ = 0 to 13.2 Gyr. 
With the resulting 200 zones and 1,321 timesteps (one extra so that age = 0 
stars are included), we let~\vice~form~$n$ = 8 stellar populations per zone per 
timestep, resulting in 2,113,600 total stellar populations with predicted 
masses and abundances. 
We set the SFR to zero beyond~\rgal~= 15.5 kpc; stellar populations 
do form beyond this radius and are part of the computational overhead, but they 
have zero mass and thus do not contribute to the chemical evolution in our 
models. 
This results in 1,627,472 stellar populations with~\textit{non-zero} masses and 
abundances, comparable to the total number of disc particles in our sample 
from~\hsim. 
These simulations run in~$\sim$5 hours on a single core with a 3 GHz 
processor and take up~$\sim$235 MB of disc space per output, including the 
extra data that we record for each stellar population's analogue star particle. 
We have also run variations with~$n$ = 2 stellar populations per 
zone per timestep, finding similar results in all cases. 
\par 
We have run~\vice~for all four of our SFHs, all four migration 
models, and all four variations in~$\tau_\text{mol}$ noted 
in~\S~\ref{sec:methods:sfe} --- a total of 64 simulations, as well as a variety 
of other test cases. 
For many of our results, only the SFH variations have a substantial impact. 
We discuss the impact of migration or~$\tau_\text{mol}$ variations where they 
are relevant.

\section{Comparison to Observations} 
\label{sec:obs_comp} 

% fig 7 
\begin{figure*} 
\centering 
\includegraphics[scale = 0.28]{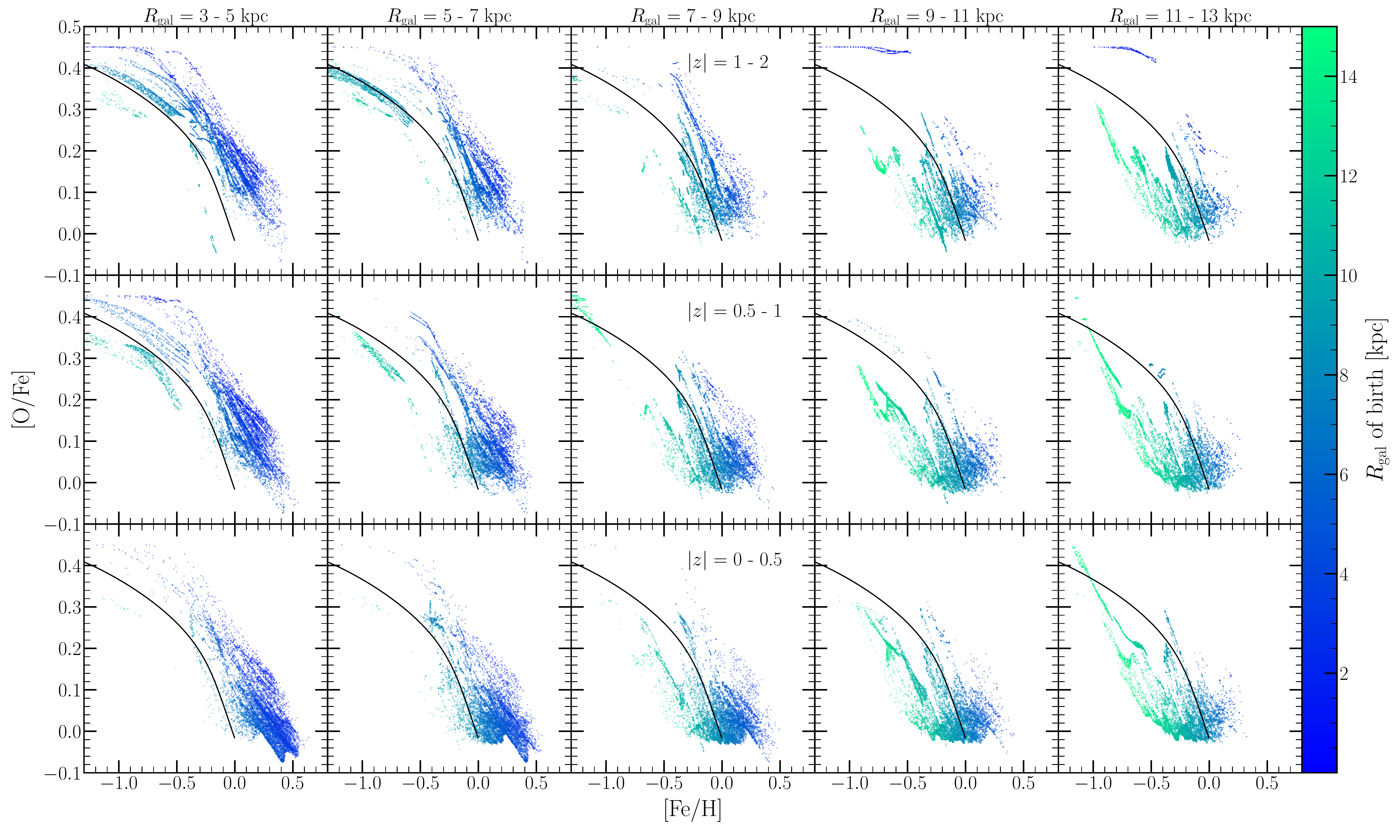} 
\caption{[O/Fe]-[Fe/H] diagrams for 15 Galactic regions spanning five bins in 
$R_\text{gal}$ and three in~$\left|z\right|$. Each region has its own panel, 
with radial bins shown in columns denoted at the top, and with bins in 
$\left|z\right|$ shown in rows denoted in text in the middle column. For each 
region, we plot~$N$~= 10,000 points sampled from our simulated stellar 
populations predicted by our fiducial model, where the probability of 
sampling is proportional to the present-day mass of each stellar population. 
In all panels, points are colour-coded according to the Galactocentric radius 
of birth of the stellar population. For reference, we plot in a solid black 
line in all panels the gas-phase [O/Fe]-[Fe/H] track predicted by the same 
SFH in the~$R_\text{gal}$ = 8 kpc annulus, but with the post-processing 
migration model; this curve is the same in all panels. } 
\label{fig:ofe_feh_diagram} 
\end{figure*} 

We begin the comparison of our model predictions to observational data with the 
distribution of stellar populations in the [O/Fe]-[Fe/H] plane. We separate 
stars into bins based on their present-day Galactic regions defined by five 
bins in~$R_\text{gal}$ (3 - 5, 5 - 7, 7 - 9, 9 - 11, and 11 - 13 kpc) and three 
bins in~$\left|z\right|$ (0 - 0.5, 0.5 - 1, and 1 - 2 kpc). Within each of the 
resulting 15 regions, we sample 10,000 stars at random from our baseline 
inside-out SFH model. Since stars in~\texttt{VICE} are stand-ins for entire 
stellar populations, we let the probability of sampling one of them be 
proportional to its present day mass. We plot the results of this procedure in 
Fig.~\ref{fig:ofe_feh_diagram}, colour-coding each stellar population by its 
birth radius; for visual reference, we also plot the gas-phase track which 
resulted from the~$R_\text{gal}$ = 8 kpc annulus with the post-processing 
migration model in a black solid line in all panels. 
\par 
In Fig.~\ref{fig:ofe_feh_diagram}, we note that high-$\alpha$~sequence stars 
are predicted to be the dominant population at small~$R_\text{gal}$ and high 
$\left|z\right|$; conversely, the low-$\alpha$~population dominates the 
statistics at large~$R_\text{gal}$~and low~$\left|z\right|$. This is consistent 
with the observational results of~\citet{Hayden2015}, who present a density map 
in the [$\alpha$/M]-[M/H] plane for the same Galactic regions (see their Fig. 
4).\footnote{
	In~\citet{Hayden2015}, [M/H] represents an overall scaling of elements with 
	solar mixture, and [$\alpha$/M] represents a scaling of~$\alpha$-elements 
	with respect to others. To a good approximation they are proxies for 
	[Fe/H] and [O/Fe], respectively, and we will treat them as such in this 
	paper. 
} 
Furthermore, the locus of the low-$\alpha$ sequence shifts from super-solar 
[Fe/H] to sub-solar [Fe/H] with increasing~$R_\text{gal}$, a shift which is 
expected given the abundance gradient that we have built into our models (see 
discussion in~\S~\ref{sec:methods:outflows}). 
The colour-coding of the points 
shows that the width of the low-$\alpha$~sequence arises out of 
stellar migration: low-$\alpha$ stars with high [Fe/H] formed in the inner 
Galaxy, and those with low [Fe/H] formed in the outer Galaxy. 
The low-$\alpha$ locus thus represents a superposition of populations achieved 
by radial migration rather than an evolutionary sequence, the interpretation 
proposed by, e.g.,~\citet{Schoenrich2009a} and~\citet{Nidever2014}. 
\par 
In Fig.~\ref{fig:ofe_feh_diagram} one can clearly see the imprint of 
evolutionary tracks from different radii, appearing in the same present-day 
$R_\text{gal}$ bin because of radial mixing. 
Though this is to some extent a consequence of the discretization of the 
Galaxy disc in our model, it also arises out of correlated fluctuations in the 
SN Ia rate (see discussion in~\S~\ref{sec:obs_comp:gradient}). 
Observational scatter would wash out the appearance of distinct tracks, and 
intrinsic scatter might blur them if our chemical evolution model was less 
idealized. 
Our model reproduces several of the qualitative trends found by 
\citet{Hayden2015}, but the distribution is less obviously bimodal. 
We quantify this point in~\S~\ref{sec:obs_comp:ofe_dists} below. 
If we remake Fig.~\ref{fig:ofe_feh_diagram} for any of our other SFHs, or for 
our alternative migration or star formation efficiency prescriptions, the 
appearance is qualitatively similar. There are significant quantitative 
differences in some cases, which we discuss in subsequent subsections. 

\subsection{Abundance Gradients} 
\label{sec:obs_comp:gradient} 

\begin{figure*} 
\centering 
\includegraphics[scale = 0.42]{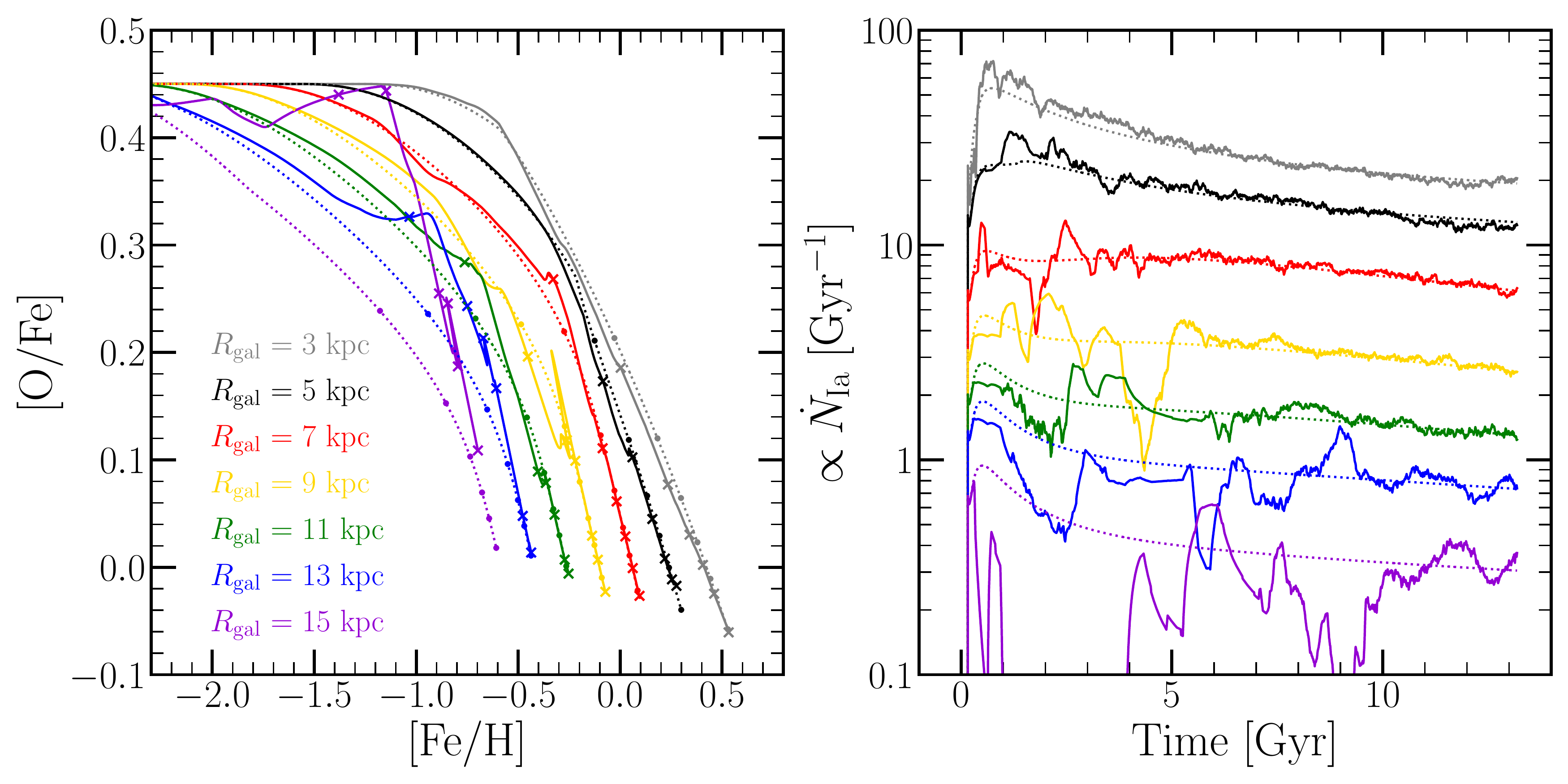} 
\caption{\textbf{Left}: Gas phase evolutionary tracks in the [O/Fe]-[Fe/H] 
plane for our inside-out SFH with either post-processing (dotted lines) or 
diffusion (solid lines) migration models. We plot tracks for seven of our 
$\Delta\rgal$ = 100 pc rings, 
colour-coded according to their Galactocentric radius and denoted by the 
legend in the lower-left. We mark simulation times of 2, 4, 6, 8, 10, and 13.2 
Gyr in X's for the diffusion model and points for the post-processing model. 
\textbf{Right}: As a function of simulation time, a proxy for the SN Ia rate 
using the total time-derivative of the Fe mass in a given annulus, calculated 
by subtracting the contribution from recycling and CCSN enrichment and adding 
back that lost to star formation and outflows. We show these rates for the 
same rings as in the left-hand panel, multiplying them by various prefactors 
to improve clarity. } 
\label{fig:tracks} 
\end{figure*}

The left panel of Fig.~\ref{fig:tracks} shows gas phase [O/Fe]-[Fe/H] 
evolutionary tracks at several radii assuming the inside-out SFH, using either 
our post-processing (dotted) or diffusion (solid) migration prescriptions 
(see~\S~\ref{sec:methods:migration} and Fig.~\ref{fig:migration_schema}). 
For the post-processing model the tracks are smooth and simply follow the 
predictions of one-zone GCE models with the parameters appropriate to each 
radius --- with post-processing, radial migration has no impact on the gas-phase 
evolution. 
However, some of the tracks in the diffusion model are notably 
different, especially at early times and large radii. 
These differences arise because radial migration can transport stars 
significantly over the timescale of SN Ia enrichment, as discussed in~\S 
\ref{sec:methods:h277}.
\par 
To demonstrate this point, the right panel of Fig.~\ref{fig:tracks} plots a 
proxy of the SN Ia rate\footnote{
	From our~\vice~outputs, the total time derivative of the Fe mass in a given 
	annulus can be obtained by its change across a single timestep. By then 
	subtracting the CCSN contribution (known exactly given our adopted yield 
	and the SFR), approximately correcting for recycling, and adding back that 
	which was lost to star formation and outflows, we obtain a simple estimate 
	of Fe production by SNe Ia. 
} as a function of time for the same 0.1 kpc rings plotted in the left panel. 
At large radii the sources for the diffusion model exhibit large fluctuations 
relative to the smooth predictions of the underlying one-zone models, as 
migration boosts or depletes the predicted number of SN Ia progenitors. 
The deficits or excesses in SN Ia rate in turn drive upward or downward 
deviations in [O/Fe] relative to the smooth model tracks in the left panel. 
As an extreme example, in the 15 kpc annulus the SN Ia rate is nearly zero for 
the first~$\sim$3 Gyr, and its resulting [O/Fe] track is nearly flat over this 
interval, even though the smooth model track has dropped from [O/Fe] = 0.45 to 
[O/Fe]~$\approx$ 0.2. 
Although we illustrate the SN Ia rates for only a handful of rings, we have 
found that the variations between nearby rings are highly correlated in our 
models, suggesting that regions of the Galaxy move coherently in 
[O/Fe]-[Fe/H] space; this lends further insight into the streak-like appearance 
of some sub-populations in Fig.~\ref{fig:ofe_feh_diagram}. 

% fig 8

% fig 9 
\begin{figure*} 
\centering 
\includegraphics[scale = 0.32]{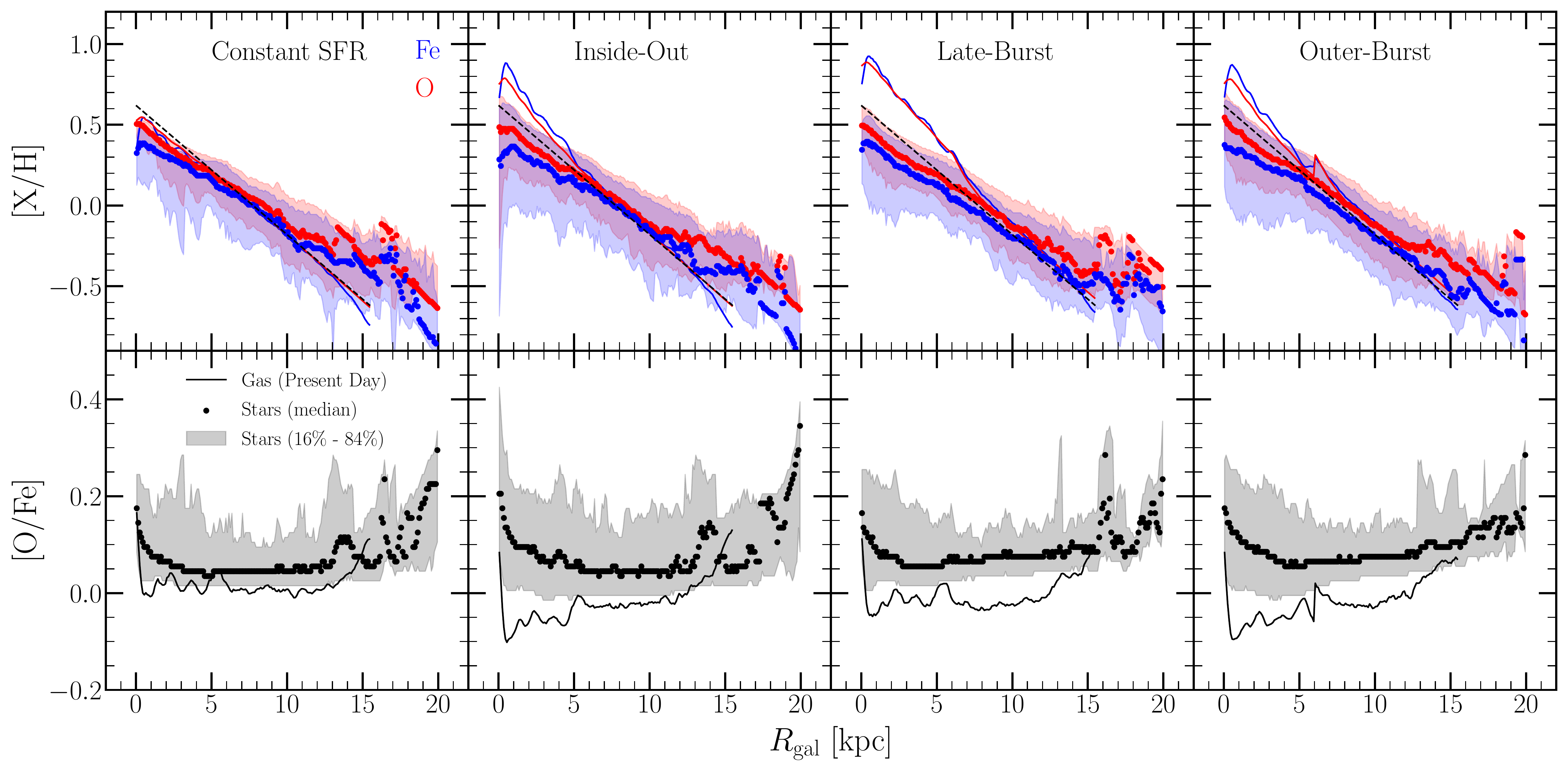} 
\caption{Radial abundance gradients in [O/H] (top, red), [Fe/H] (top, blue), 
and [O/Fe] (bottom) for our four SFHs with diffusion migration - constant (far 
left), inside-out (left-middle), late-burst (right-middle), and outer-burst 
(far right). We plot the gas-phase abundance at the present day as a function 
of Galactocentric radius in solid lines. Points denote the median of the stellar 
MDF of the 100-pc width ring at a given radius, with shaded regions marking 
the 16th and 84th percentiles thereof. Black lines in the top panels denote our 
target [$\alpha$/H] gradient of mode([$\alpha$/H]) = +0.3 at~$R_\text{gal}$~= 
4 kpc with a slope of -0.08 kpc$^{-1}$. } 
\label{fig:metallicity_gradient} 
\end{figure*} 

Because we are analyzing a single hydrodynamic simulation, we cannot say 
whether the systematic depression of the SN Ia rate at large radii and early 
times is a general expectation or a consequence of the specific dynamical 
history of this galaxy. 
However, greater fluctuations at large~\rgal~and small~$t$ are a natural 
consequence of the low SFR. These fluctuations can have a significant impact 
on [O/Fe]-[Fe/H] evolution even if their sign varies from galaxy to galaxy. 
The fact that >10\% of events are seen at >10 kpc from their host galaxies in 
the ASAS-SN bright SN catalog~\citep{Holoien2019} adds qualitative 
observational support to the argument that SN Ia progenitors may often form at 
significantly different radii than where their explosions are observed. 
We will show below that these SN Ia rate fluctuations lead to the formation of 
$\alpha$-enhanced intermediate age populations that do not arise in our 
post-processing radial migration model (see Fig.~\ref{fig:age_alpha}). 
\par 
Fig.~\ref{fig:metallicity_gradient} plots the radial gradients of gas phase 
and stellar abundances, [O/H], [Fe/H], and [O/Fe], for our four SFH models. 
For constant SFR, the gas phase abundances closely track the target gradient 
that we used to set our~$\eta(\rgal)$ profile (see equation~\ref{eq:eta_rgal}). 
For a declining SFR the equilibrium abundance is higher than that assumed for 
equation~\refp{eq:eta_rgal}~\citep[see][]{Weinberg2017}, and in the inside-out 
and outer-burst models the gas phase abundances rise above the target gradient 
at small~\rgal~where the decline is fastest. 
In the late-burst model the inner Galaxy abundances are further boosted by 
enhanced late-time star formation, and a similar effect is seen at~\rgal~= 6 
kpc in the outer-burst model. 
In accretion-induced starbursts, re-enrichment can briefly produce 
super-equilibrium abundances in the gas-phase, which then decay back to the 
equilibrium abundance as the SFR declines~\citep{Johnson2020}. 
\par 
The stellar metallicity gradients are shallower than in the gas phase, and 
they are similar in the four models. 
The 16th-84th percentile range at each radius is large, typically 0.3-0.5 
dex, and typically larger in [Fe/H] than [O/H]. 
The mode of the stellar metallicity distribution is a noisy quantity in our 
0.1-kpc rings, so points in Fig.~\ref{fig:metallicity_gradient} show the 
median of the distributions. 
The trends for the mode are slightly steeper, as expected given the change in 
shape of the metallicity distribution functions (see~\S 
\ref{sec:obs_comp:mdfs}), and closer to the target gradient shown by the black 
solid line. We do not include observational data in this figure, but the target 
gradient itself is observationally motivated, so Fig. 
\ref{fig:metallicity_gradient} implies that our models, by design, give a 
reasonable match to Milky Way abundance gradients. 
\par 
Median [O/Fe] values are close to solar at nearly all radii, rising at the 
smallest~\rgal~and in some models at the largest~\rgal. However, the spread in 
[O/Fe] is large at all radii, and typical values depend on~\absz~and [Fe/H] as 
shown in Fig.~\ref{fig:ofe_feh_diagram}. 
We present a comparison to observations in~\S~\ref{sec:obs_comp:ofe_dists} 
below. 
The larger values of [O/Fe] beyond~\rgal~= 15.5 kpc are expected, since this is 
the radius at which we shut off star formation. In our models, all stellar 
populations at these radii migrated there, so they tend to be old and therefore 
$\alpha$-enhanced. 
The idea that outer disc populations are dominated by migration was proposed 
by~\citet{Roskar2008b} based on simulation predictions. 
\citet{RadburnSmith2012} present observational evidence for this prediction in 
the observations of the NGC 7793 disc. 
Although we are not modeling the bulge in this paper, our model predicts the 
disc stars in these regions to have higher [O/Fe] than at larger~\rgal, in 
qualitative agreement with observations (see discussion in, e.g., 
\citealp{Duong2019, Griffith2021}). 

\subsection{Metallicity Distribution Functions} 
\label{sec:obs_comp:mdfs} 

Metallicity distribution functions (MDFs) and their variations with Galactic 
region are a core prediction of GCE models. 
We have computed distributions of [Fe/H] and [O/H] using abundances from the 
16th data release~\citep[DR16;][]{Ahumada2020, Joensson2020} of APOGEE 
\citep{Majewski2017}. 
Abundances are determined by the APOGEE Stellar Parameters and Chemical 
Abundances Pipeline (ASPCAP;~\citealp{Holtzman2015, GarciaPerez2016}). 
We restrict our sample to stars with effective temperatures of 4000 K 
$\leq T_\text{eff} \leq$~4600 K, surface gravities of 1.0~$\leq \log g \leq$ 
2.5, and signal-to-noise ratios of at least 100. 
These cuts ensure that our sample consists of stars on the upper red giant 
branch, luminous enough to cover all regions of the disc, while excluding red 
clump stars to avoid possible abundance offsets between these stars and stars 
on the giant branch. 
Our observational results are similar to those shown by~\citet{Hayden2015}, 
but we use a larger data set, more recent APOGEE observations, and [O/H] and 
[Fe/H] rather than [$\alpha$/H] and [M/H]. 
\par
\begin{figure} 
\centering 
\includegraphics[scale = 0.34]{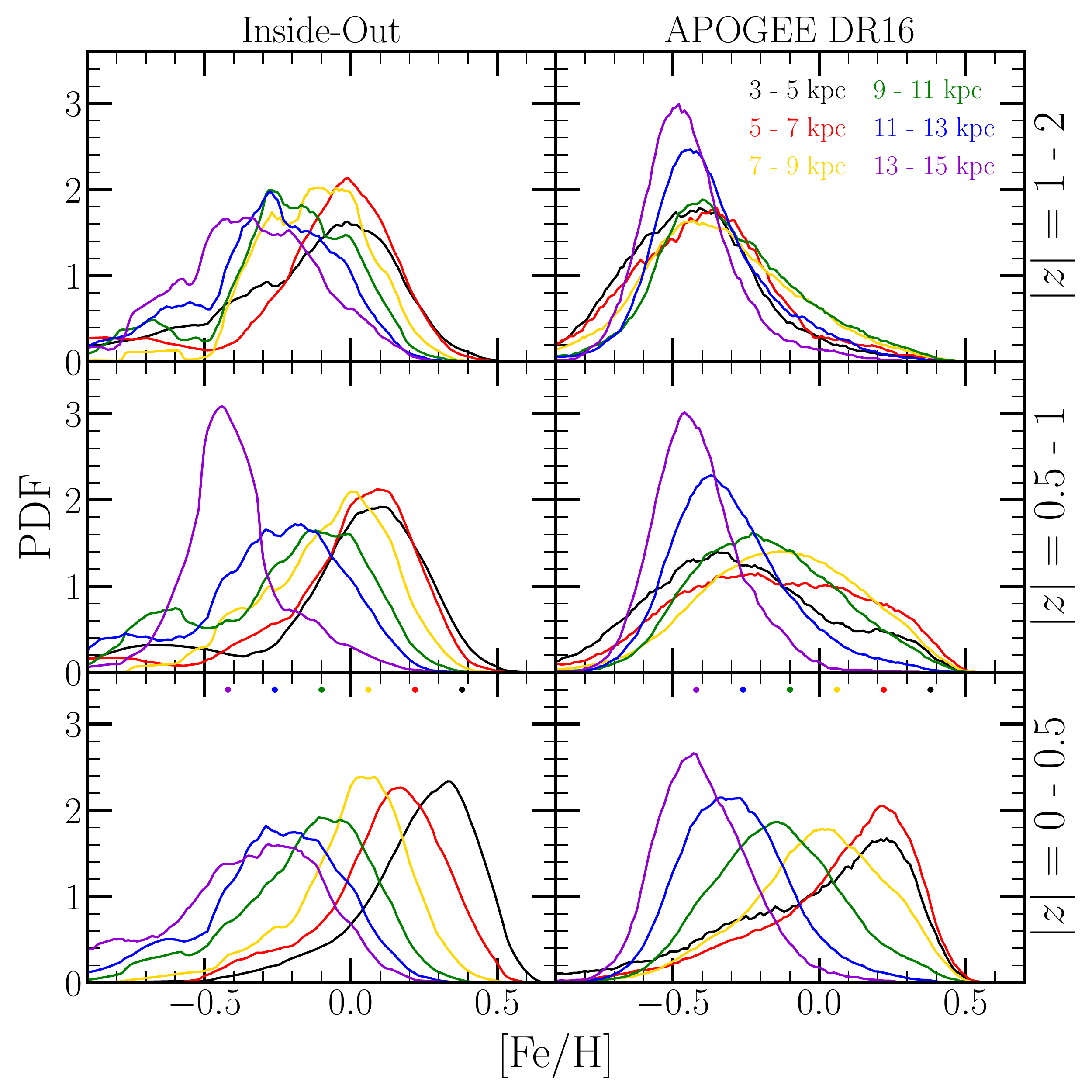} 
\caption{Metallicity distribution functions (MDFs) in [Fe/H] predicted by our 
fiducial, inside-out model (left) and as observed in APOGEE DR16 (right), for 
stars and simulated stellar populations with present-day~$\left|z\right|$~= 0 - 
0.5 kpc (bottom), 0.5 - 1 kpc (middle), and 1 - 2 kpc (top). MDFs are shown in 
bins of Galactocentric radius: 3 - 5 kpc (black), 5 - 7 kpc (red), 7 - 9 kpc 
(yellow), 9 - 11 kpc (green), 11 - 13 kpc (blue), and 13 - 15 kpc (purple). 
The points near the top of the bottom panels denote what the mode abundance 
would be if it followed out target gradient of [Fe/H] = +0.3 at~$R_\text{gal}$ 
= 4 kpc with a slope of -0.08 kpc$^{-1}$ exactly, assuming the inner radius of 
each bin (i.e. there is no point plotted for 15 kpc). All distributions are 
smoothed with a box-car width of [Fe/H]~$\pm$~0.1 for clarity. } 
\label{fig:mdf_3panel_fe} 
\end{figure}

 
The left and right panels of Fig.~\ref{fig:mdf_3panel_fe} show [Fe/H] 
distributions from our fiducial inside-out model and the APOGEE data, 
respectively, in bins of~\rgal~and~\absz. 
The observed midplane distributions (\absz~$\leq$~0.5 kpc) show the striking 
phenomenon first noted by~\citet{Hayden2015}: a shift from a skew-negative form 
in the inner Galaxy to a skew-positive form in the outer Galaxy, with a roughly 
symmetric [Fe/H] distribution at the solar circle. 
The simulation reproduces this behaviour, confirming that realistic radial 
migration can explain the radial dependence of the MDF shape as conjectured by 
\citet{Hayden2015} and illustrated in a more idealized simulation by 
\citet{Loebman2016}. 
\par 
In detail, the MDFs are not as smooth at large~\rgal~as in the observations; 
they are not as skewed in the inner and outer disc either. 
Dots in the lower left panel mark the target metallicities implied by our 
$\eta(\rgal)$ prescription in equation~\refp{eq:eta_rgal}. 
The modes of the stellar MDFs track these targets quite closely, except at 
\rgal~= 13 - 15 kpc, where the migrated population is so large compared to the 
in-situ population that it reshapes the peak of the MDF as well as the tails. 
The model also predicts a continuing increase of the mode metallicity down to 
\rgal~= 3 - 5 kpc, while the observed mode is the same at 3 - 5 kpc and 5 - 7 
kpc. 
We do not view this as a serious discrepancy, as it could be reduced by a
moderate adjustment of the~$\eta(\rgal)$ recipe so that the metallicity 
gradient in these regions is flat. 
We have computed such a comparison case and find that it does indeed reproduce 
this observational result, though it still underpredicts the skewness at 
small~\rgal. 
Alternatively, the flattening of the observed gradient could be a consequence 
of more aggressive quenching of star formation than assumed in our models. 
The surface density of star formation~$\dot{\Sigma}_\star$ in the Milky Way 
is known to reach a maximum at~\rgal~$\approx$ 4 kpc and decline by a factor 
of a few at smaller radii (see Fig. 1 of~\citealp{Peek2009} and Fig. 2 
of~\citealp{Fraternali2012} and data therein). 
Early quenching could cutoff the MDF at high [O/H] and [Fe/H] if it happens 
before the ISM reaches equilibrium abundance. 
Visual inspection of Fig.~\ref{fig:tracks} suggests that this process should 
occur around~$T \approx$ 6 - 8 Gyr if the MDF is to peak at 
[Fe/H]~$\approx$~+0.2 - 0.3. 

% fig 10 

% fig 11 
\begin{figure} 
\centering 
\includegraphics[scale = 0.34]{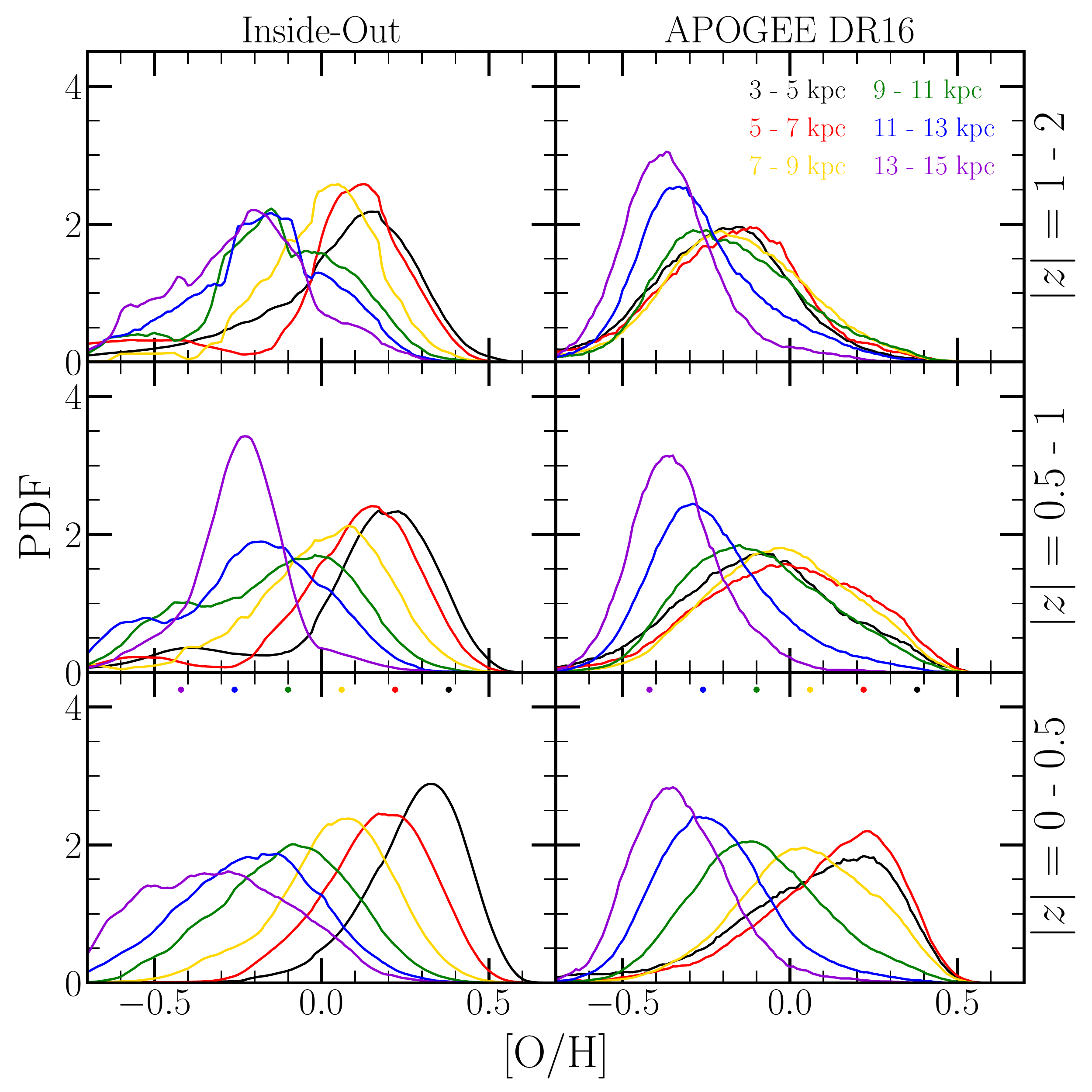} 
\caption{The same as Fig.~\ref{fig:mdf_3panel_fe}, but for [O/H]. } 
\label{fig:mdf_3panel_o} 
\end{figure} 

Going up from the midplane, the observed MDFs for the four inner annuli 
(\rgal~< 11 kpc) shift to lower average metallicity, and they converge in 
location and in shape, being roughly symmetric at~\absz~= 0.5 - 1 kpc and 
mildly skew-positive at~\absz~= 1 - 2 kpc. 
Qualitatively, the simulation reproduces this shift of mean metallicity, change 
of shape, and convergence of distributions at different~\rgal, although the 
model does not account for the entirety of the effect. 
We consider this a significant success of this simulation-based approach 
because our GCE model is constructed and tuned in~\rgal~alone, so trends with 
\absz~follow from the combinations of age-metallicity trends, age-velocity 
trends arising from ``upside-down'' disc formation and dynamical heating, and 
correlations between radial migration and vertical energy. 
\citet{Freudenburg2017} showed that chemical evolution in a vertically settling 
gas disc could explain the MDF trend with~\absz~seen by APOGEE in the inner 
disc, and here we see similar behaviour in a more fully ab initio model. 
\citet{Bird2020} showed that the dynamics of the~\hsim~simulation leads to good 
agreement with the observed age-velocity relation, and here we show that 
this success extends qualitatively to the vertical trends of chemical 
abundances. 
\par 
Quantitatively, there are significant differences between the predicted and 
observed distributions above the midplane. 
The model MDFs from~\absz~= 0.5 - 1 kpc are narrower and more skewed than the 
observed MDFs, with higher median [Fe/H]. 
At~\absz~= 1 - 2 kpc, the predicted MDFs are not as strikingly converged as the 
observed MDFs. 
In both the data and the model, the high-\absz~MDFs for the 11 - 13 and 13 - 15 
kpc annuli remain closer to their midplane counterparts. 
The model [O/Fe]-[Fe/H] distributions in the outer Galaxy appear to show the 
imprint of a few large migration episodes (see Fig.~\ref{fig:ofe_feh_diagram}). 
The smoothness of the observed MDFs at these radii and their similarity across 
\absz~suggests a more vigorous stirring. 
\par 
Fig.~\ref{fig:mdf_3panel_o} plots distributions of [O/H] instead of [Fe/H]. 
The appearance is quite similar to Fig.~\ref{fig:mdf_3panel_fe}, which is 
unsurprising but non-trivial given the different timescales of CCSN and SN Ia 
enrichment. 
The agreement and disagreement between the model and data are similar, though 
the discrepancy for~\absz~= 0.5 - 1 kpc is somewhat clearer in [O/H]. 
The model's outer Galaxy MDFs are less irregular in [O/H] than in [Fe/H], an 
indication that some of the structure in the [Fe/H] distributions is caused by 
the large fluctuations in the SN Ia rate as seen in Fig.~\ref{fig:tracks}. 
We have confirmed this conjecture by computing [Fe/H] MDFs for the 
post-processing radial migration prescription, finding that they are indeed 
more smooth at~\rgal~= 11 kpc. 
\par 
The MDF predictions for our other SFH scenarios - constant SFR, late-burst, and 
outer-burst - are different in detail, but they show the same qualitative 
trends as those in Figs.~\ref{fig:mdf_3panel_fe} and~\ref{fig:mdf_3panel_o}. 
Our findings on the radial and vertical trends of the MDF are also similar 
to those of~\citet{Loebman2016}, who use a galaxy simulation evolved from 
rotating gas in a dark matter halo rather than cosmological initial conditions. 
The qualitative similarity of our results across different models implies that 
the radial and vertical trends are a generic effect of radial migration, 
upside-down disc formation, and dynamical heating in a galaxy with realistic 
abundance gradients and time evolution.

\subsection{[O/Fe] distributions in Bins of [Fe/H]} 
\label{sec:obs_comp:ofe_dists} 

% fig 12 
\begin{figure*} 
\centering 
\includegraphics[scale = 0.32]{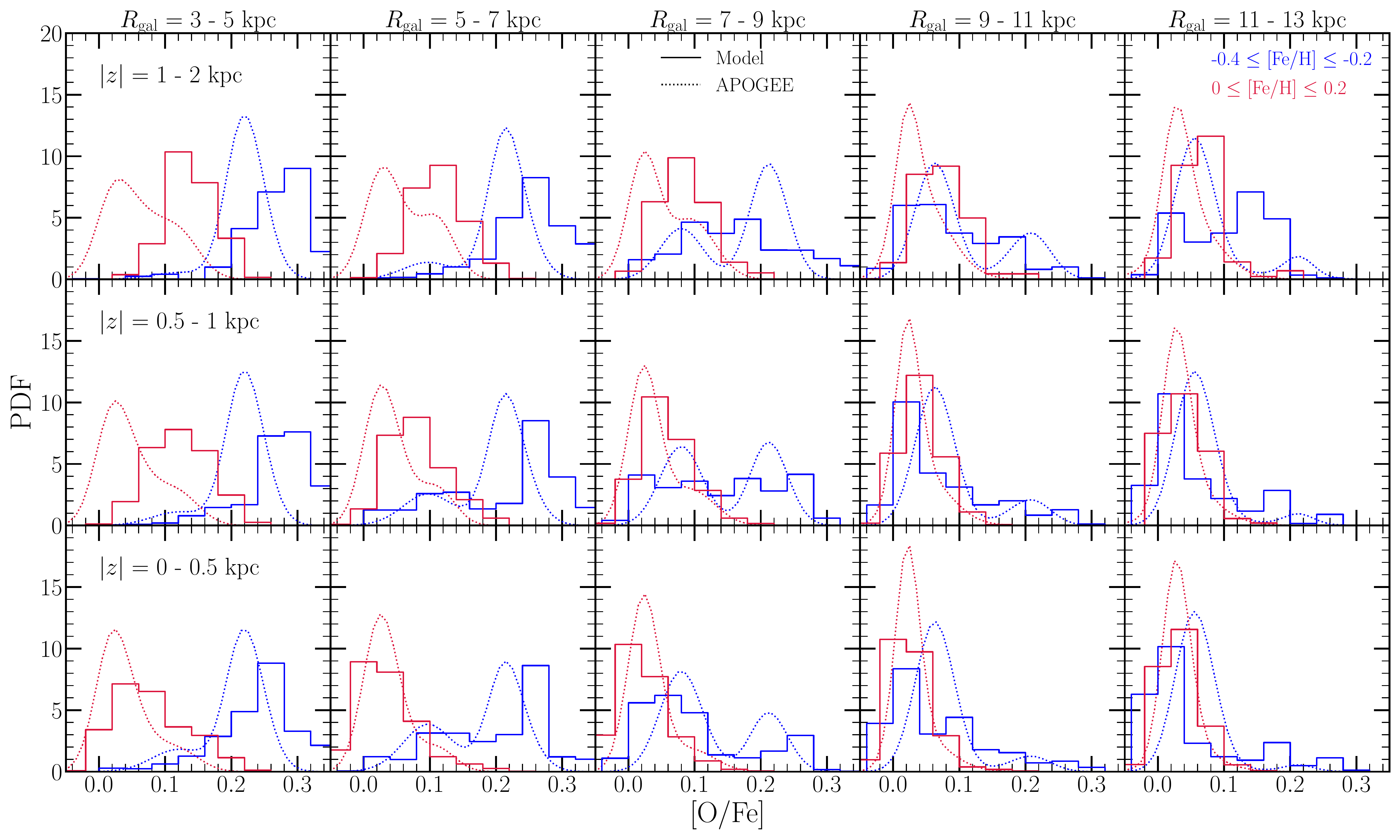} 
\caption{Predicted distributions in [O/Fe] in 15 Galactic regions and in two 
bins in [Fe/H]. Columns correspond to bins in~$R_\text{gal}$, denoted at the 
top of each column. Rows correspond to bins in~$\left|z\right|$, denoted in 
text in the left-hand column. Distributions are color-coded according to the 
[Fe/H] the sample is drawn from, denoted by the legend in the upper right 
panel. Solid lines represent that predicted by our inside-out SFH in 
$\Delta$[O/Fe] = 0.04 bins, while dashed lines correspond to the fits to the 
APOGEE DR16 data presented in~\citet{Vincenzo2021a}, which quantify the 
intrinsic distributions accounting for observational uncertainties and the 
APOGEE selection function. } 
\label{fig:ofe_mdfs_insideout} 
\end{figure*} 

Fig.~\ref{fig:ofe_mdfs_insideout} compares our predicted [O/Fe] distributions 
in Galactic zones to those recently published by 
\citet{Vincenzo2021a}.\footnote{
	\citet{Vincenzo2021a} present [Mg/Fe] distributions in their figures, but 
	they have quantified the [O/Fe] as well, and we use the latter here. 
}
As in our~\S~\ref{sec:obs_comp:mdfs}, they use APOGEE DR16 with similar cuts on 
effective temperature, surface gravity, and signal-to-noise (see their~\S~2). 
They correct for observational scatter as well as age-dependent and (more 
importantly)~\absz-dependent selection in APOGEE to infer the intrinsic 
distribution of [O/Fe] in bins of [Fe/H] that would be found for an unbiased 
sample of long-lived disc stars. 
At a given~\rgal~and [Fe/H],~\citet{Vincenzo2021a} fit a model comprised of two 
Gaussians in [O/Fe], one each for the high-$\alpha$ and low-$\alpha$ 
populations.
The~\absz-dependence of the distribution follows from the empirical scale 
heights of these two populations, taken from~\citet{Bovy2016}. 
Dotted curves in Fig.~\ref{fig:ofe_mdfs_insideout} show these models integrated 
over the corresponding ranges in~\absz. 

Solid histograms in Fig.~\ref{fig:ofe_mdfs_insideout} show distributions in 
0.4-dex bins of [O/Fe] for our fiducial inside-out model. 
For clarity we have chosen to focus our comparison on two bins of [Fe/H], one 
just above solar metallicity (0.0~$\leq$ [Fe/H]~$\leq$ 0.2) and one at 
sub-solar metallicity (-0.4~$\leq$ [Fe/H]~$\leq$ -0.2) where the bimodality of 
[O/Fe] is most pronounced. Results for our other models are qualitatively 
similar, and these two [Fe/H] bins illustrate the range of model successes and 
failure. 
\par 
Beginning with the [Fe/H] = 0 - 0.2 bin, we see that the model roughly 
reproduces the observed width of the [O/Fe] distribution. 
In the midplane (\absz~$\leq$ 0.5 kpc) zones it predicts the correct 
skew-positive shape, though the peak of the observed distribution is sharper 
than the model prediction. However, the model histograms shift towards higher 
[O/Fe] with increasing~\absz~while the peak of the observationally inferred 
distributions stays fixed, a discrepancy that is most obvious in the~\rgal~= 
3 - 5 kpc and 5 - 7 kpc bins. The constancy of the observed peak is partly 
a consequence of the~\citet{Vincenzo2021a} fitting procedure, which assumes 
that the location of the high-$\alpha$ and low-$\alpha$ Gaussians stay 
constant at a given~\rgal~and [Fe/H] and only their relative amplitudes change 
with~\absz. We have gone back to the raw data histograms fit by 
\citet{Vincenzo2021a}, and while they do allow some increase in model [O/Fe] 
with~\absz, they do not allow a shift as large as that predicted by our model. 
At~\absz~= 1 - 2 and~\rgal~= 3 - 5 and 5 - 7 kpc, the number of stars 
contributing to the fit is 31 and 17 respectively, so while the shape of the 
distribution is not well constrained, the centroid is robustly determined. 
\par 
This discrepancy could reflect differences in the dynamical heating history 
of the~\hsim~simulation and the Milky Way. 
With its most recent major merger occurring at~$z \approx$ 3, this galaxy was 
previously selected for investigation for its quiescent merger history 
\citep[e.g.][]{Zolotov2012}. N-body models for the tidal disruption of the 
Sagittarius dwarf galaxy suggest repeated pericentric passages at 1 - 2 Gyr 
intervals~\citep{Law2010}, which could trigger episodes of infall and star 
formation~\citep[e.g.][]{RuizLara2020}. 
These pericentric passages might also heat low-$\alpha$ disc populations to 
higher~\absz, an effect absent in~\hsim. 
Alternatively, we have investigated the impact of changing our~$\eta(\rgal)$ 
prescription to become constant within~\rgal~= 5 kpc, as discussed previously 
in~\S~\ref{sec:obs_comp:mdfs} in the context of the [Fe/H] gradient. 
This change also dampens the predicted trend of [O/Fe] with~\absz~by bringing 
the inner Galaxy's [O/Fe]-[Fe/H] tracks closer together (see 
Figs.~\ref{fig:ofe_feh_diagram} and~\ref{fig:tracks}). 
We find that this change can account for some but not all of the discrepancy, 
shifting the peak of the distribution down by~$\sim$0.05 dex in the upper left 
panel of Fig.~\ref{fig:ofe_mdfs_insideout}. 
\par 
Turning to the -0.4~$\leq$ [Fe/H]~$\leq$ -0.2 bin, the model shows partial but 
by no means complete success. 
It does reproduce the breadth of the [O/Fe] distribution at this intermediate 
metallicity, and in nearly all~\rgal-\absz~zones it predicts skewness of the 
correct sign. 
At~\rgal~= 3 - 5 kpc, the model predicts mode([O/Fe])~$\approx$ 0.3, while 
the observed mode is at [O/Fe]~$\approx$ 0.22. 
This discrepancy is affected by our choice of CCSN yields, which produces an 
[O/Fe] plateau at +0.45. 
While this value is consistent with some observational data (e.g. the 
\citealp*{Ramirez2013} data set modeled by~\citealp{Andrews2017}), APOGEE 
measurements place the plateau at [O/Fe]~$\approx$ +0.3, so it is not 
surprising that our model overpredicts [O/Fe] at low metallicity. 
Unfortunately, uncertainties in the observed abundance scales and the 
theoretical CCSN elemental yields remain an obstacle to sharp GCE model tests. 
\par 
The most significant discrepancy with data is for~\rgal~= 7 - 9 kpc, where the 
observations show a clearly bimodal [O/Fe] distribution in all three~\absz~ 
ranges but the model predicts bimodality only near the midplane. 
This bimodality is also evident in the raw data histograms prior to model 
fitting and correction for selection effects (see Figs. 10 and 11 of 
\citealp{Vincenzo2021a}). 
The idea that radial migration could give rise to an [$\alpha$/Fe] dichotomy 
was proposed by~\citet{Schoenrich2009a}, who noted that the evolutionary tracks 
at any given~\rgal~would produce most stars at low [O/Fe] and that radial 
mixing of these populations would produce a low-$\alpha$ ``sequence'' that is 
a superposition of these evolutionary endpoints. 
\citet{Nidever2014} explored this superposition scenario in the context of 
APOGEE observations, and~\citet{Sharma2020} have recently implemented a 
detailed parameterized scenario matched to APOGEE. 
Although the superposition effect clearly operates in our model, as shown in 
Fig.~\ref{fig:ofe_feh_diagram}, the model produces too many stars at 
intermediate [O/Fe], so there is no clear minimum between the high-$\alpha$ 
and low-$\alpha$ peaks. 
As argued by~\citet{Vincenzo2021a}, the generic problem is that one-zone models 
with smooth evolutionary histories always produce most stars near the low 
[$\alpha$/Fe] endpoints and a much smaller peak at the high-$\alpha$ 
plateau, so one cannot superpose such models in a way that produces strong 
bimodality. 
The~\citet{Sharma2020} model seems like a possible counter-example, but they 
adopt parameterized evolutionary tracks and do not demonstrate that these can 
be drawn from a self-consistent chemical evolution model. 
\par 
The simplest way to produce stronger [$\alpha$/Fe] bimodality in our models 
would be to adopt a two-phase star formation model like that envisioned in 
two-infall scenarios of chemical evolution~\citep[e.g.][]{Chiappini1997, 
Chiappini2001, Romano2010, Grisoni2017, Noguchi2018, Palla2020, Spitoni2016, 
Spitoni2018, Spitoni2019a, Spitoni2020, Spitoni2021}. 
These scenarios turn down the SFR while the ISM abundances pass through the 
intermediate [$\alpha$/Fe] regime. 
Other scenarios such as large early starbursts~\citep{Clarke2019} or reverse 
[Fe/H] evolution with late-time outflows~\citep{Weinberg2017} could also 
enhance bimodality, but the late starbursts considered here are insufficient 
on their own. 
We will explore these alternative scenarios for the origin of bimodality in 
future work. 

\subsection{The Age-[$\alpha$/Fe] Relation} 
\label{sec:obs_comp:age_alpha} 
In this section, we assess the predicted age-[O/Fe] relations of our models, 
using the results of~\citet{Feuillet2019} as the observational benchmark. 
Their stellar age measurements are based on isochrone matching, 
using APOGEE DR14 stars for which parallax measurements are available 
from Gaia~\citep{Abolfathi2014, GaiaDR2}. 
With their spatial and quality cuts, the final sample consisted of 77,562 
stars. 
In bins of [O/Fe], they assume a Gaussian distribution of log age, fitting the 
mean (or equivalently, the median) and standard deviation to the stars in that 
bin. 
The choice of a Gaussian distribution is driven by simplicity, enabling 
estimates of mean and spread for data with significant age errors, but because 
our model predicts non-Gaussian log age distributions at fixed [O/Fe], the 
comparison to data is not free of nuance. 
We base most of our model comparisons below on the mass-weighted median age, 
simply denoting the 50th percentile of the mass-weighted age distribution of 
some subsample. 
\par 

% fig 13 
\begin{figure*} 
\centering 
\includegraphics[scale = 0.34]{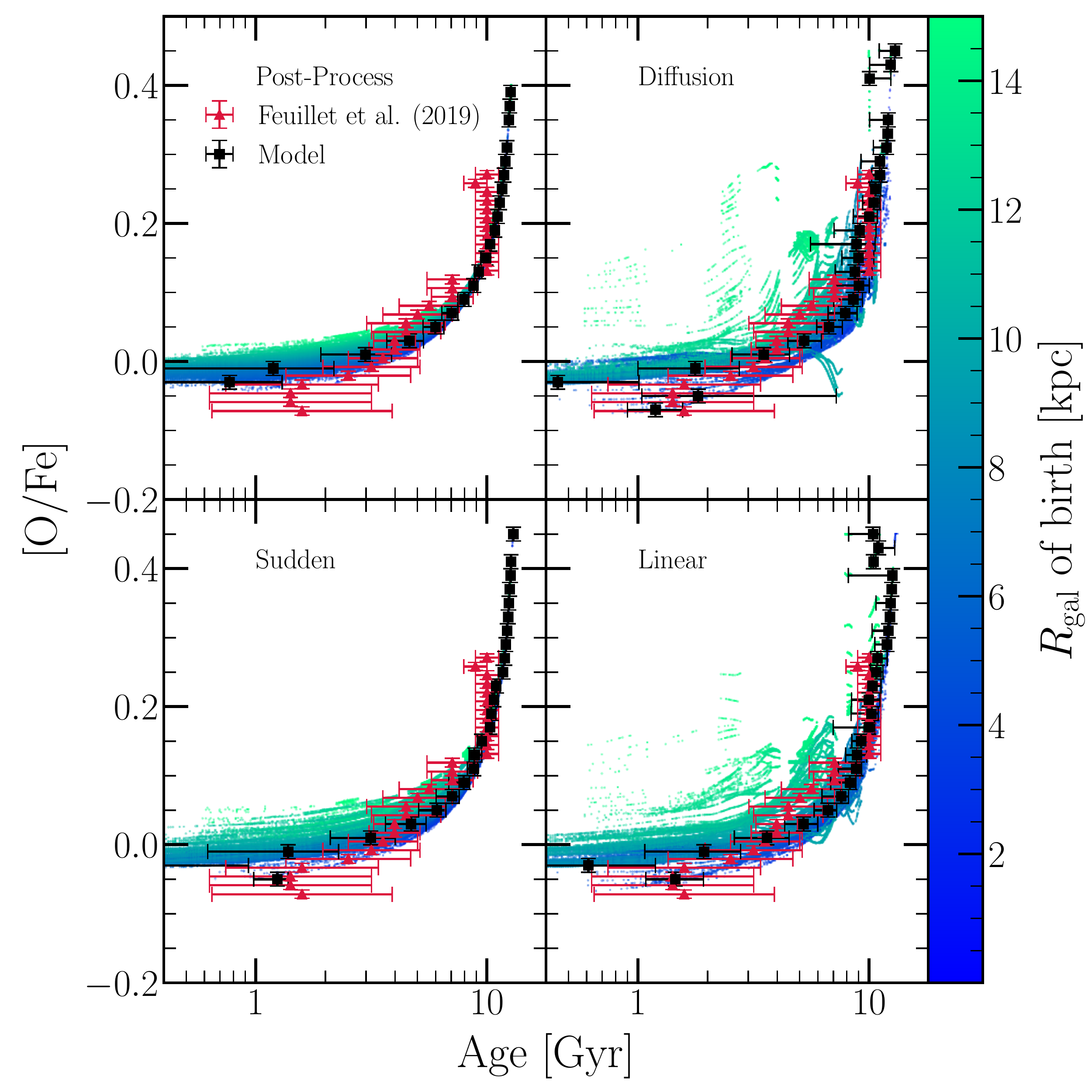} 
\includegraphics[scale = 0.34]{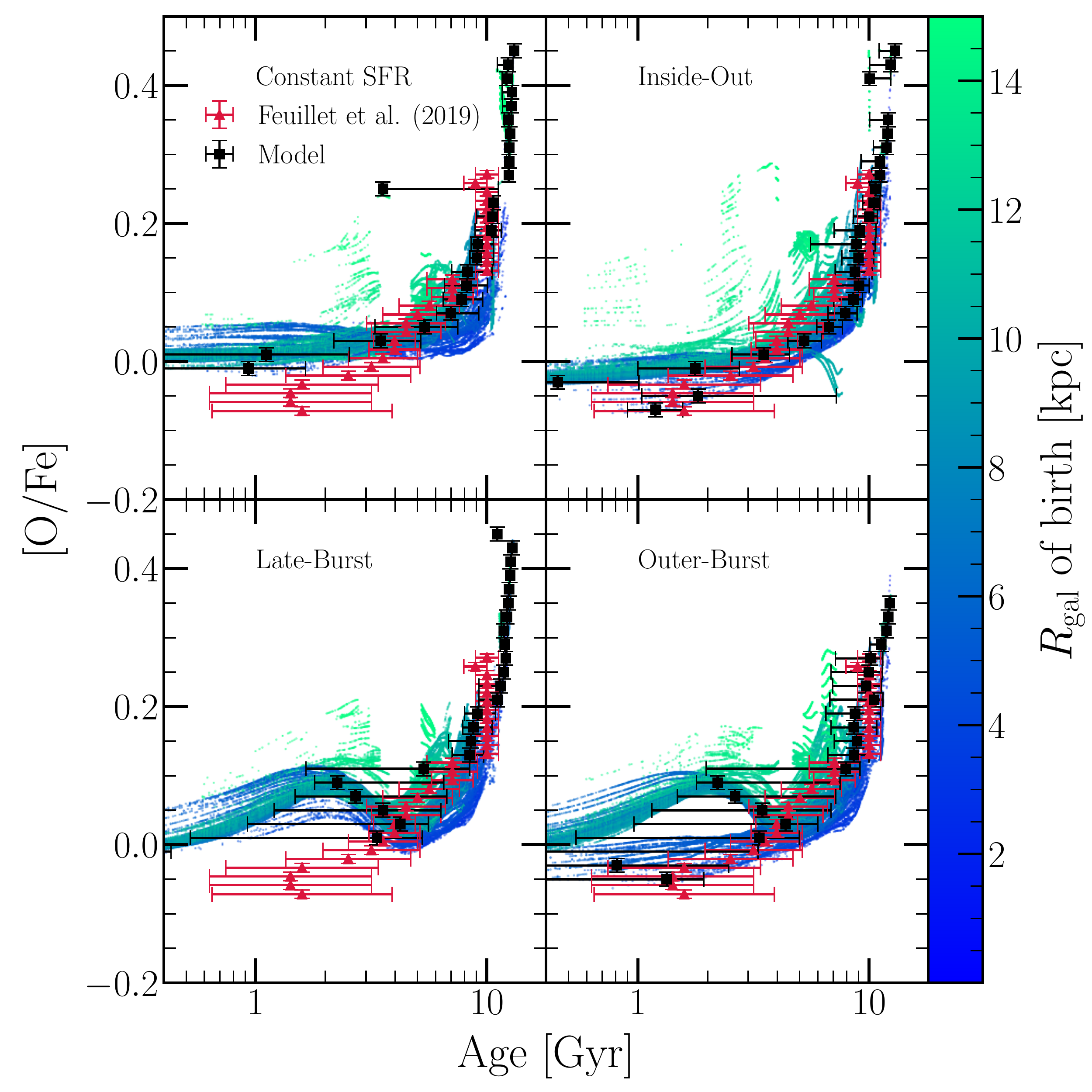} 
\caption{
\textbf{Left}: A comparison of the predicted age-[O/Fe] relation for the solar 
neighbourhood ($R_\text{gal}$ = 7 - 9 kpc and~$\left|z\right|$ = 0 - 0.5 kpc) 
between the post-processing (upper left), diffusion (upper right), sudden 
(lower left), and linear (lower right) migration models, assuming our 
inside-out SFH. 
\textbf{Right}: The same as the left-hand panels, instead comparing the impact 
of our constant (upper left), inside-out (upper right), late-burst (lower left), 
and outer-burst (lower right) SFHs, assuming diffusion migration. In all panels, 
red triangles and error bars denote the observed median age and dispersion 
thereof in bins of [O/Fe] as reported by~\citet{Feuillet2019}; here we include 
only their bins containing at least 15 stars. Black squares denote the 
mass-weighted median age in 0.02-dex bins in [O/Fe] predicted by our models, 
with error bars denoting the 16th and 84th percentiles of the mass-weighted 
age distribution in those bins. Points in the background denote each individual 
stellar population from the model with a final position in the solar 
neighbourhood, colour-coded according to their Galactocentric radius of birth. 
}
\label{fig:age_alpha} 
\end{figure*} 

In the left-hand set of panels of Fig.~\ref{fig:age_alpha}, we compare the 
age-[O/Fe] relation in the solar neighbourhood ($R_\text{gal}$~= 7 - 9 kpc and 
$\left|z\right|\leq$~0.5 kpc)\footnote{
	In order to avoid confusion, we distinguish between solar ``annulus'' and 
	solar ``neighbourhood'' by defining the solar neighbourhood to be the 
	solar annulus population at~\absz~$\leq$~0.5 kpc. This definition should 
	approximate a sample within a spherical radius of~$\sim$0.5 kpc around the 
	Sun, similar to typical observational definitions. 
} predicted by our four migration models to the 
\citet{Feuillet2019} measurements, shown in red triangles. 
We mark the mass-weighted median age in bins of [O/Fe] in each model with black 
squares, and plot for reference in the background each individual stellar 
population in the solar neighbourhood, colour-coded according to its 
Galactocentric radius of birth. 
\par 
The median age-[O/Fe] trend is similar for all four migration prescriptions, 
and the model prediction is in reasonable agreement with the APOGEE 
measurements. 
From the colour-coding in the post-processing model, one can see that the 
predicted relation is insensitive to the chemical evolution parameters across 
the range represented in our Galactic radial zones, differing only in the 
precise value of [O/Fe] reached at late times. 
Nonetheless, these differences are large enough that radial mixing causes a 
large spread in age at low values of [O/Fe], where the median trend itself 
becomes shallow. 
Nearly all high [O/Fe] stars are old. 
The spread is similar, but not identical, among the four migration 
prescriptions, and it is in reasonable agreement with the observationally 
inferred spread. 
This agreement is a significant success of the migration predicted by the
\hsim~simulation in concert with our GCE model. 
% The observed [O/Fe] distribution extends to values below -0.05 that are not 
% present in our model. This difference could reflect missing stochasticity in 
% our model, or slightly incorrect yield or SFH choices, or observational errors 
% in some [O/Fe] measurements. 
\par 
Although their median trends and characteristic spreads are similar, the 
diffusion and linear migration models show a marked diffrence from the 
post-processing and sudden models, predicting populations of young 
($\lesssim$ 4 Gyr) and intermediate age (4 - 7 Gyr)~$\alpha$-enhanced stars 
([O/Fe]~$\approx$ +0.1 - 0.2) in the solar neighbourhood, which formed at large 
\rgal~with [O/Fe] values well above the main trend for their age. 
They arise from the large fluctuations in the SN Ia rate at large~\rgal~shown 
in Fig.~\ref{fig:tracks} and discussed in~\S~\ref{sec:obs_comp:gradient}, 
which occur when stellar populations migrate to different radii before 
producing most of their SNe Ia. 
When the SN Ia Fe production rate fluctuates downward, [O/Fe] fluctuates 
upward in the ISM, and the stars that form from it inherit such a composition. 
These stars then migrate to the solar annulus. 
The presence of young and intermediate-age~$\alpha$-enhanced stars in APOGEE 
has been demonstrated using ages based on carbon-to-nitrogen ratios 
\citep{Martig2016} calibrated against the asteroseismic ages of the APOKASC 
catalog~\citep{Pinsonneault2014}, and with the asteroseismic ages directly 
\citep{SilvaAguirre2018}. 
\par 
\citet{SilvaAguirre2018} demonstrate that these stars have kinematics similar 
to the rest of the high-$\alpha$ population, and they suggest that they are 
in fact old stars ``rejuvenated'' by mergers or mass transfer events. 
In a sample of 51 young,~$\alpha$-rich red giants,~\citet{Hekker2019} 
demonstrate that a portion of these stars have carbon-to-nitrogen ratios 
consistent with mass transfer events, but that others do not, indicating that 
they are either truly young stars or the result of mergers on the main 
sequence. 
\citet{Weinberg2017} and~\citet{Johnson2020} proposed that young~$\alpha$-rich 
stars could arise in bursts of star formation that enhance the rate of CCSN 
enrichment. The explanation suggested here is in some sense the opposite: the 
stars are not so much~$\alpha$-rich as they are Fe-poor because of the 
deficit in SNe Ia events. 
In our diffusion model,~$\sim$0.2\% of solar neighbourhood stars with 0.1~$\leq$ 
[O/Fe]~$\leq$ 0.2 have ages below 4 Gyr, but~$\sim$13.9\% have ages below 7 Gyr. 
\par 
The right panels of Fig.~\ref{fig:age_alpha} compare the model predictions of 
our four different SFHs, all assuming the diffusion migration prescription, 
with the same plotting scheme and colour-coding as in the left panels. 
The predicted trend for the constant SFR model is similar to that of the 
inside-out model, but the agreement with data is slightly worse because with a 
constant SFR the evolutionary tracks do not extend below solar [O/Fe]. 
The starburst models, on the other hand, predict a 0.05 - 0.1 dex upward 
fluctuation in [O/Fe] at ages of 1 - 3 Gyr due to the perturbed ratio of 
core-collapse to Type Ia supernova rates~\citep{Johnson2020}. 
In the outer-burst model, stars formed at~\rgal~> 6 kpc follow this perturbed 
track while stars from the inner Galaxy follow the original track. 
These models are motivated by observational results which provide empirical 
evidence for elevated recent star formation~\citep{Mor2019, Isern2019}. 
However, in conjunction with our chemical evolution prescriptions they produce 
clear disagreement with the age-[O/Fe] distributions of~\citet{Feuillet2019} 
and~\citet{Miglio2021}. 
This disagreement could indicate that the starbursts were more limited in 
Galactic radius than we have assumed, or that some other ingredient is missing 
from our models. 
The discrepancy in these panels also implies that if young~$\alpha$-rich stars 
do arise in a starburst, then it must remain sufficiently localized that the 
resultant stellar populations remain outliers from an otherwise monotonic 
age-[$\alpha$/Fe] relation. 

% fig 14 
\begin{figure*} 
\centering 
\includegraphics[scale = 0.32]{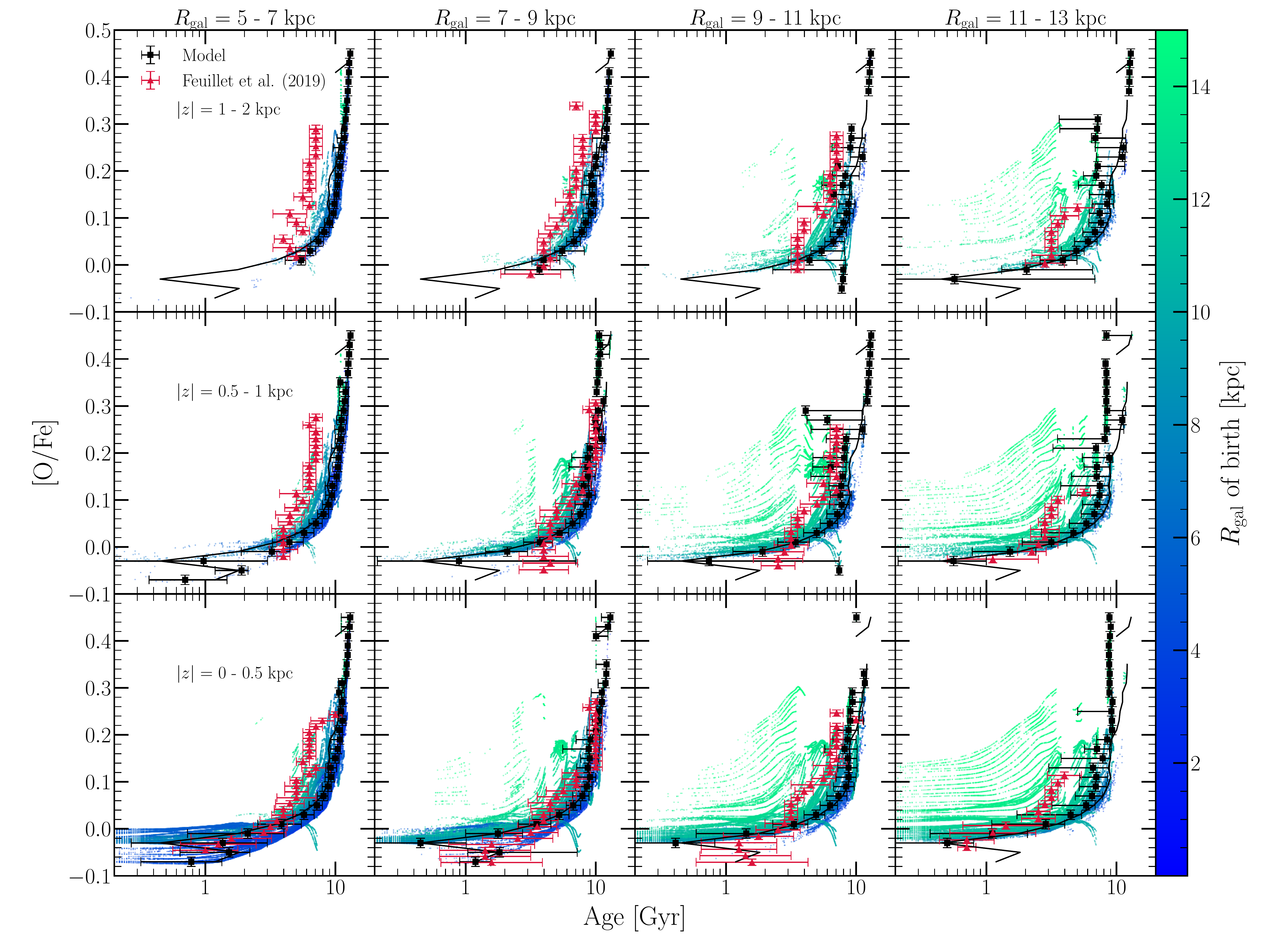} 
\caption{The age-[O/Fe] relation in 12 Galactic regions predicted by our 
inside-out SFH. Bins in Galactocentric radius are shown in columns, and labeled 
at the top. 
Bins in the disc midplane distance~\absz~are shown in rows, noted in the 
left-hand column. Red triangles, black squares, 
error bars, and background points are as in Fig.~\ref{fig:age_alpha} for the 
corresponding Galactic region. The solid black line connects the black squares 
in the bottom, left-middle panel, and is replicated elsewhere for reference. } 
\label{fig:age_alpha_regions} 
\end{figure*} 

Fig.~\ref{fig:age_alpha_regions} extends the comparison of our baseline, 
inside-out model and the observations of~\citet{Feuillet2019} to other 
Galactic regions. 
The predicted median age-[O/Fe] trend is nearly independent of location, as we 
can see by comparing the black model points in each panel to the black line, 
which shows the solar neighbourhood prediction. 
~\citet{Feuillet2019} report ages for~$\alpha$-rich stars that are younger at 
large~$R_\text{gal}$ and high $\left|z\right|$, though in most cases only by 
$\sim$20\%. 
Our model predicts a larger population of intermediate-age~$\alpha$-rich stars 
at large~\rgal, but this effect is not large enough to reproduce the observed 
trend. 
None of our model variants reproduce this trend of age-[O/Fe] offset with~\absz, 
so if the observational result is correct then it points to a missing 
ingredient in the model. 
As with the overpredicted [O/Fe] ratios at 
small~\rgal~(\S~\ref{sec:obs_comp:ofe_dists}), dynamical stirring of younger 
disc populations to higher~\absz~by the Sagittarius dwarf could play a role in 
resolving this discrepancy. 
However, Fig.~\ref{fig:age_alpha_regions} shows that the model does not have a 
reservoir of younger high-$\alpha$ stars at low-\absz~available to be stirred, 
so some additional change would likely be needed. 
Furthermore, our model predicts the 16th-84th percentile ranges of the age 
distribution at fixed [O/Fe] to increase for [O/Fe]~$\gtrsim$ +0.1 stars, at 
least in part due to the higher frequency of young and intermediate-age 
$\alpha$-rich stars. 
This increase in the intrinsic scatter of the relation is simply a consequence 
of the SN Ia rate variability having a higher amplitude at large~\rgal~(see 
discussion in~\S~\ref{sec:obs_comp:gradient}). 
\par 
In this paper, Fe is the only element that we consider with a delayed 
nucleosynthetic source. 
Nonetheless, we expect stellar migration to induce similar variability and 
scatter for other elements with delayed sources. 
Elements such as strontium (Sr) are produced by the $s$-process in AGB stars on 
a timescale intermediate between CCSN and SN Ia enrichment (see Fig. 5 of 
\citealp{Johnson2020}), and elements with a large contribution from low mass 
AGB stars could trace longer timescales. 
The DTD for $r$-process elements such as europium (Eu) is uncertain because the 
relative importance of prompt sources (e.g., collapsars) and delayed sources 
(e.g., neutron star mergers) remains poorly constrained~\citep{Cote2019, 
Mishenina2019, Siegel2019, Vincenzo2021b}. 
Using hydrodynamical simulations from the Auriga project~\citep{Grand2017}, 
\citet{vandeVoort2020} indeed find that the intrinsic scatter in $r$-process 
abundances increases for models with longer characteristic delay times. 
Future comparisons of observed trends between age and abundance ratios with the 
predictions of models like those presented here could better isolate missing 
ingredients of the models, as well as potentially improving understanding of 
the nucleosynthetic sources. 
% \par 
% {\color{red} (The next paragraph probably moving to conclusions) I added a 
% couple sentences to the prior paragraph that, with what is currently stated in 
% the conclusions, may make this redundant. -James}  
% In our models, radial mixing of populations with different evolutionary tracks 
% is the only effect that creates a spread of abundance ([Fe/H] and [O/Fe]) at 
% fixed age at a given final~\rgal. 
% The trend of evolutionary tracks with birth radius is fairly smooth, though 
% the impact of migration on SN Ia rates creates stochastic variations that 
% give rise to the young and intermediate-age~$\alpha$-rich stars in our models. 
% Incomplete mixing of the ISM in azimuth and~\absz~\citep{Krumholz2018b} is an 
% additional source of stochasticity in chemical evolution that is currently not 
% represented in our models. 
% This stochasticity will be larger for elements produced by rare events, such as 
% neutron star mergers, and there have been some recent theoretical 
% investigations of predicted scatter in~$r$-process abundances from this source 
% \citep[e.g.][]{vandeVoort2020}. 
% Observationally, the correlation of residual abundances (i.e., of deviations 
% from mean trends) may be a more sensitive probe of stochasticity and mixing 
% than variance of abundance ratios~\citep{Ting2021}. Characterizing the physics 
% of element mixing in the ISM and the stochastic effects on chemical evolution 
% is an important frontier for models and observations. 

\subsection{The Age-Metallicity Relation} 
\label{sec:obs_comp:amr} 

% fig 15 
\begin{figure} 
\centering 
\includegraphics[scale = 0.35]{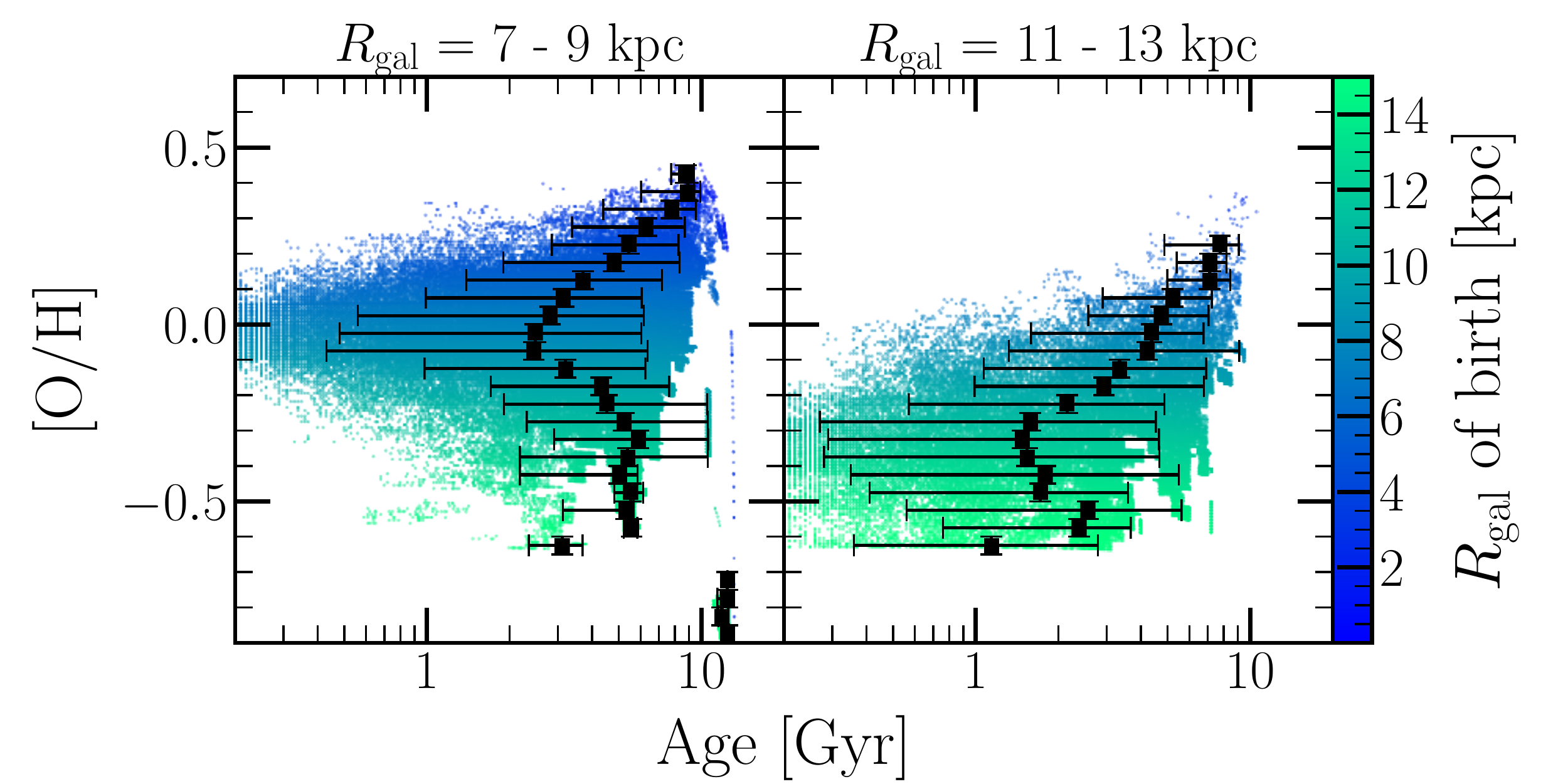} 
\caption{The age-[O/H] relation predicted by our constant SFR model for 
$R_\text{gal}$~= 7 - 9 kpc (left) and 11 - 13 kpc (right). Each panel plots 
only the~$\left|z\right|\leq$~0.5 kpc population. The colored points in the 
background and the black squares with error bars are as in Fig. 
\ref{fig:age_alpha}, but with the model prediction quantified in bins 
of~$\Delta$[O/H] = 0.05. } 
\label{fig:age_oh_static} 
\end{figure} 

Although the AMR is usually formulated in terms of 
[Fe/H], it is also interesting to quantify the age-[O/H] relation because it 
is not affected by SN Ia enrichment. 
The extent to which the two AMRs differ indicates the extent to which the 
delayed timescale and impact of migration on SN Ia enrichment are important in 
shaping the age-[Fe/H] relation. 
Fig.~\ref{fig:age_oh_static} presents the age-[O/H] relation predicted by our 
constant SFR model for the~$\left|z\right|\leq$~0.5 kpc population at 
$R_\text{gal}$ = 7 - 9 and 11 - 13 kpc. 
The black squares with error bars denote the mass-weighted median age as 
in~\S~\ref{sec:obs_comp:age_alpha}, and we again plot the individual stellar 
populations in the background for reference, colour-coded according to their 
Galactocentric radius of birth. 
We omit the~\citet{Feuillet2019} measurements from this figure but present 
observational comparisons for our more empirically motivated inside-out SFH 
model below. 
The constant SFR model allows us to isolate the effects of stellar migration 
from the time-dependent SFR. 
\par 
The intrinsic scatter in the observed AMR has previously been interpreted as 
evidence for radial mixing~\citep{Edvardsson1993, Sellwood2002}. 
In the solar neighbourhood,~\citet{Feuillet2018} demonstrate that the most 
metal-rich stars tend to be significantly older than solar metallicity stars. 
Using the \citet{Weinberg2017} analytic models of one-zone chemical evolution 
coupled to a simplified recipe for radial mixing, they argue that this 
non-monotonic trend is the result of old stars born at small~\rgal~where the 
equilibrium abundance is high. 
Only the old stars are able to migrate to the solar neighbourhood due to the 
time required for such a change in their orbital radius. 
\par 
Fig.~\ref{fig:age_oh_static} demonstrates this behaviour for our 
simulation-motivated radial migration prescription. Although it is 
observationally advantageous to measure age in bins of abundance because the 
abundance measurements are much more precise, it is conceptually easier to 
understand the model behaviour in terms of abundance spread at a given age. 
In both regions plotted, the youngest stars form with a composition inherited 
from the local ISM, which in this model is reflective of the late-time 
equilibrium abundance. 
At a given age, only stars born in a well-defined region of the 
Galaxy will have had adequate time to migrate to a given present-day radius. 
Since a range of~\rgal~maps directly to a range of metallicity once 
the ISM is close to the equilibrium abundance, this necessitates a maximum 
width of the [O/H] distribution at fixed age. 
With increasing age, the [O/H] distribution then gets wider because any given 
present-day radius samples stars formed at a wider range of~\rgal. 
We see this effect in Fig.~\ref{fig:age_oh_static}, with the colour-coding of 
the background points making it clear that stellar migration 
is the culprit. 
In the outer Galaxy, virtually all old stars added by migration lie above the 
local equilibrium metallicity. 
In the~\rgal~= 11 - 13 kpc region, our constant SFR model predicts a nearly 
monotonic increase of median population age with increasing [O/H], from 
[O/H] = -0.5 to +0.3. 
This trend is entirely backwards from that expected by simple one-zone models 
of GCE. 

% fig 16 
\begin{figure} 
\centering 
\includegraphics[scale = 0.45]{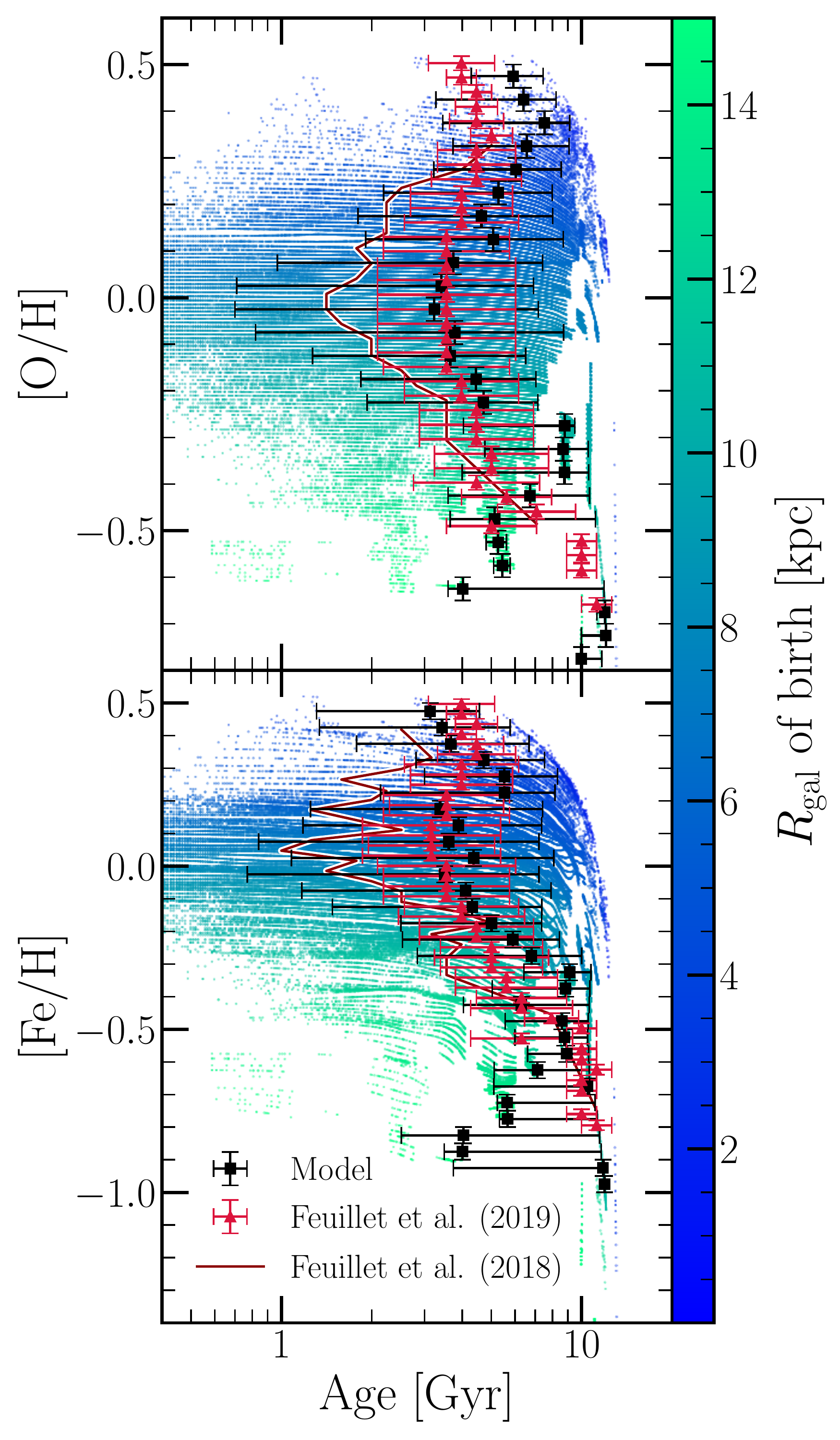} 
\caption{The age-[O/H] (top) and age-[Fe/H] (bottom) relations for the solar 
neighbourhood (i.e.~$R_\text{gal}$ = 7 - 9 kpc,~$\left|z\right|\leq$~0.5 kpc) 
as predicted by the fiducial model. Red triangles, black squares, error bars, 
and background points are as in Fig.~\ref{fig:age_alpha}, but with the model 
prediction quantified in bins of~$\Delta$[O/H] =~$\Delta$[Fe/H] = 0.05. For 
comparison, we plot the~\citet{Feuillet2018} measurements in a dark red line, 
omitting the associated uncertainties for visual clarity. } 
\label{fig:amr_solar_annulus} 
\end{figure} 

Fig.~\ref{fig:amr_solar_annulus} compares the age-[O/H] (top) and 
age-[Fe/H] (bottom) relations predicted by our inside-out SFH model for the 
solar neighbourhood (\rgal~= 7 - 9 kpc and \absz~$\leq$ 0.5 kpc) to the 
measurements from \citet{Feuillet2019}. 
For reference, we add the solid dark red line, which denotes the AMR measured 
by~\citet{Feuillet2018}. 
Although our model shows reasonable agreement with the~\citet{Feuillet2019} 
measurements in the solar annulus,~\citet{Feuillet2018} report ages for solar 
metallicity that are considerably younger ($\sim$1 - 2 Gyr as opposed to 
$\sim$3 - 4 Gyr). 
The origin of this difference is not clear, nor is it clear which AMR 
measurement is more reliable (D. Feuillet, private communication). 
Predictions for the inside-out SFH are similar to those shown previously 
for the constant SFR model, demonstrating that radial migration plays a larger 
role than the detailed SFH in determining the qualitative form of the AMR. 
The age-[O/H] and age-[Fe/H] relations are similar for both observations and 
models, a non-trivial result given the different time profile of SN Ia 
enrichment. 
Agreement between the model and the data is somewhat better for [O/H] than for 
[Fe/H]. 
The model predicts a large spread in age at fixed metallicity, comparable to 
but somewhat larger than the spread esimated by~\citet{Feuillet2019}. 
\par
Fig.~\ref{fig:amr_insideout_vs_lateburst_fe} compares the predictions of the 
inside-out (top) and late-burst (bottom) SFH models to the~\citet{Feuillet2019} 
age-[Fe/H] relation in four~\rgal~annuli, always with~\absz~$\leq$ 0.5 kpc. 
While neither model reproduces the data perfectly, the agreement is better 
overall for the late-burst model, a difference that is more evident from 
looking at all four annuli rather than the solar neighbourhood alone. 
The late-burst of star formation reduces the median age of [Fe/H]~$\approx$ 0 
stars, typically by 1 - 2 Gyr. 
The median age of the highest metallicity stars increases because the 
enhanced pristine gas accretion that fuels the starburst (see Fig. 
\ref{fig:evol}) dilutes the ISM metallicity, an effect that is evident in the 
tracks for a given birth~\rgal~in the lower panels. 
Stars from the inner Galaxy that form during the burst do so at roughly solar 
metallicity rather than appearing in the [Fe/H] = 0.2 - 0.5 bins. 
With reduced median age at [Fe/H]~$\approx$ 0 and increased median at 
[Fe/H] > 0.2, the late-burst model more consistently reproduces the C-shaped 
AMR that is the most striking qualitative feature of the observations. 
The CCSN and SN Ia associated with the burst also boost the evolutionary 
tracks of the inner Galaxy to higher [Fe/H] at late times, so the predicted 
age distribution at high [Fe/H] is bimodal, especially at~\rgal~= 5 - 7 kpc. 
The~\citet{Feuillet2019} results do not support this dichotomy, but they 
model the observations with the assumption of a Gaussian log age distribution, 
so it is worth returning to the data with the possibility of bimodal age 
distributions in mind. 
\par 
The outer-burst model mitigates this issue since it lacks the burst at small 
\rgal, predicting ages for the most metal-rich stars more in line with the 
\citet{Feuillet2019} measurements. 
We illustrate this difference in Fig.~\ref{fig:age_oh_comparison} in 
Appendix~\ref{sec:age_oh_relation} for the age-[O/H] relation. 
The detailed form of the AMR in our models is sensitive to the nature of the 
recent starburst, and its true form was likely more complicated than our 
parameterization. 
Fig.~\ref{fig:age_oh_comparison} also includes a comparison of the inside-out 
and late-burst model predictions, now without the effect of SN Ia enrichment. 
The late-burst model improves upon the inside-out model in age-[O/H] for the 
same reasons as in age-[Fe/H], but perhaps more clearly when looking 
bin-by-bin. 
% We do not believe these conclusions are impacted by further discrepancies 
% between any individual model and the observational sample, because our models 
% are not designed to reproduce the data exactly. 
\par 
The late-burst model is motivated by the star formation histories inferred 
empirically by~\citet{Isern2019} and~\citet{Mor2019}. 
Adding this late time enhancement to star formation improves agreement with the 
\citet{Feuillet2019} data, but it is not enough to reproduce the very young 
(< 2 Gyr) median ages of solar metallicity stars found by~\citet{Feuillet2018}. 
If future analyses confirm the 2018 ages rather than the 2019 ages, they would 
require a later and more extreme starburst to explain them. 
As discussed in~\S~\ref{sec:obs_comp:age_alpha}, the late-burst 
\textit{worsens} agreement with the~\citet{Feuillet2018, Feuillet2019} 
age-[O/Fe] relation because of the increased [O/Fe] at late times (see Fig. 
\ref{fig:age_alpha}). 
We do not see an obvious way to achieve good simultaneous agreement with the 
observationally inferred SFH, the age-[O/Fe] relation, and the age-[Fe/H] 
relation. 
Possibly a well-tuned choice of yields and the temporal and radial range of 
the burst could achieve such agreement; it is possible that the latter is 
related to the spatial dependence of the age-[O/Fe] relation seen in 
the~\citet{Feuillet2019} data (see Fig.~\ref{fig:age_alpha_regions}). 
Alternatively, an outflow prescription that preferentially removes CCSN 
ejecta rather than ambient ISM~\citep[see, e.g.,][]{Vincenzo2016, Chisholm2018, 
Christensen2018} could mitigate the increase in [O/Fe] during the burst (such a 
prescription is physically plausible, but it would require a wholesale 
recalibration of our model parameters). 
For now we conclude that radial migration can naturally explain otherwise 
puzzling features of the AMR and that evidence for enhanced late-time star 
formation in the Milky Way is ambiguous. 
Age-abundance relations are clearly a powerful diagnostic of GCE models, and 
observational investigations that demonstrate consistency across age 
determination methods, probe the shape of the age distribution at fixed 
abundance, and extend still further across the disc are highly desirable.

\section{Conclusions} 
\label{sec:conclusions} 

% fig 17 
\begin{figure*} 
\centering 
\includegraphics[scale = 0.35]{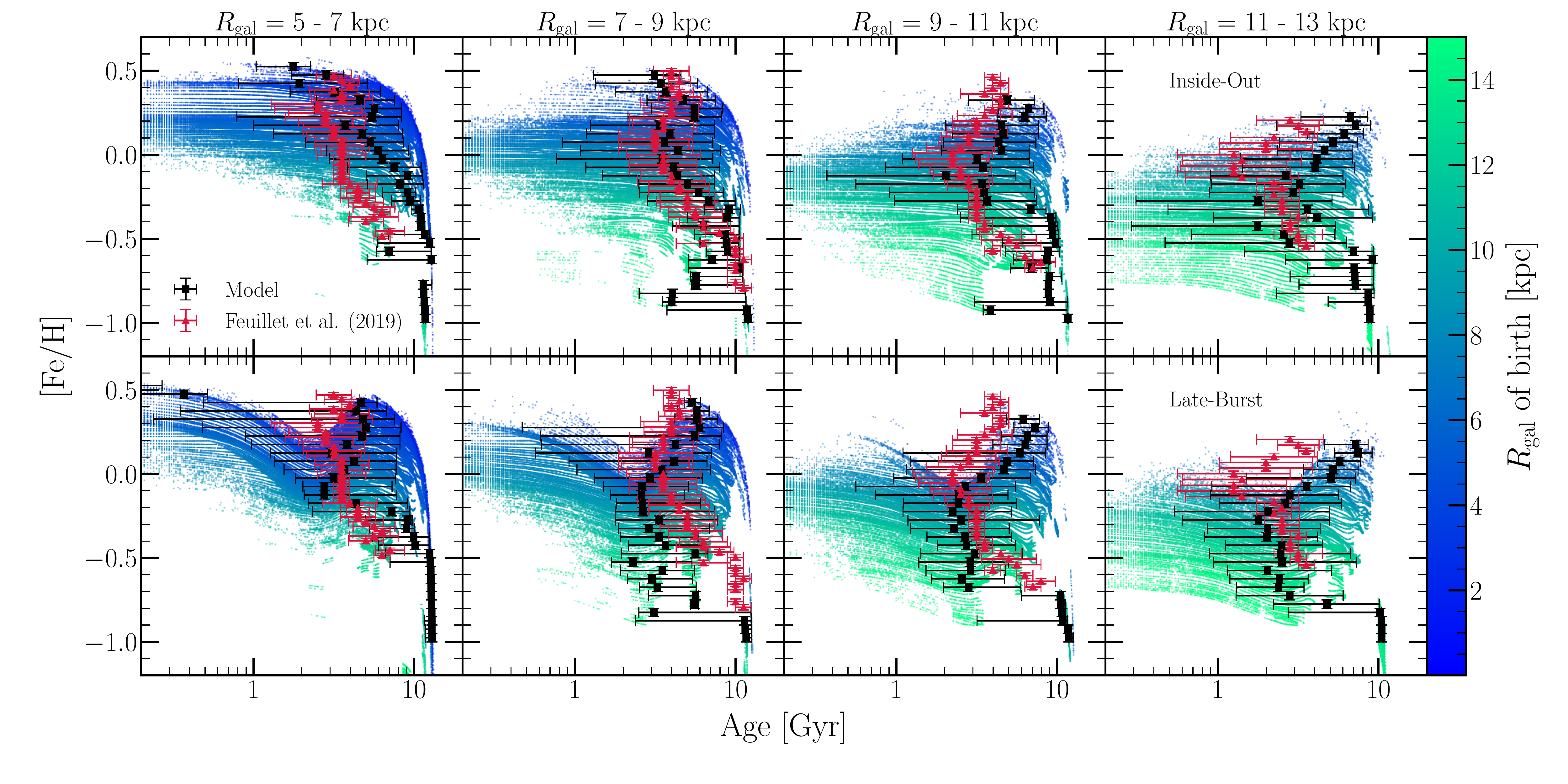} 
\caption{The age-[Fe/H] relation predicted by our inside-out (top) and 
late-burst (bottom) SFH models for~$R_\text{gal}$ = 5 - 7 kpc (left), 7 - 9 kpc 
(left middle), 9 - 11 kpc (right middle), and 11 - 13 kpc (right). Each panel 
shows only the~$\left|z\right|\leq$~0.5 kpc population. Red triangles, black 
squares, error bars, and coloured points are as in Fig.~\ref{fig:age_alpha} for 
the corresponding Galactic region, but with the model prediction quantified in 
bins of~$\Delta$[Fe/H] = 0.05. } 
\label{fig:amr_insideout_vs_lateburst_fe} 
\end{figure*} 

In this paper, we have modeled the Milky Way as a series of concentric rings 
with~$\Delta R_\text{gal}$~= 100 pc width, describing each ring as a 
conventional one-zone chemical evolution model and allowing the exchange of 
stellar populations between zones to include the impact of stellar migration in 
the model Galaxy. 
Though there have been other studies that employ a similar methodology 
\citep[e.g.][]{Schoenrich2009a, Schoenrich2009b, Kubryk2015a, Kubryk2015b, 
Sharma2020}, ours and the~\citet{Minchev2013, Minchev2014, Minchev2017} model 
are the only ones that make use of a hydrodynamical simulation to describe 
radial mixing. 
In our case we use the~\hsim~zoom-in simulation of a Milky Way-like disc galaxy 
formed from cosmological initial conditions~\citep{Christensen2012, 
Zolotov2012, Loebman2012, Brooks2014, Bird2020}. 
The simulation provides a prescription for radial migration and vertical 
structure with no free parameters, though there is still a choice to be made 
about how to model the time dependence of migration for a given stellar 
population (see~\S~\ref{sec:methods:migration}). 
In our fiducial prescription, a stellar population traverses the~$\Delta\rgal$ 
between its birth and final radius with a~$\sqrt{t}$ time-dependence 
characteristic of diffusion, as in~\citet{Frankel2018, Frankel2020}. 
While CCSN nucleosynthetic products are deposited instantaneously in the 
population's birth annulus, its SN Ia iron production follows a~$t^{-1.1}$ 
delay-time distribution and is therefore spread across all the annuli that it 
traverses. 
\par 
We adopt supernova yields from oxygen and iron based on a combination of 
theoretical and empirical constraints. 
We base our star formation law on the observed 
$\dot{\Sigma}_\star - \Sigma_\text{g}$ relation in local spirals 
\citep{Bigiel2010, Leroy2013, Krumholz2018a}, with the redshift dependence 
suggested by the observations of~\citet{Tacconi2018}. 
We choose a radially dependent outflow mass loading factor~$\eta(\rgal)$ to 
reproduce observations of the Galactic disc metallicity 
gradient~\citep[e.g.][]{Frinchaboy2013, Hayden2014, Weinberg2019}. 
Our fiducial model adopts an inside-out SFH with e-folding timescales 
calibrated to the~\citet{Sanchez2020} age gradients of low redshift disc 
galaxies (see discussion in~\S~\ref{sec:methods:sfhs}). 
Motivated by the observational results of~\citet{Mor2019} and~\citet{Isern2019}, 
we construct models that exhibit a recent burst in star formation on top of the 
baseline model, as well as a constant SFH model for comparison purposes. 
\par 
We have compared our fiducial inside-out SFH model and its variants to a 
variety of observations, most of them derived from the SDSS APOGEE survey 
\citep{Majewski2017}, finding a number of qualitative successes but also some 
significant qualitative discrepancies. 
\begin{enumerate} 

	\item[\textbf{1.}] The relative number of high-$\alpha$ and low-$\alpha$ 
	stars and the \feh~distribution of these two populations changes 
	systematically with~\rgal~and~\absz, in qualitative agreement with the 
	findings of~\citet{Nidever2014} and~\citet{Hayden2015}. 
	See Fig.~\ref{fig:ofe_feh_diagram}. 

	\item[\textbf{2.}] The~\feh~and~\oh~distributions of stars near the 
	Galactic plane (\absz~$\leq$ 0.5 kpc) change shape, from negatively skewed 
	at small~\rgal~ to roughly symmetric in the solar neighbourhood to 
	positively skewed in the outer Galaxy, in agreement with the findings of 
	\citet{Hayden2015} and with our new measurements based on APOGEE DR16. 
	The influence of radial migration on MDF shape agrees with the simplified 
	model presented by~\citet{Hayden2015} and with the numerical simulation 
	results of~\citet{Loebman2016}. 
	See Figs.~\ref{fig:mdf_3panel_fe} and~\ref{fig:mdf_3panel_o}. 

	\item[\textbf{3.}] Moving up from the midplane, the~\feh~and~\oh~MDFs 
	become more symmetric and less dependent on~\rgal, again in agreement with 
	the observational findings of~\citet{Hayden2015} and the simulation results 
	of~\citet{Loebman2016}. 
	However, the high~\absz~MDFs are not a perfect match to the APOGEE data, 
	especially at~\absz~= 0.5 - 1 kpc where they remain too skewed and too
	\rgal-dependent. 
	See Figs.~\ref{fig:mdf_3panel_fe} and~\ref{fig:mdf_3panel_o}. 

	\item[\textbf{4.}] The distributions of~\ofe~in bins of~\feh~are broad, and 
	their skewness and width change with~\rgal~and~\absz~in qualitative 
	agreement with the APOGEE-based measurements of~\citet{Vincenzo2021a}. 
	However, the model~\ofe~distributions at sub-solar~\feh~do not reproduce 
	the pronounced bimodality found by~\citet{Vincenzo2021a}, and the 
	centroid of the model~\ofe~distribution at super-solar~\feh~shifts upwards 
	with increasing~\absz, a trend not seen in the data. 
	See Fig.~\ref{fig:ofe_mdfs_insideout}. 

	\item[\textbf{5.}] The trend of median stellar age in bins of~\ofe~agrees 
	with the measurements of~\citet{Feuillet2019} in the solar neighbourhood. 
	The width of the log(age) distribution is narrow at high~\ofe~and broad 
	near solar~\ofe, again in agreement with the data. The model predicts a 
	median age-\ofe~relation that is nearly constant over the range 
	\rgal~= 5 - 13 kpc and~\absz~= 0 - 2 kpc, but~\citet{Feuillet2019} find a 
	$\sim$20\% reduction in the median age of high~\ofe~stars at high~\absz. 
	See Figs.~\ref{fig:age_alpha} and~\ref{fig:age_alpha_regions}. 

	\item[\textbf{6.}] While most stars with~\ofe~$\geq$ 0.1 are old, the model 
	predicts a population of young and intermediate-age $\alpha$-rich stars. 
	These stars form in the outer Galaxy (\rgal~> 10 kpc) at times when the SN 
	Ia rate, and thus the iron enrichment, has fluctuated to low values because 
	stellar populations have migrated away before most of their SN Ia have time 
	to explode (see Fig.~\ref{fig:tracks}). 
	This mechanism, which is only realized because we track SN Ia enrichment 
	through rings as populations migrate (\S~\ref{sec:methods:migration}), 
	offers a novel explanation for the existence of young and intermediate-age 
	$\alpha$-rich stars seen in APOGEE~\citep{Chiappini2015, Martig2015, 
	Martig2016, Warfield2021}. 
	See Figs.~\ref{fig:age_alpha} and~\ref{fig:age_alpha_regions}. 

	\item[\textbf{7.}] In the solar neighbourhood, the predicted distribution 
	of stellar age at solar~\feh~or~\oh~is broad, and the trend of median age 
	with metallicity is non-monotonic, with both sub-solar and super-solar 
	metallicity stars being older on average than solar metallicity stars. 
	These predictions agree with the observational results of 
	\citet{Feuillet2019}, though the age scatter is larger than 
	\citet{Feuillet2019} infer, and the agreement of median trends is better 
	for~\oh~than for~\feh. 
	The old population at super-solar metallicity has migrated from the inner 
	Galaxy, as suggested by~\citet{Feuillet2018, Feuillet2019}, and migration 
	produces a non-monotonic age-metallicity relation throughout the disc. 
	The agreement between predicted and observed age-\feh~realtions is 
	noticeably worse at~\rgal~= 5 - 7 kpc and somewhat worse at~\rgal~= 11 - 13 
	kpc. 
	See Figs.~\ref{fig:age_oh_static} -~\ref{fig:amr_insideout_vs_lateburst_fe}. 

	\item[\textbf{8.}] The models with late bursts of star formation, either 
	throughout the disc or at~\rgal~> 6 kpc only (see Fig.~\ref{fig:evol}), 
	achieve better agreement with the~\citet{Feuillet2019} age-\feh~relation 
	over a range of~\rgal. 
	In particular, these models better reproduce the young median ages of solar 
	metallicity stars and the C-shaped form of the observed relation. 
	However, they predict a~$\sim$0.1-dex uptick of~\ofe~at ages of~$\sim$2 Gyr 
	that is not seen in the data. 
	The SFH of these models is empirically motivated~\citep{Mor2019, Isern2019}, 
	and we do not know if there is some variant of our implementation that 
	would preserve their improved agreement with age-\feh~while mitigating 
	their mismatch to age-\ofe. 
	For the other measures listed above, the predictions of these models are 
	qualitatively similar to those of the inside-out model. 
	See Figs.~\ref{fig:age_alpha_regions} and 
	\ref{fig:amr_insideout_vs_lateburst_fe}. 

\end{enumerate} 

All of these predictions are affected by radial migration, and those involving 
vertical trends also inherit the simulation's predicted correlations between 
final~\absz~and the age, birth radius, and final radius of a stellar 
population. 
We regard the overall level of agreement with many distinctive features of the 
Milky Way disc's abundance structure as a significant success of the models. 
However, at least within the framework explored here, it appears that models 
with smooth star formation are not able to explain the pronounced bimodality 
of the observed~\afe~distribution. As discussed by~\citet{Vincenzo2021a}, 
we expect that this problem is generic: a one-zone model with a smooth SFH 
produces an~\afe~distribution that peaks at low values, so it is difficult 
to create a superposition of such models that has a bimodal distribution. 
\par 
The most widely explored solution to this problem involves a two-phase SFH, 
with gas accretion resetting the ISM to low metallicity in between the two 
epochs. 
Versions of this scenario arise in two-infall GCE models 
\citep[e.g.][]{Chiappini1997, Spitoni2019a, Khoperskov2021} and in cosmological 
simulations that give rise to bimodal~\afe~\citep{Mackereth2017, Grand2018, 
Buck2020b}. 
An alternative scenario proposed by~\citet{Clarke2019}, motivated by 
hydrodynamic simulations, attributes the low-$\alpha$ sequence to an 
evolutionary track with low star formation efficiency and the high-$\alpha$ 
sequence to clumpy bursts of star formation that self-enrich with $\alpha$ 
elements. 
In a third, possibly fanciful scenario proposed by~\citet{Weinberg2017}, 
increased outflow efficiency at late times causes the low-$\alpha$ population 
to evolve ``backwards'' to lower~\feh, after formation of the high-$\alpha$ 
sequence. 
Radial migration is likely to reshape the predictions of any of these scenarios, 
even if it does not fully explain bimodality on its own. 
These scenarios can be easily realized within our modeling framework by 
changing the SFH, star formation efficiency, and outflow parameterizations. 
We intend to explore them in future work, seeking observable signatures that 
can distinguish these alternative explanations for one of the most striking 
features of the disc abundance distribution. 
\par 
The computational speed of our hybrid chemical evolution methodology makes it a 
valuable complement to calculating chemical evolution within full hydrodynamic 
cosmological simulations~\citep[e.g.][]{Mackereth2017, Grand2018, Naiman2018, 
Buck2020b, Vincenzo2020}. 
For a given cosmological simulation, we can consider many different choices of 
yields and chemical evolution parameters, varying them individually to isolate 
physical effects and exploring parameter space to identify good fits, 
degeneracies, and persistent discrepancies with data. 
There are many obvious directions to go in extending this approach. 
One is to apply it to additional cosmological simulations to understand the 
impact of different dynamical histories, and to add the ability to model 
stellar populations accreted from satellites. 
A second is to consider additional elements that probe different 
nucleosynthetic pathways; many of these are already incorporated in~\vice, and 
it is easy to add other elements and sources to models. 
A third is to include treatment of radial gas flows and fountains, both of 
which have been explored in more idealized GCE models (e.g.,~\citealp{Lacey1985, 
Bilitewski2012, Kubryk2015a, Kubryk2015b};~\citealp*{Spitoni2013}; 
\citealp{Pezzulli2016, Sharda2021a, Sharda2021b}). 
More ambitious is to implement treatment of stochastic enrichment and 
incomplete ISM mixing~\citep[e.g.][]{Montes2016, Krumholz2018b, Beniamini2020}, 
which have so far been little explored in the context of Milky Way disc 
evolution but which are likely important in understanding the detailed 
correlations of elemenal abundances~\citep{Ting2021}. 
As multi-element spectroscopic surveys grow even faster in scope and precision, 
efficient and flexible theoretical models will be essential for extracting the 
lessons they have to teach about the origin of elements and the history of the 
Milky Way.

\section{Acknowledgements} 
\label{sec:acknowledgements} 

% fig 18 
\begin{figure*} 
\includegraphics[scale = 0.32]{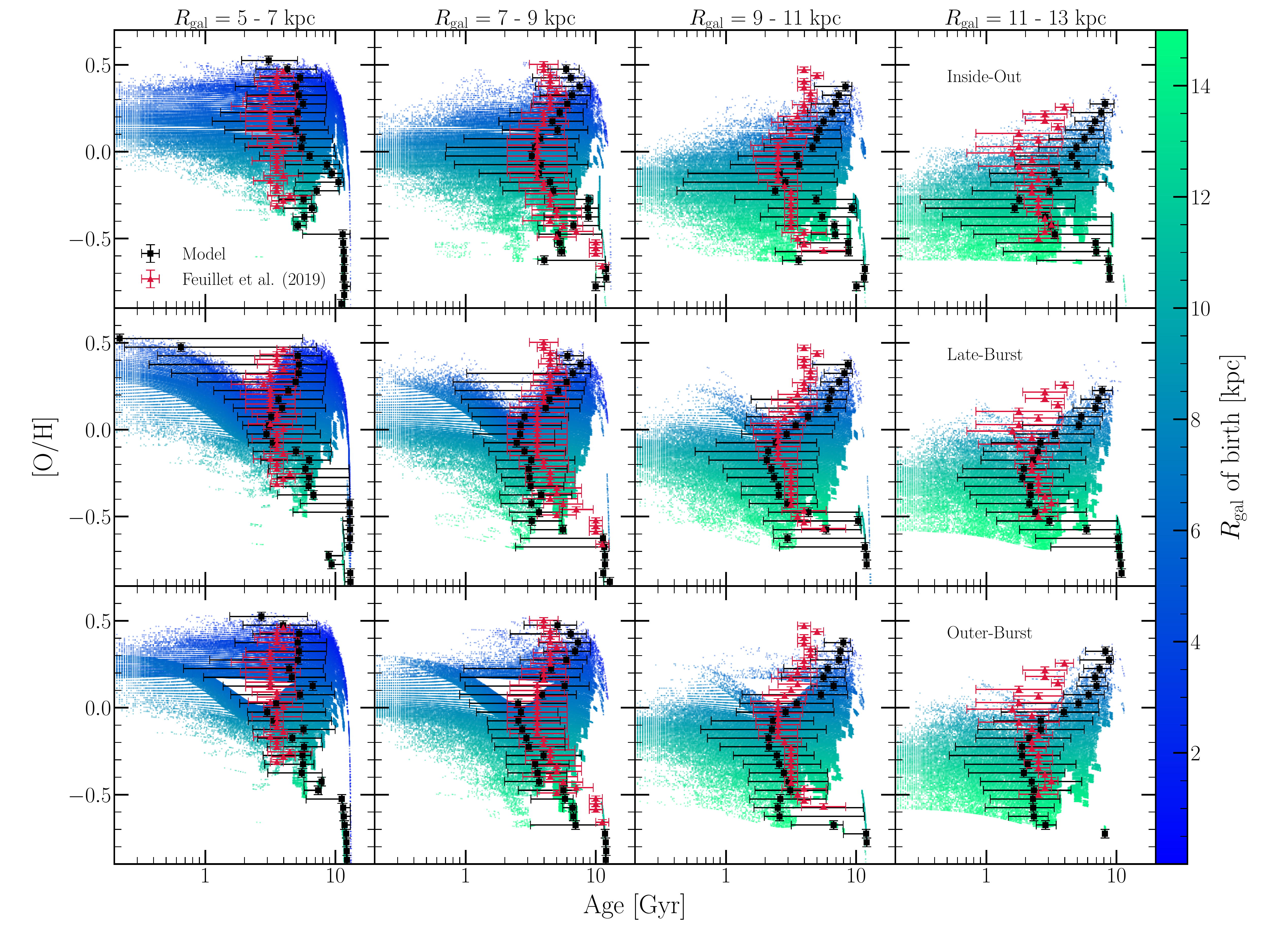} 
\caption{The same as Fig.~\ref{fig:amr_insideout_vs_lateburst_fe}, but for 
\oh, and with an additional comparison to the outer-burst model in the 
bottom row.} 
\label{fig:age_oh_comparison} 
\end{figure*} 

We acknowledge valuable discussion with Jennifer Johnson, Adam Leroy, Grace 
Olivier, Amy Sardone, Jiayi Sun, Todd Thompson, and other members of the Ohio 
State Astronomy Gas, Galaxies, and Feedback group, as well as various 
participants of the SDSS Gotham 2020 virtual collaboration meeting and the 
virtual 236th meeting of the American Astronomical Society. 
We thank Diane Feuillet for providing the results of~\citet{Feuillet2019} in 
digital form. 
This work was supported in part by National Science Foundation grant 
AST-1909841. 
D.H.W. is grateful for the hospitality of the Institute for Advanced Study and 
the support of the W.M. Keck Foundation and the Hendricks Foundation. 
F.V. acknowledges the support of a Fellowship from the Center for Cosmology and 
AstroParticle Physics at the Ohio State University. 
\par 
In this paper we have made use of data from SDSS-IV APOGEE-2 DR16. 
Funding for the Sloan Digital Sky 
Survey IV has been provided by the 
Alfred P. Sloan Foundation, the U.S. 
Department of Energy Office of 
Science, and the Participating 
Institutions. 
\par 
SDSS-IV acknowledges support and 
resources from the Center for High 
Performance Computing  at the 
University of Utah. The SDSS 
website is~\url{www.sdss.org}.
\par 
SDSS-IV is managed by the 
Astrophysical Research Consortium 
for the Participating Institutions 
of the SDSS Collaboration including 
the Brazilian Participation Group, 
the Carnegie Institution for Science, 
Carnegie Mellon University, Center for 
Astrophysics | Harvard \& 
Smithsonian, the Chilean Participation 
Group, the French Participation Group, 
Instituto de Astrof\'isica de 
Canarias, The Johns Hopkins 
University, Kavli Institute for the 
Physics and Mathematics of the 
Universe (IPMU) / University of 
Tokyo, the Korean Participation Group, 
Lawrence Berkeley National Laboratory, 
Leibniz Institut f\"ur Astrophysik 
Potsdam (AIP),  Max-Planck-Institut 
f\"ur Astronomie (MPIA Heidelberg), 
Max-Planck-Institut f\"ur 
Astrophysik (MPA Garching), 
Max-Planck-Institut f\"ur 
Extraterrestrische Physik (MPE), 
National Astronomical Observatories of 
China, New Mexico State University, 
New York University, University of 
Notre Dame, Observat\'ario 
Nacional / MCTI, The Ohio State 
University, Pennsylvania State 
University, Shanghai 
Astronomical Observatory, United 
Kingdom Participation Group, 
Universidad Nacional Aut\'onoma 
de M\'exico, University of Arizona, 
University of Colorado Boulder, 
University of Oxford, University of 
Portsmouth, University of Utah, 
University of Virginia, University 
of Washington, University of 
Wisconsin, Vanderbilt University, 
and Yale University.

\textit{Software}: Matplotlib~\citep{Matplotlib}; Astropy~\citep{Astropy2013, 
Astropy2018}; NumPy~\citep{NumPy}.

\section{Data Availability} 
\label{sec:data_availability} 

\vice~is open-source software, and the code that computes abundances for our 
models is publicly available as well. 
In its GitHub repository, we provide detailed instructions on how to run our 
models and variations thereof; information on their computational overhead and 
other aspects of~\vice~can be found in Appendix~\ref{sec:vice}. 
All observational data appearing in this paper is publicly available, and is 
also included with the source code for our models and figures. 
Our sample of star particles from~\hsim, including those with bulge and 
halo-like kinematics, is available through~\vice, which will download the files 
automatically the first time it needs them. The rest of the data 
from~\texttt{h277} can be accessed at~\url{https://nbody.shop/data.html}.

\begin{appendices}

\section{The Age-\oh~Relation} 
\label{sec:age_oh_relation} 

Fig.~\ref{fig:age_oh_comparison} presents a comparison of the age-\oh~ 
relation predicted by our inside-out, late-burst, and outer-burst SFHs to 
the~\citet{Feuillet2019} measurements in the same Galactic regions as in 
Fig.~\ref{fig:amr_insideout_vs_lateburst_fe}. 
The age-\oh~relation shows a smoother population-averaged trend than the 
age-\feh~relation (see Fig.~\ref{fig:amr_insideout_vs_lateburst_fe}). 
Affected by the variability in Type Ia supernova rates discussed in 
\S~\ref{sec:obs_comp:gradient}, the gas-phase Fe abundance at fixed radius 
fluctuates as a function of simulation time, resulting in higher intrinsic 
scatter in the age-\feh~relation than in age-\oh. 
We can make similar arguments about the age-\oh~relation as we do for 
age-\feh~in~\S~\ref{sec:obs_comp:amr}: the late-burst model better reproduces 
the C-shaped nature of the AMR throughout the disc, particularly beyond the 
solar neighbourhood. 
The bin-by-bin comparison is also somewhat more convincing in age-\oh~than in 
age-\feh. 
The late-burst model improves the agreement in all annuli with the exception of 
the solar annulus, where it potentially worsens the agreement slightly, but 
both models adequately reproduce the data there anyway. 
% \par 
In comparing the late-burst model to the outer-burst model, it is clear the 
outer-burst model mitigates the very young ages of the most metal-rich stars 
at~\rgal~= 5 - 7 kpc seen in the late-burst model; this is a consequence of 
the starburst producing a bimodal age distribution at these abundances (see 
discussion in~\S~\ref{sec:obs_comp:amr}). 
% No model illustrated here reproduces the data perfectly, but we have made 
% simple assumptions regarding the enhancement in the SFR. 
% Although these assumptions are nonetheless motivated by the observations of 
% \citet{Mor2019} and~\citet{Isern2019}, the detailed form of the AMR in these 
% models is sensitive to both the strength and time-dependence of the recent SFH 
% as a function of both Galactocentric radius and time. 
% Ascertaining a best-fit set of parameters to describe the burst would require 
% running many variations of our models, perhaps within a more sophisticated 
% mathematical framework. 

\section{Normalizing the Star Formation History} 
\label{sec:normalize_sfh} 

In this appendix, we derive a prescription for calculating the prefactors of an 
adopted star formation history (SFH) for each annulus in our models. As 
mentioned in~\S~\ref{sec:methods:sfhs}, this procedure requires a unitless 
description of the time-dependence of the SFH in each annulus, denoted 
$f(t|R_\text{gal})$, and a unitless description of the radial dependence 
of the stellar surface density, denoted~$g(R_\text{gal})$. 
By additionally selecting a 
total stellar mass of the present day model Galaxy, the solution to the 
detailed form of the star formation history 
$\dot{\Sigma}_\star(t, R_\text{gal})$ is unique. With this approach, we assume 
that stellar migration does not significantly impact the form of 
$g(R_\text{gal})$, an assumption we demonstrate to be accurate 
in~\S~\ref{sec:methods:surface_density_gradient}. 
\par 
The surface density of star formation with units of mass per area per time can 
be expressed in terms of~$f(t|R_\text{gal})$ as: 
\begin{equation} 
\dot{\Sigma}_\star(t, R_\text{gal}) = \dot{\Sigma}_\star(t = 0, R_\text{gal}) 
f(t|R_\text{gal}) 
\label{eq:sfh_terms_of_f} 
\end{equation} 
and the present-day radial surface density gradient with units of mass per area 
as: 
\begin{equation} 
\Sigma_\star(R_\text{gal}) = \Sigma_\star(R_\text{gal} = 0) g(R_\text{gal}). 
\label{eq:sigma_terms_of_g}  
\end{equation} 
The integral of~$\dot{\Sigma}_\star$ with time should yield the surface density 
gradient at a given radius~$R_\text{gal}$, up to a prefactor accounting for the 
return of stellar envelopes to the interstellar medium (ISM): 
\begin{subequations}\begin{align} 
\Sigma_\star(R_\text{gal}) &= (1 - r)\int_0^T \dot{\Sigma}_\star(t, 
R_\text{gal}) dt 
\\ 
&= (1 - r) \dot{\Sigma}_\star(t = 0, R_\text{gal})\int_0^T f(t|R_\text{gal}) dt 
\\ 
\dot{\Sigma}_\star(t = 0, R_\text{gal}) &= \Sigma_\star(R_\text{gal}) 
\left[(1 - r) \int_0^T f(t|R_\text{gal})dt\right]^{-1} 
\\ 
&= \Sigma_\star(R_\text{gal} = 0)g(R_\text{gal}) \left[(1 - r) \int_0^T 
f(t|R_\text{gal})\right]^{-1} 
\label{eq:intermediate_1} 
\end{align}\end{subequations} 
where~$(1 - r) \approx 0.6$~is an adequate approximation for a 
\citetalias{Kroupa2001} IMF~\citep[][see discussion in 
their~\S~2.2]{Weinberg2017}. 
This expression relates the two unkowns introduced by 
equations~\refp{eq:sfh_terms_of_f} and~\refp{eq:sigma_terms_of_g}. We continue 
by asserting that the integral of the stellar surface density over the area of 
the disc should be equal to the present-day stellar mass of the Milky Way: 
\begin{subequations}\begin{align} 
M_\star^\text{MW} &= \int_0^R \Sigma_\star(R_\text{gal}) 2\pi R_\text{gal} 
dR_\text{gal} 
\\ 
&= \Sigma_\star(R_\text{gal} = 0) \int_0^R g(R_\text{gal}) 2\pi R_\text{gal} 
dR_\text{gal} 
\\ 
\Sigma_\star(R_\text{gal} = 0) &= M_\star^\text{MW} \left[\int_0^R 
g(R_\text{gal}) 2\pi R_\text{gal}dR_\text{gal}\right]^{-1}. 
\label{eq:intermediate_2} 
\end{align}\end{subequations} 
\par 
Now plugging equation~\refp{eq:intermediate_2} into equation 
\refp{eq:intermediate_1}, and then that into equation~\refp{eq:sfh_terms_of_f} 
yields the desired result: 
\begin{subequations}\begin{align} 
&\dot{\Sigma}_\star(t, R_\text{gal}) = Af(t|R_\text{gal})g(R_\text{gal}) 
\\ 
&A = M_\star^\text{MW}\left[(1 - r) \int_0^R g(R_\text{gal})2\pi R_\text{gal} 
dR_\text{gal} \int_0^T f(t|R_\text{gal})dt\right]^{-1}, 
\end{align}\end{subequations} 
where the upper limits should be the maximum radius of star formation and the 
end time of the simulation (15.5 kpc and 13.2 Gyr in this paper). This result 
makes intuitive sense, simply stating that the required normalization of 
$f(t|R_\text{gal})$~is specified by two things: the total stellar mass of the 
Galaxy and how steeply the stellar density falls with increasing radius. As 
long as the assumption that stellar migration does not significantly alter the 
form of~$g(R_\text{gal})$~is not violated, this procedure can be used to 
calculate prefactors for future models of disc galaxies.

\section{\texttt{VICE}} 
\label{sec:vice} 

\vice~is an open-source~\texttt{python} package available for Linux and Mac OS 
X.\footnote{ 
	Install (PyPI): \url{https://pypi.org/project/vice} \\ 
	Documentation: \url{https://vice-astro.readthedocs.io} \\ 
	Source Code: \url{https://github.com/giganano/VICE.git} 
} 
Windows users should install and use~\vice~entirely within the Windows Subsystem 
for Linux. 
It requires~\texttt{python >=}~3.6 and can be installed in a terminal via 
\texttt{pip install vice}, after which~\texttt{vice -{}-docs} will launch a web 
browser to the documentation at \url{https://vice-astro.readthedocs.io}. 
\texttt{vice -{}-tutorial} will also 
launch a web browser, but to a jupyter notebook in the GitHub repository 
intended to familiarize first-time users with~\vice's API. 
\texttt{Python} code that runs the simulations presented in this paper is 
included as supplementary material in the GitHub repository; they can be run 
from a terminal without modifying the source code. 
When taking into account time-dependent stellar migration,~\vice~requires 
$\sim$2.5 CPU-hours per chemical element to compute masses, abundances, and 
initial and final Galactic regions for~$\sim$1.6 million stellar populations 
spanning a little over 1,300 timesteps. 
This estimate was made using a single core with a 3 GHz processor. 
Although their outputs require only~$\sim$235 MB of disc space each, our models 
as computed in this paper can require up to~$\sim$3 GB of RAM at any given 
time owing to the number of timesteps and stellar populations used, but these 
can be adjusted via command line arguments. 
Beyond what has been presented in this paper,~\vice's capabilities include 
user-defined, arbitrary functions of time describing star formation and infall 
histories, star formation laws, outflow prescriptions, SN Ia delay-time 
distributions, and element-by-element infall metallicities. 
It allows the IMF to be a user-defined function of stellar mass. 
It will compute yields from supernova and asymptotic giant branch star 
nucleosynthesis studies, but allows the user to specify arbitrary mathematical 
forms for use in chemical evolution models. 
A complete breakdown of its abilities can be found in the documentation. 
While providing this level of versatility in a~\texttt{python} package, 
\vice~also enjoys a backend implemented in ANSI/ISO~\texttt{C}, providing it 
with the powerful computing speeds of a compiled library; with typical 
parameters, yield calculations and one-zone model integrations require only a 
fraction of a second.

\end{appendices} 

\bibliographystyle{mnras} 
\bibliography{ms} 

\label{lastpage} 
\end{document}